\documentclass[fleqn,usenatbib]{mnras}

\usepackage{newtxtext,newtxmath}

\usepackage[T1]{fontenc}

\DeclareRobustCommand{\VAN}[3]{#2}
\let\VANthebibliography\thebibliography
\def\thebibliography{\DeclareRobustCommand{\VAN}[3]{##3}\VANthebibliography}

\usepackage{graphicx}	
\usepackage{amsmath}	
\usepackage{enumitem}

\usepackage{color,soul}
\usepackage{multirow}
\usepackage{lineno}
\usepackage{threeparttable}

\title[Red quasars show steep radio spectral slopes]{A striking excess of red quasars with steep radio spectral slopes: a dusty blow-out phase revealed through AGN-driven shocks?}

\author[C. L. Sargent et al.]{C. L. Sargent,$^{1}$\thanks{e-mail: ciera.l.sargent@durham.ac.uk} D. M. Alexander,$^{1}$ C. L. Greenwell,$^{1, 2}$ V. A. Fawcett,$^{2,3}$ L. K. Morabito,$^{1,4}$  C. M. Harrison,$^{2}$\and
M. Meenakshi$^{5}$ and R. C. Hickox$^{6}$\\
$^{1}$Centre for Extragalactic Astronomy, Department of Physics, Durham University, Durham, DH1 3LE, UK\\
$^{2}$School of Mathematics, Statistics and Physics, Newcastle University, Newcastle upon Tyne, NE1 7RU, UK\\
$^{3}$European Southern Observatory, Karl-Schwarzschild-Strasse 2, 85748 Garching bei München, Germany\\
$^{4}$Institute for Computational Cosmology, Department of Physics, Durham University, Durham, DH1 3LE, UK\\
$^{5}$Leibniz Institute for Astrophysics, An der Sternwarte 16, D-14482 Potsdam, Germany\\
$^{6}$Department of Physics and Astronomy, Dartmouth College, 6127 Wilder Laboratory, Hanover, NH 03755, USA
}

\date{Accepted XXX. Received YYY; in original form ZZZ}

\pubyear{\the\year{}}

\begin{document}
\label{firstpage}
\pagerange{\pageref{firstpage}--\pageref{lastpage}}
\maketitle
\begin{abstract} Red quasars exhibit a higher incidence of compact (galaxy-scale or smaller) radio emission than blue quasars, arising from systems near the radio-loud/radio-quiet threshold. In this paper we select quasars from SDSS ($0.2 <z <2.4$), and use archival radio data (FIRST, VLASS, LoTSS) to visually determine the radio morphologies of 573 red quasars compared to a control sample of 1278 typical blue quasars. We find an excess of steep-slope radio emission ($\alpha_{1.4-3\text{ GHz}}\sim-1$, where $S_\nu \propto \nu^\alpha$) from red quasars with compact ($<6''$) radio morphologies over 144 MHz, 1.4 GHz, and 3 GHz. This excess steep radio emission signature is not seen in normal blue quasars (radio compact or extended) or red quasars with extended low-frequency radio emission, which instead show a broad range of radio spectral slopes consistent with a range of different physical processes. We show that the strength of the excess steep spectral slope component increases with dust extinction, along with an overall increase in the radio-detection fraction. We argue that this excess steep-slope radio emission is due to shocks between quasar-driven winds/jets and the dusty nuclear-host galaxy environment. The majority ($86^{+5}_{-21}\%$) of the dustiest quasars ($E(B-V)>0.4$ mag) with steep slopes have radio luminosities consistent with the prediction from a wind-shock model with wind efficiencies of up to 7\%. This agrees with the scenario where these compact red quasars are undergoing a “dusty blow-out” phase, where compact jets and/or AGN-driven winds interact with a dusty ISM, causing shocks, leading to steep spectral slopes and enhanced radio detection rates.

\end{abstract}




\section{Introduction}
Supermassive black holes (SMBHs), with masses up to billions of times that of our Sun,  reside at the centres of nearly all massive galaxies in the local Universe \citep{Kormendy2013CoevolutionGalaxies}. These black holes grow through the accretion of cold gas from their host galaxies and, during periods of intense accretion, release vast amounts of energy and are observed as Active Galactic Nuclei (AGN; \citealp{Alexander2012WhatHoles,Alexander2025WhatProgress}). In extreme cases, AGN can become so luminous that they outshine their host galaxies, emitting radiation across the entire electromagnetic spectrum. Their emission is characterised by a distinctive spectral energy distribution (SED), which enables their identification through various techniques across a wide wavelength range \citep{Padovani2017ActiveName,Hickox2018ObscuredNuclei}. This extreme energy output from AGN is believed to exert a significant influence on the evolution of their host galaxies through a process known as AGN feedback. This AGN feedback can manifest in several ways. Radiation pressure can act directly on the surrounding gas; powerful winds launched by intense accretion-disc radiation can shock gas and drive outflows across the host galaxy; and relativistic jets can inject energy into the interstellar medium, regulate star formation, and even impact gas on scales larger than the host galaxy \citep{Fabian2012ObservationalFeedback,Harrison2024ObservationalInterpretation}. All forms of AGN feedback are expected to play a significant role in redistributing, heating, or expelling the host galaxy gas. Therefore, understanding AGN is essential for developing a comprehensive picture of galaxy formation and evolution.

Quasars, or Quasi-Stellar Objects (QSOs), are some of the most luminous AGN, with bolometric luminosities typically exceeding \( L_{\text{bol}} > 10^{45} \) erg s\(^{-1}\), and represent a phase of rapid black hole growth. Whilst QSOs are normally associated with an unobscured view of the accretion disc, theoretical models suggest that the majority of black hole growth occurs in an obscured phase \citep{Fabian1999TheHoles, Hopkins2006TheUniverse,Ishibashi2016AGN-starburstFeedback,Hickox2018ObscuredNuclei,Blecha2018TheStudy}. In the classical AGN unification model (e.g., \citealp{Antonucci1993UnifiedQuasars.,Urry1995UnifiedNuclei,Netzer2015RevisitingNuclei}), the observed level of obscuration towards a QSO is due to our relative line of sight through a dusty torus that surrounds the black hole and accretion disc. Unobscured, or Type 1 QSOs, are usually optically identified due to a direct view of the accretion disc and broad line region, unobscured by the dusty torus. These unobscured QSOs are most commonly observed to have blue optical colours, a result of the strong accretion disc emission peaking in the ultraviolet. In obscured, or Type 2 QSOs, our line of sight intercepts the torus, blocking the central engine and broad line region, thus making them difficult to identify in the optical due to host galaxy dilution of the continuum and lack of broad emission lines. They can however be identified by narrow optical emission lines or from strong mid-infrared (MIR) emission from the reprocessing of the accretion disc emission by the dusty torus (e.g., \citealp{Reyes2008SpaceQuasars,Andonie2022ASystems}).

Among unobscured/Type 1 QSOs, a small subset have red optical-infrared colours, due to a moderate column of dust along the line of sight, which attenuates the shorter-wavelength emission \citep{Webster1995EvidenceQuasars,Richards2003RedSurvey, Kim2018WhatAnalysis, Fawcett2022FundamentalX-shooter}. These dust-reddened QSOs appear to span a range of reddening levels and thus a variety of optical, near-IR, and MIR selection methods are used to identify them (e.g., \citealp{Banerji2012HeavilyEvolution,Klindt2019FundamentalOrientation,Glikman2022TheRegime})\footnote{In this work we consider optically selected red QSOs which generally have extinction values $E(B-V) \approx 0.05-1$ mag. Assuming an SMC extinction curve, this is an $A_V\sim 0.1-2.7$ mag.}. The nature of this dust, and whether it is on host-galaxy or nuclear scales is debated, with the AGN unification model implying that red QSOs are just typical unobscured/blue QSOs observed at a slight angle through the dusty torus. An alternative scenario, supported by  simulations of galaxy evolution, suggests that red QSOs may represent a short-lived, transitional phase. In this interpretation, red QSOs are undergoing a “dusty blow-out” stage in which feedback from the AGN, through the driving of winds and/or jets, expels the obscuring gas and dust. This process decreases the obscuration over time to ultimately reveal the accretion disc, and hence a bluer QSO, and may even quench star formation within the host galaxy \citep{Sanders1988UltraluminousQuasars,Hopkins2006TheUniverse,Hopkins2008AActivity,Alexander2025WhatProgress}.

A key piece of empirical evidence that red QSOs are not just typical blue QSOs observed at a slight angle comes from their radio properties. Over the last few years, many studies covering a range of radio luminosities, frequencies, and spatial resolutions (e.g., \citealp{Klindt2019FundamentalOrientation,Fawcett2020FundamentalQuasars,Rosario2020FundamentalLoTSS,Rivera2021TheReddening,Glikman2022TheRegime,Fawcett2023AQSOs,Petley2024HowRate,Yue2024AQuasars}) have found that red QSOs show a factor of $2-3$ times higher radio detection fractions compared to equally luminous normal QSOs at the same redshifts, which is not predicted by the standard orientation model. Since radio emission is not attenuated by dust, differences in the radio emission between these two populations could not only be due to differences in dust along the line of sight. In fact, on the basis of the standard orientation model, an enhancement of radio emission would be expected from the blue QSOs (i.e., the opposite to the obtained result) due to Doppler boosting of the jet emission from the face-on systems. It is therefore difficult to reconcile the enhanced radio emission in red QSOs with an orientation model.

Key recent work has found an intrinsic link between the amount of line-of-sight dust extinction and the radio emission in QSOs, suggesting a causal connection between the dust and radio emission, which could be due to low-powered jets or winds causing shocks in a dusty environment \citep{Fawcett2023AQSOs,rivera24,Petley2024HowRate}. The enhanced radio emission observed in red QSOs has been found to predominately originate on scales within the host galaxy \citep{Fawcett2020FundamentalQuasars,Rosario2021FundamentalE-MERLIN} and is primarily associated with systems near the radio-loud/radio-quiet threshold (i.e., moderate radio luminosities relative to the AGN; \citealp{Klindt2019FundamentalOrientation,Fawcett2020FundamentalQuasars,Rosario2020FundamentalLoTSS}).\footnote{\label{foot:radioR}Radio loudness is the ratio of the radio power of the quasar to the intrinsic accretion power. This is defined in \citet{Klindt2019FundamentalOrientation} as $\mathcal{R}= \log_{10}(L_{1.4\text{ GHz}}/L_{6{\mu \text{m}}})$, with the radio-loud/radio-quiet threshold defined as $\mathcal{R}=-4.2$, which is the definition we take in this work. \citet{Klindt2019FundamentalOrientation} demonstrated that this is equivalent to the $\mathcal{R}= \frac{L_{5\text{ GHz}}}{L_{4400\text{\AA}}}>10$ cut that is classically used \citep{Kellermann1989VLASurvey}.} 

Using sub-arcsec resolution e-MERLIN observations of QSOs with unresolved radio emission from FIRST, \citet{Rosario2021FundamentalE-MERLIN} found that the majority of the high-frequency radio emission from both red and blue QSOs occurs on $<2$ kpc scales, but that the red QSOs also show an additional enhancement of extended kpc-scale radio emission over host-galaxy scales. This excess diminishes on larger scales ($>$\,10\,kpc), suggesting large-scale, powerful radio jets are not responsible for the radio enhancement in red QSOs. However, we note that the scale of the radio excess may depend on the frequency of the radio emission, with \citet{rivera24} and Sweijen et al. (\textit{in prep.}) finding that the majority of the low-frequency radio emission from red QSOs occurs over host-galaxy--halo scales ($\sim2-80$ kpc), which may be due to the greater sensitivity of the low-frequency radio emission to older electron populations, and therefore potentially to past episodes of activity (i.e., relic emission; \citealp{Morganti2024WhatSurveys}). A suite of analyses have found no significant differences in the star-formation properties between red and blue QSOs \citep{Fawcett2020FundamentalQuasars,Rosario2020FundamentalLoTSS,Rivera2021TheReddening,Andonie2022ASystems,Yue2024AQuasars}, suggesting the excess radio emission from red QSOs is likely due to AGN-related processes, such as low-power jets or shocks from radiatively driven winds. However, the exact origin of this radio emission -whether dominated by jets, winds, or a combination of both- remains unclear.

One approach to constrain the physical processes driving the radio excess in red QSOs is by analysing their radio SEDs, particularly the spectral shape. In QSOs, radio emission typically arises from synchrotron radiation produced by relativistic electrons in magnetic fields, following a power-law spectrum $S_\nu \propto \nu^\alpha$, where $\alpha$ is the spectral index, and $S_\nu$ is the flux density at frequency $\nu$. Synchrotron emission may originate from relativistic jets or shocks driven by AGN winds \citep{Zakamska2014QuasarQuasars,Nims2015ObservationalNuclei,Panessa2019TheNuclei,Harrison2024ObservationalInterpretation}. The radio spectral slope provides insight into the emission mechanism: optically thin synchrotron emission yields steep spectra ($\alpha < -0.5$), while flatter or inverted spectra ($\alpha > -0.5$) can indicate absorption processes such as synchrotron self-absorption or free–free absorption, often seen in compact sources \citep{Laor2019WhatQuasars}. Theoretical work suggests AGN-driven winds interacting with the ISM can cause shocks that produce steep slopes (e.g. $\alpha \sim -1$; \citealt{Nims2015ObservationalNuclei,Yamada2024DecipheringNuclei,Xia2025RadioObservations}), while Meenakshi et al. (\textit{in prep.}) further predicts that AGN-driven winds are capable of producing steeper spectral indices than those from jets. Comparing the radio spectral slopes of red QSOs with the typical QSO population can therefore help identify the origin of the excess radio emission from red QSOs.

Previous work by \citet{Fawcett2025ConnectionQSOs}, on the broadband (144 MHz $-$ 3 GHz) radio SEDs of 19 red QSOs and 19 control QSOs, found that dusty QSOs, defined as having $ E(B-V) > 0.1$ mag ($A_V > 0.4$ mag), are more likely to exhibit steep spectral slopes, and are less likely to show peaked SEDs than their non-dusty counterparts. Additionally, a tentative correlation was found between the steepness of the spectral slope and the level of dust extinction, further suggesting that the amount of obscuration is intrinsically linked to the observed radio emission. In this work, we extend this analysis to a significantly larger sample, enabling a statistical population study of both spectral slopes and multi-frequency radio morphologies of red QSOs. Whilst prior studies have primarily focused on radio morphologies at a single frequency (e.g., \citealp{Klindt2019FundamentalOrientation, Rosario2020FundamentalLoTSS}), the extent of the emission can depend on the radio frequency.

In this work, we use data from three wide-area radio surveys over 144 MHz -- 3 GHz: the LOw-Frequency ARray \citep[LOFAR;][]{vanHaarlem2013LOFAR:ARray} Two-metre Sky Survey \citep[LoTSS;][]{LoTSS_catalog,Shimwell2019TheRelease,LoTSSDR2} at 144 MHz, the Very Large Array (VLA) Faint Images of the Radio Sky at Twenty-Centimeters \citep[FIRST;][]{FIRST1} survey at 1.4 GHz, and the VLA Sky Survey \citep[VLASS;][]{Lacy_2020} at 3 GHz. We study the radio spectral slopes of red and typical QSOs, and classify and compare their radio morphologies across these three frequencies. In Section~\ref{sec:method} we describe the datasets utilised,  the sample selection, and our visual inspection approach to classify radio morphologies. In Section~\ref{sec:results} we present our results and in Section~\ref{sec:discussion} we discuss the origin of the radio emission, by drawing on both observational evidence and comparing simulation studies of both jet- and wind-driven mechanisms. Throughout this work we use the convention $S_\nu \propto \nu^\alpha$ to define the spectral index $\alpha$. We assume a standard flat-$\Lambda$CDM cosmology with $H_0 = 70$kms$^{-1}$Mpc$^{-1}$, $\Omega_M = 0.3$, and $\Omega_{\Lambda} = 0.7$ \citep{Aghanim2020PlanckParameters}. When calculating uncertainties on fractions, we report errors as $1\sigma$ binomial uncertainties, following the methodology in \citet{Cameron2011OnApproach}, unless stated otherwise.


\section{Datasets and Methods}
\label{sec:method}
\noindent In this work we explore the differences in the radio morphologies and radio spectral slopes of SDSS optically selected QSOs at $0.2 < z < 2.4$ to explore the origin of the excess radio emission from red QSOs. In Section \ref{sec:datasets} we describe the multi-wavelength datasets utilised, including the optical sample selection in Section \ref{sec:optical}, and our luminosity-redshift matching approach in Section \ref{sec:L6z}. In Section \ref{sec:EBV} we describe a procedure to estimate the $E(B-V)$ values of our sample. The key advance in this paper over our previous work is the multi frequency radio morphology and spectral slope comparisons. In Section \ref{sec:radio_data} we outline the radio surveys used and in Section \ref{sec:selection} we explain our matching and subsample selection procedures. We describe our multi-frequency radio visual inspection and morphology classification process in Section \ref{sec:VI}, which is undertaken with two aims:
\begin{enumerate}[align=parleft,left=0pt,labelindent=0pt,labelsep=0pt]
\item To study the incidence of different radio morphologies at low and high radio frequencies between red QSOs and blue QSOs, with a particular focus on sources that are classified as compact using FIRST at 1.4 GHz, as has been done in previous work, (e.g., \citealp{Klindt2019FundamentalOrientation}), but do not appear compact in either lower or higher frequency radio images.
    \item To define a "radio-compact" sample, i.e., QSOs with unresolved radio emission at all frequencies studied in this paper, for our radio spectral slope analyses. This is to ensure that we are tracing the same component of radio emission at all frequencies.  Previous work has also shown that these radio compact sources are where red QSOs and blue QSOs display the most significant differences in their radio properties \citep{Klindt2019FundamentalOrientation,Fawcett2020FundamentalQuasars,Rosario2020FundamentalLoTSS}.
\end{enumerate}

\subsection{Datasets}
\label{sec:datasets}
\subsubsection{Colour selected QSO sample: optical and infrared data}

\label{sec:optical}
\noindent Since the optical colours of red QSOs are known to be correlated with modest levels of dust extinction along the line of sight \citep{Klindt2019FundamentalOrientation, Fawcett2022FundamentalX-shooter}, selecting QSOs based on their colour (quantified by the difference in magnitude between two photometric bands) can be used as a proxy for the amount of dust reddening. We use the colour-selected QSO sample produced by \citet{Fawcett2020FundamentalQuasars} (based on the QSO selection approach originally developed by \citealp{Klindt2019FundamentalOrientation}) which uses the SDSS DR14 Quasar catalogue  \citep{Paris2018TheRelease} in combination with mid-infrared (MIR) data from the \textit{Wide-Field Infrared Survey Explorer} \citep[\textit{WISE};][]{Wright2010ThePerformance}. Full details on the selection approach can be found in Section 2 of \citet{Fawcett2020FundamentalQuasars}, which we briefly summarise here.

The colour-selected QSO sample was produced by selecting QSOs from the SDSS DR14 Quasar catalogue  with redshifts $0.2<z< 2.4$, and MIR counterparts with a SNR $>2$ in the All-Sky \textit{WISE} Source catalogue  (ALL-WISE) W1, W2 and W3 bands, within a cross-matching radius of $2\farcs7$. This resulted in a full parent sample of 218,\,747 QSOs, each with g$^*$ (4770 \AA) and $i^*$ (7625 \AA) band galactic extinction-corrected photometry, and with MIR data that allows the rest-frame 6 $\mu$m luminosity ($L_{6\rm \mu m}$)\footnote{Strictly, we are using $\lambda L_{6\rm \mu m}$ (erg s$^{-1}$), not the luminosity density $L_{6\rm \mu m}$ (erg s$^{-1}$ {\AA}$^{-1}$), but we will use the shorthand $L_{6\rm \mu m}$ throughout the paper.} to be calculated, which is used as a measure of the intrinsic power of the AGN. 

\begin{figure*}
\centering
\includegraphics[width=0.9\textwidth]{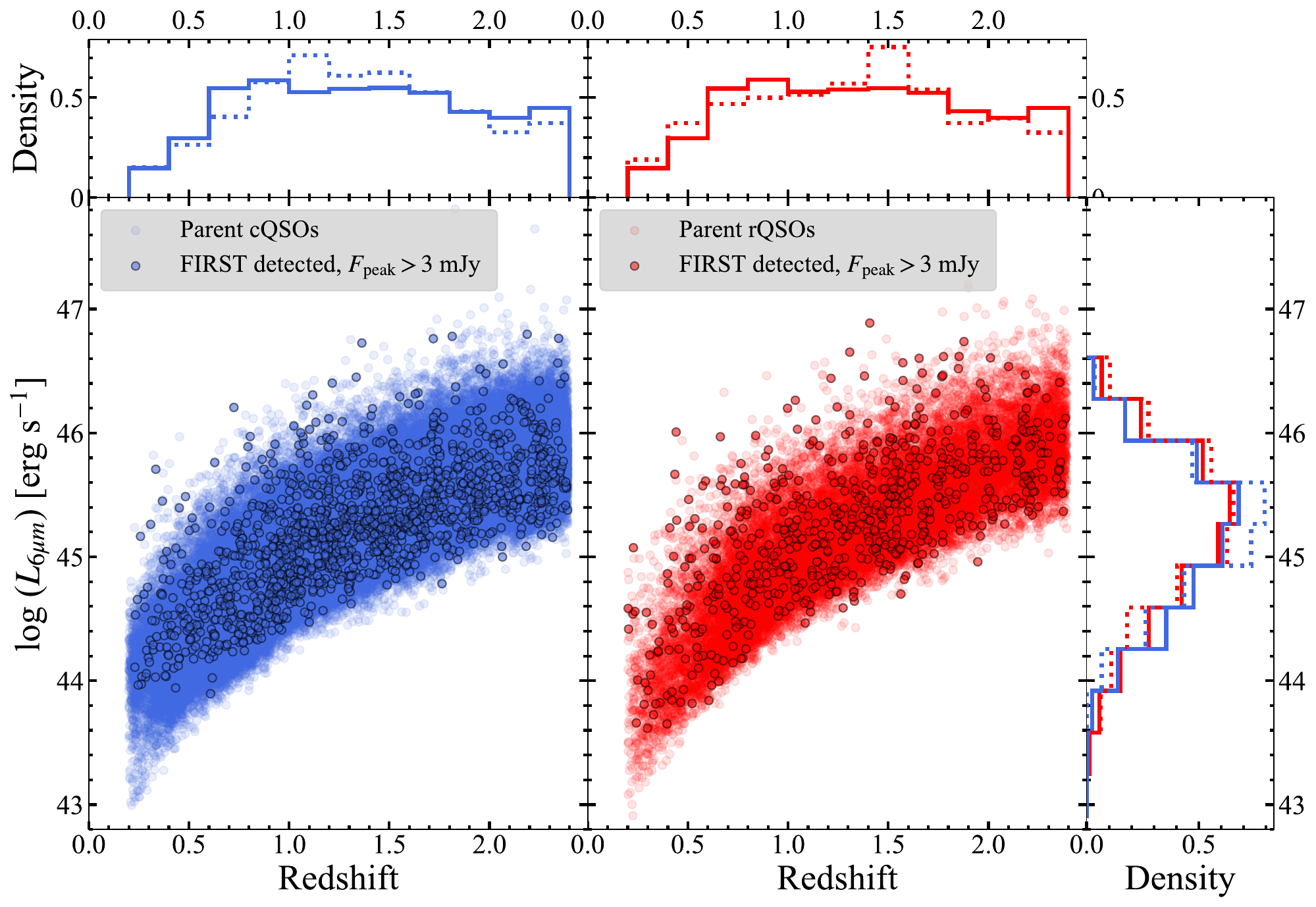}
\caption{Rest-frame 6$\rm \mu m$ luminosity versus redshift for our parent samples of cQSOs (left panel, blue points) and rQSOs (right panel, red points) before $L_{6\mu \text{m}}-z$ matching. Overlaid on both panels (outlined points) are the sources that are detected in the FIRST radio catalogue, with a 1.4 GHz peak flux density, $F_{\text{peak}}>3$ mJy. The solid (dotted) line histograms indicate the distributions in 6$\rm \mu m$ luminosity and redshift for the parent (radio detected) samples.}
\label{fig:L6}
\end{figure*}

For the majority of the analysis in this work (Sections~\ref{sec:morph_results} - \ref{sec:slopes}), we compare red QSOs (rQSOs) and a control sample of QSOs (cQSOs). To do this, the colour-selected QSOs are sorted into contiguous redshift bins of 1000 sources. Within each bin the QSOs with $g^* - i^*$ colours above the 90th percentile of the observed SDSS $g^* - i^*$ distribution are defined as the red QSOs (rQSOs) and those within the middle 50th percentile as the control QSOs (cQSOs), which represent a sample of typical unobscured QSOs. This produces a redshift-sensitive colour selected sample of 218,747 QSOs of which 21,800 are classified as rQSOs and 109,000 are classified as cQSOs. Figure~\ref{fig:L6} shows the $L_{6\mu \text{m}}$ as a function of redshift for the rQSO and cQSO parent samples. The advantage of this approach, where the control and red QSOs are drawn from the same parent sample, is that we can directly compare their radio properties with minimal selection biases.

\subsubsection{Luminosity-redshift matching}
\label{sec:L6z}
To ensure any differences observed in the radio properties of the rQSOs and cQSOs are not driven by differences in bolometric luminosity or redshift, we match our rQSOs and cQSOs in rest-frame $L_{6\mu \text{m}}$ and redshift. Previous work (e.g., \citealp{Klindt2019FundamentalOrientation,Rosario2020FundamentalLoTSS}) has found no significant differences in the results between $L_{6\mu \text{m}}-z$ matched samples and unmatched samples, implying that luminosity and redshift are not significant drivers of the differences in the radio properties of rQSOs and cQSOs.

Nevertheless, as a check, we follow a similar procedure, where for each rQSO we randomly choose two unique cQSOs within a redshift tolerance of 0.05 and a $L_{6\mu \text{m}}$ tolerance of 0.1 dex, and remove any rQSOs without matched cQSOs. Since this is a Monte Carlo process, we repeat this procedure for 500 independent iterations to minimise the impact of random variance on the results. In all iterations, $>99\%$ of the rQSO parent sample is matched to two unique cQSOs. Unless stated otherwise, the average results for our $L_{6\mu \text{m}}-z$ matched samples are consistent with the results of our main (unmatched) samples, and we present the results for the unmatched samples in the rest of this work as they provide better source statistics.

\subsubsection{Dust extinction estimates}
\label{sec:EBV}
Since recent work \citep{Fawcett2023AQSOs,rivera24,Fawcett2025ConnectionQSOs} has shown that the radio properties of QSOs are correlated with the line of sight dust extinction, we also use $E(B-V)$ to quantify the level of dust for some of our analyses (Section~\ref{sec:radio_dust}). We estimate the level of dust extinction by calculating $\Delta(g^*-i^*)$ (the deviation from the median $g^*-i^*$ value in 16 redshift bins) and comparing to the average $E(B-V)$ value from Dark Energy Spectroscopic Instrument (DESI) spectral fitting in 19 $\Delta(g^*-i^*)$ bins.

\citet{Fawcett2023AQSOs} quantified the amount of dust extinction for $\sim34,000$ DESI QSOs by fitting a blue QSO spectral template with varying levels of dust extinction to the individual spectra, following the same method as in \citet{Fawcett2022FundamentalX-shooter} (see Section 3 therein for a full discussion). Overall 11,702 QSOs had a $g^* - i^*$ value\footnote{Since the DESI Legacy Survey DR9 \citep{legacy} utilised did not include $i$-band fluxes, these QSOs were matched to their SDSS counterparts from the DR16 Quasar catalogue  \citep{Lyke2020TheRelease}, in order to calculate a $g^*-i^*$ value and thus a $\Delta(g^*-i^*)$ value, of which 19,368 QSOs had a match. Of these, 11,702 QSOs had a good $E(B-V)$ fit from the DESI spectral fitting.} and a good $E(B-V)$ fit from the DESI spectral fitting, for which the median $E(B-V)$ was calculated in bins of $\Delta(g^*-i^*)$ in increments of 0.2 and $z$ in increments of 0.125, with a minimum requirement of at least 30 sources in the given bin. We thus bin \textit{our} QSO parent samples into the same $\Delta(g^*-i^*)$ and $z$ bins, and take the median $E(B-V)$ value from the DESI spectral fits as an estimate of the $E(B-V)$ for our QSOs. 

To check the robustness of this approach, we match the DESI QSOs to our SDSS QSO parent sample, and find 5,903 direct matches with a good DESI $E(B-V)$ fit. The mean residual, calculated as $E(B-V)_\text{fit}$--estimated $E(B-V)_\text{binning}$, is 0.014, indicating a slight average underestimation of the dust extinction. We find 85\% of residuals are within $^{+0.08}_{-0.03}$ of the median residual and, thus, we adopt a minimum bin width of 0.1 for $E(B-V)$ in our analyses. Based on this analysis, the rQSOs show extinction values of $E(B-V) \sim 0.1 - 0.5$ mag, and the cQSOs show extinction values from $E(B-V) \sim 0 - 0.1$ mag, consistent with previous work \citep{Fawcett2022FundamentalX-shooter,rivera24}.

\subsubsection{Radio data}
\label{sec:radio_data}
\noindent Since the aim of this work is to compare the differences in the radio properties of the population of rQSOs versus cQSOs, we need to retain a relatively large sample of rQSOs and cQSOs to reliably test the differences with good statistics, and therefore we need to use large area radio surveys that cover the SDSS field. Additionally, for our radio spectral slope analysis, the radio data need to be of similar spatial resolutions,  to ensure that we are capturing the integrated radio emission over the same physical size scale, and different observing frequencies. Therefore, we use 1.4 GHz radio data from the Very Large Array (VLA) Faint Images of the Radio Sky at Twenty-Centimeters \citep[FIRST;][]{FIRST1} survey, along with 144 MHz radio data from the LOw-Frequency ARray \citep[LOFAR;][]{vanHaarlem2013LOFAR:ARray}. We also use 3 GHz radio data from the VLA Sky Survey \citep[VLASS;][]{Lacy_2020}, which has a factor $\sim2\times$ higher resolution than FIRST and LOFAR, but for our spectral slope analyses (see Section~\ref{sec:slopes}), we only consider compact radio morphologies in VLASS, since it has been shown that the flux biases introduced from the different resolutions only become significant for extended radio morphologies \citep{Kukreti2023IonisedCycle}.

FIRST was a VLA 1.4 GHz radio survey from $1993-2011$ that observed $\sim10,575~\text{ deg}^2$ of the sky (covering the SDSS region) at a spatial resolution of 5$''$. 
The FIRST catalogue  contains~$946,432$ radio sources above the $\sim$1~mJy detection threshold of the survey and has a typical rms sensitivity of 0.15~mJy. 
We also make use of the FIRST cutout server\footnote{See \href{https://third.ucllnl.org/cgi-bin/firstcutout}{https://third.ucllnl.org/cgi-bin/firstcutout} for the FIRST cut-out server.} when obtaining the images for our visual inspection.

The LOFAR Two-metre Sky Survey \citep[LoTSS;][]{LoTSS_catalog,Shimwell2019TheRelease,LoTSSDR2} is an ongoing project aiming to survey the whole northern sky (above declination, $\delta >0^\circ $) at $120-168$ MHz (referred to as 144 MHz) at a spatial resolution of 6$''$, similar to FIRST. The current data release, LoTSS Data Release 2 \citep[DR2;][]{LoTSSDR2} covers $\sim$27\% of the northern sky, an area of $\sim5,700 \text{ deg}^2$, with a median rms sensitivity of 83~$\mu$Jy/beam (corresponding to 0.017 mJy/beam at 1.4 GHz, and 0.010 mJy/beam at 3 GHz for $\alpha=-0.7$), and so is, equivalently, $\sim 10 \times$ deeper than FIRST and VLASS. The catalogue  derived from the DR2 images contains~$4,396,228$ radio sources\footnote{Additionally, optical identifications of a large fraction of these radio sources has been carried out by \citet{LoTSSopt}, resulting in a catalogue of~$4,116,934$ radio sources in the area with good optical data, of which 85\% have an optical or infrared identification from the DESI Legacy Imaging Surveys \citep{legacy} and/or un\textit{WISE} \citep{unWISE}. We make use of this catalogue when checking for associated sources in our visual inspection (see Appendix~\ref{sec:app_VI}).} which we use when matching to our QSOs. Additionally, we use the LoTSS cutout images\footnote{See \href{https://lofar-surveys.org/cutout$\_$api$\_$details.html}{https://lofar-surveys.org/cutout$\_$api$\_$details.html} for the LoTSS DR2 cut-out server.} when searching for undetected sources and obtaining images for our visual inspection.

VLASS is an ongoing project with the VLA to observe the entire sky above declination, $\delta > -40^\circ$ ($\sim33,885~\text{deg}^2$) at 2$-$4 GHz (referred to as 3 GHz), at a spatial resolution of 2$\farcs5$. The entire region is being surveyed in three epochs (each with a 1$\sigma$ sensitivity of $\sim$0.12 mJy - comparable to FIRST), $\sim$ two years apart, to enable the detection of transient sources. Quick Look (QL) images for all of Epoch 1, observed from 2017 to 2019, and Epoch 2, observed from 2020 to 2022, are now available. A number of issues have been identified with the QL images, including but not limited to a systematic underestimation of the peak\footnote{The peak flux densities are predicted to be underestimated by $\sim$15\% and $\sim$8\% for the first half of Epoch 1 (known as Epoch 1.1) and second half of Epoch 1 (Epoch 1.2), respectively (see \citealp{Lacy_2020} for details).} and total flux densities in Epoch 1. In this work, we make use only of the VLASS Epoch 2 data which does not suffer from the same flux underestimations.

Despite the limitations of the QL images, we can have some confidence in the radio properties when used in combination with the other radio data from the different surveys. Catalogues based on the available QL images have been produced by the Canadian Initiative for Radio Astronomy Data Analysis \citep[CIRADA;][]{gordon2021}, which includes a `Component table' of~$2,995,025$ radio detections at 5$\sigma$ above the image mean, which we use for matching to our QSO samples. As for the other radio surveys, we also make use of the VLASS cut-out service\footnote{See \href{cutouts.cirada.ca}{cutouts.cirada.ca} for the VLASS cut-out server.} when searching for undetected sources and obtaining images for our visual inspection.

\subsection{Radio sample selection}
\label{sec:selection}

\begin{figure*}
\centering
\includegraphics[width=0.8\textwidth]{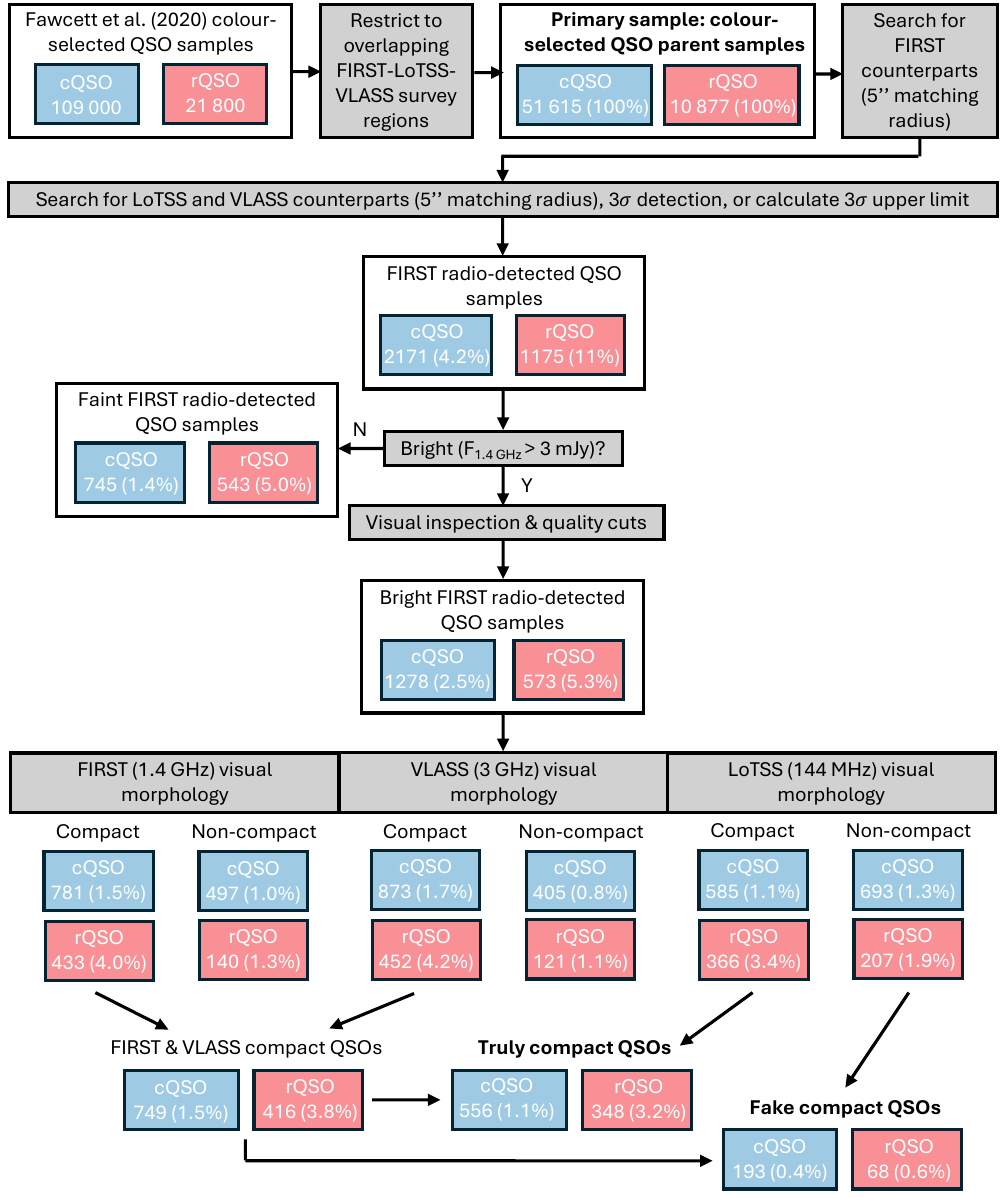}
\caption{A flowchart showing our sample selection process. We start with the colour-selected QSO samples from \citet{Fawcett2020FundamentalQuasars}, and restrict them to QSOs falling within the area covered by all three radio surveys in this work (FIRST, LoTSS, VLASS), to define our colour-selected parent samples. All percentages given are relative to these parent samples. We then search for a FIRST counterpart to our QSOs within 5$''$ of the SDSS position, and then search for VLASS and LoTSS detections also within 5$''$ of the SDSS position, or calculate an upper limit. We then classify QSOs with a 1.4 GHz peak flux density $>3$ mJy as bright, and visually inspect the radio images of these bright FIRST-detected QSOs in each survey, independently classifying each as either compact or non-compact in each of the three radio surveys. A more detailed description of our visual inspection procedure is shown in Appendix~\ref{sec:app_VI}, Figure~\ref{fig:VI_flowchart}. Once visually classified we group our QSOs into (1) those that are compact at low-frequency (144 MHz) and high frequency (1.4 and 3 GHz), defined as the truly compact QSOs, and (2) those that are compact at high-frequency, but non-compact at low-frequency, defined as the fake compact QSOs. These definitions are discussed further in Section~\ref{sec:multi_radio}.}
\label{fig:flowhcart}
\end{figure*}

\noindent Figure \ref{fig:flowhcart} shows a flowchart that summarises the approach taken to define our various subsamples. Starting from the \citet{Fawcett2020FundamentalQuasars} colour-selected QSO samples described in Section \ref{sec:optical}, we initially restrict to QSOs that lie within the overlapping sky region of all three radio surveys to create our colour-selected QSO parent samples. This is done by using multi-order coverage maps (MOCs) from FIRST \citep{moc} and LoTSS\footnote{See \href{https://lofar-surveys.org/public/DR2/catalogue s/dr2-moc.moc}{https://lofar-surveys.org/public/DR2/catalogue s/dr2-moc.moc} for the LoTSS DR2 MOC.} to determine whether a given sky coordinate falls within the given MOC. VLASS covers the SDSS region (and hence our QSO samples) entirely.

We define our radio-detected sample based on FIRST\footnote{The majority of previous red QSO studies have been performed at 1.4 GHz (typically using FIRST e.g., \citealp{Klindt2019FundamentalOrientation,Glikman2022TheRegime}) including demonstrating that the excess radio emission is due to FIRST compact radio sources.}, requiring a FIRST detection (i.e., included in the catalogue of sources above $\sim$\,1~mJy). We perform a cross match between our QSO parent samples and the FIRST catalogue  using a matching radius of 5$''$, resulting in 2171 FIRST radio-detected cQSOs ($4.2 \pm 0.1$\% of parent cQSOs) and 1175 FIRST radio-detected rQSOs ($10.8 \pm 0.3$\% of parent rQSOs). Since FIRST is the least sensitive (but of similar relative sensitivity to VLASS) of the three radio surveys, this approach reduces the number of LoTSS and VLASS flux density upper limits in our sample.

\subsubsection{LoTSS cross-matching}
We then performed a cross match of these FIRST-detected QSOs to the LoTSS DR2 catalogue, again with a search radius of 5$''$ between the SDSS and LoTSS positions, and find 1894/2171 (87\%) of the cQSOs and 1049/1175 (89\%) of the rQSOs have at least one match within this radius. For the undetected/unmatched sources, we performed our own search for 3$\sigma$ LoTSS detections, with the approach outlined in Appendix~\ref{sec:undetected}. Among the QSOs brighter than 3~mJy in FIRST, where we can perform basic morphological analyses, we find that 1324/1426 (93\%) of the cQSOs and 602/632 (95\%) of the rQSOs are also detected in LoTSS at $>3\sigma$ (the majority are $>5\sigma$) significance, with $3\sigma$ upper limits calculated for the remaining 102 cQSOs and 30 rQSOs.

Given the better sensitivity and lower frequency of LoTSS when compared to FIRST, we might expect a higher LoTSS detection rate for our FIRST-detected QSOs. However, due to this increased sensitivity and low frequency, LoTSS contains more extended and diffuse radio sources, which, given our search radius of $5''$, could be missed. The radio position in LoTSS is given by a flux weighted position, therefore an extended radio galaxy with lobes brighter than the core could be missed with this search radius and thus considered "undetected" by our above procedure. These types of sources are identified in (and provides some of the motivation for) our visual inspection procedure outlined in Section~\ref{sec:VI}. 

\subsubsection{VLASS cross-matching}
We repeat the same procedure for VLASS, finding catalogue matches for the FIRST detected QSOs within 5$''$ of the SDSS positions for 1877/2171 (87\%) of cQSOs and 942/1175 (80\%) of rQSOs. The procedure for searching for $3\sigma$ VLASS detections is also outlined in Appendix~\ref{sec:undetected}. Among the QSOs brighter than 3~mJy in FIRST, we find that 1407/1426 (99\%) of the cQSOs and 624/632 (99\%) of the rQSOs are detected in VLASS, with 3$\sigma$ upper limits calculated for the remaining 19 cQSOs and 8 rQSOs.

\subsubsection{Final sample}
\label{sec:final_sample}
Our final sample of radio-detected QSOs consists of 2171 cQSOs and 1175 rQSOs with FIRST $5 \sigma$ detections and LoTSS and VLASS detections or upper limits. Of these, we define those as faint when the FIRST 1.4 GHz peak flux density is less than 3 mJy, resulting in 745 and 543 faint radio-detected rQSOs and cQSOs, respectively; these sources are too faint for reliable visual inspection. The faint sources are also incomplete to the full range of radio spectral slope values (see Appendix~\ref{sec:app_faint}), a key analysis in this work, hence we do not include them in the main analyses. We refer to those above 3 mJy in FIRST as our "bright" sample, which are bright enough to visually inspect and thus form the main sample utilised in this work. 

\begin{figure*}
\centering
\includegraphics[width=1\textwidth]{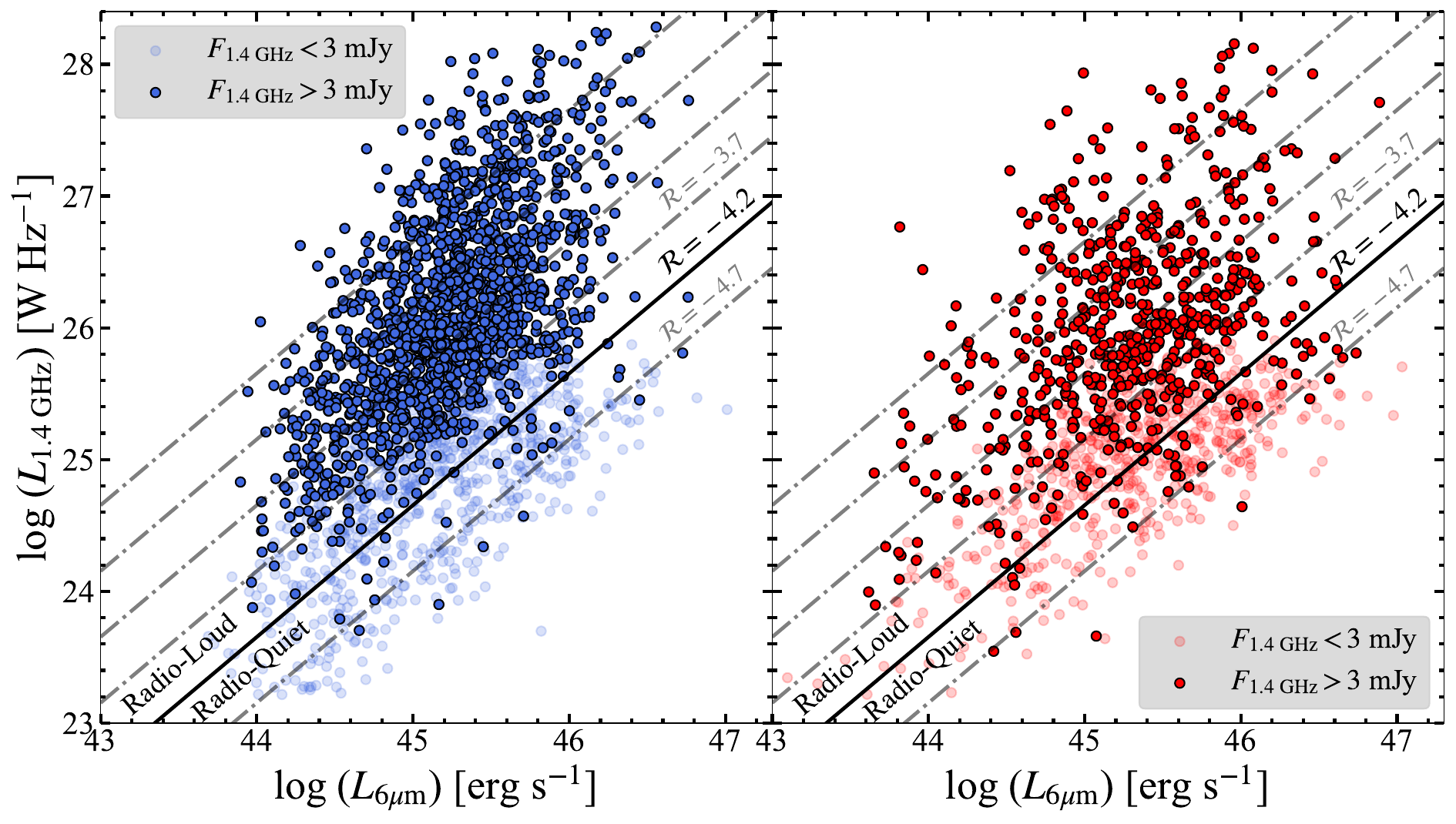}
\caption{The rest-frame 1.4 GHz radio luminosity versus the rest-frame $6\mu$m luminosity for all FIRST-detected cQSOs (left panel) and rQSOs (right panel) in coloured points. The solid filled coloured points indicate all QSOs brighter than 3 mJy in their FIRST 1.4 GHz flux density, which are the QSOs we visually inspect. The dash-dotted lines indicate tracks of radio-loudness ($\mathcal{R}= \log_{10}(L_{1.4\text{ GHz}}/L_{6{\mu \text{m}}})$) values in increments of 0.5, with the solid line indicating the typical radio-loud/radio-quiet divide of $\mathcal{R}=-4.2$.}
\label{fig:Lradio_L6}
\end{figure*}

In Figure~\ref{fig:Lradio_L6} we plot the rest-frame 1.4 GHz radio luminosity vs. the rest-frame 6$\mu$m luminosity for our samples of radio-detected rQSOs and cQSOs. The 1.4 GHz luminosities are calculated from the FIRST 1.4 GHz peak flux density and assuming a radio spectral index of $\alpha = -0.7$\footnote{We check how much the 1.4 GHz luminosities change using instead the  measured $1.4-3$ GHz spectral slopes (see Section~\ref{sec:slopes}). The median offset in the radio luminosity is just an increase of $0.1\%$ for the rQSOs and a decrease of $0.2\%$ for the cQSOs. This equates to an increase of $0.6\%$, and a decrease of $1.1\%$, in the median radio loudness for the rQSOs and cQSOs, respectively.}. The diagonal lines in Figure~\ref{fig:Lradio_L6} denote tracks of constant radio loudness, which we define in the same way as \citet{Klindt2019FundamentalOrientation} with $\mathcal{R} = \log_{10}(L_{1.4 \text{GHz}}/L_{6\mu\text{m}})$. A radio loudness value of $\mathcal{R} \approx -4.2$ is used as the division between a radio-loud and radio-quiet AGN, since it corresponds to a mechanical to radiative power ratio of $\approx 0.1\% $ (\citealp{Klindt2019FundamentalOrientation}, see also Footnote~\ref{foot:radioR}). This figure shows the limitation of our 3 mJy bright-sample cut on probing the lowest $\mathcal{R}$ values.

\subsection{Visual inspection and morphology classification}
\label{sec:VI}
The motivation for our visual inspection comes from a desire to understand the morphology of our sources across multiple radio frequencies, since previous work has found the excess radio emission in red QSOs is predominantly due to compact systems, and also to check any low significance sources for correct associations. We employ a simple classification of our radio sources: compact (unresolved at the resolution of the radio survey), and non-compact (which includes both marginally/slightly extended sources and large extended sources). Given the different frequencies and (slightly) different spatial resolutions between our three radio surveys, we determine the radio morphology of each QSO in each of the three radio images semi-independently: each survey image gets its own classification, but we use information from the other images to inform our overall radio-morphology classification.

For each QSO, we simultaneously viewed the $2' \times 2'$ cutout images in each survey, centered on the QSO optical positions. We used the measured major axes from the catalogues as a guide, where available, and centered a circle with a radius equal to the resolution of the survey on each image, to guide the eye. We classified the source in each image as either:
\begin{itemize}[align=parleft,left=0pt]
    \item \textbf{Compact}: the radio emission is point-like and contained within the radius equal to the resolution of the survey.
    \item \textbf{Non-compact:} the radio emission extends beyond the survey resolution radius.
\end{itemize}
The non-compact sources cover a range of sizes and morphologies, from only marginally extended beyond the beam size, up to very extended sources ($\sim 1$~Mpc; though these are rare). The visual inspection was carried out by four authors (CLS, DMA, CLG, LKM), and the classifications were agreed upon as a group. The visual inspection approach was motivated by that adopted in \citet{Klindt2019FundamentalOrientation}, but with a simpler classification for extended sources, and access to more frequencies. We had no information on whether the QSO was an rQSO or a cQSO throughout the visual inspection process, so as to not influence our classification.

Whilst each QSO is given a classification in each of the three radio images, we use information from each image to influence the classification in certain cases. For example, if we saw extended features in LoTSS that were associated with fainter features (or even individual bright associated sources) in FIRST and/or VLASS then we classify the source as non-compact in FIRST and/or VLASS. An example of this case is shown in Appendix~\ref{sec:app_VI} in the bottom panel of Figure~\ref{fig:VI_example}.

A detailed discussion of the various quality cuts that we applied, can be found in the flowchart and discussion in Appendix~\ref{sec:app_VI}, where we also show examples of visual inspection panels. We did not include sources in our final sample where the morphology classification was uncertain, for example due to another source in the image, or strong imaging artefacts. After the quality cuts, we retain 90\% of the original sample selected for visual inspection, and find no difference in the fractions of rQSOs and cQSOs with various quality cut flags, with 148/1426 ($10.1\%$) of the cQSOs and 59/632 ($9.3\%$) of the rQSOs removed from the sample.

Of the QSOs remaining after the visual inspection and quality cuts, 99 cQSOs and 28 rQSOs were formally LoTSS undetected ($3\sigma$ upper limits) and 18 cQSOs and 8 rQSOs were formally VLASS undetected ($3\sigma$ upper limits). After visual inspection, it became clear the majority of these "undetected" sources were not undetected but were actually extended sources with faint radio cores within 5$''$ of the optical QSO position, but bright and extended lobes clearly associated with the source beyond the 5$''$ that were thus missed in our initial matching. Overall, 96/99 (97\%) of the LoTSS "undetected" cQSOs and 26/28 (93\%) rQSOs were of this type and we classified them as LoTSS non-compact\footnote{\label{foot:LoTSS_no.}The remaining three cQSOs and two rQSOs were too faint to determine their morphologies and were assumed to have compact LoTSS morphologies, based on their compact FIRST morphologies. It is possible they could be classified as extended given a higher sensitivity image, but since the potential contamination is small we classify them as compact.}. Likewise, 11/18 (61\%) of the VLASS "undetected" cQSOs and 3/8 (28\%) rQSOs showed bright extended lobes beyond the 5$''$ matching radius and were classified as VLASS non-compact\footnote{Similar to the LoTSS case (Footnote~\ref{foot:LoTSS_no.}), the remaining seven cQSOs and five rQSOs were given a compact VLASS classification in line with their compact FIRST classification.}. The final sample used in our analyses consists of 1278 cQSOs and 573 rQSOs that are FIRST detected and bright ($F_{1.4\text{GHz}} > 3$ mJy), and have been classified as either compact or non-compact in each of the three frequencies. Of the 573 rQSOs, 571/573 ($>99$\%) were detected in LoTSS, and 568/573 (99\%) were detected in VLASS. Similarly, of the 1278 cQSOs, 1275/1278 ($>99$\%) were detected in LoTSS, and 1271/1278 (99\%) were detected in VLASS.

\section{Results}
\label{sec:results}
Utilising our FIRST-LoTSS-VLASS radio analyses of the SDSS QSOs, we explore the multi-frequency radio properties of dust-reddened QSOs with respect to a control sample of typical QSOs. In Section~\ref{sec:morph_results} we present the radio detection rates of rQSOs and cQSOs across the three radio surveys used. In-line with previous work, we find that rQSOs are consistently detected more often in the radio than cQSOs, and are more likely to have compact radio morphologies across all frequencies. In Section~\ref{sec:multi_radio} we explore differences in the radio morphologies across the three radio frequencies, focusing mostly on QSOs that are compact at the higher radio frequencies but non-compact at the lower frequency of LoTSS, which may provide important clues on the life-cycle stage of the QSO. In Section~\ref{sec:slopes} we take advantage of the multiple radio frequencies available to explore the radio spectral slopes of our different morphological classifications, finding that rQSOs are more likely to have very steep $1.4-3$ GHz radio spectral slopes than cQSOs. Finally in Section~\ref{sec:radio_dust} we consider our results as a function of dust extinction, showing that the radio properties and opacity appear to be intrinsically connected.
\subsection{Radio detection rates of rQSOs and cQSOs}
\label{sec:morph_results}

Previous work has shown that red QSOs are significantly more likely to be detected in the radio than typical blue QSOs, which could indicate different radio emission mechanisms, likely connected to the amount of dust \citep{Fawcett2023AQSOs}. Here, we extend this work by comparing detection rates and morphologies across three different radio frequencies, and present the results of our visual inspection as outlined in Section~\ref{sec:VI}. For FIRST-detected QSOs with \( F_{1.4\text{GHz}} > 3 \, \text{mJy} \), we visually examined the images from the three radio surveys and classified each source as either compact or non-compact: Table~\ref{table:morph_stats} summarizes the numbers and fractions of the rQSO and cQSO parent samples in each classification and the corresponding enhancement in the radio detection rates of rQSOs relative to cQSOs. 

From Table~\ref{table:morph_stats} we can see that the rQSOs with $F_{1.4\text{ GHz}} >3$ mJy show enhancements of factors of $\approx 2.5 - 3.0$ among the compact sources across all three radio surveys\footnote{Among faint radio-detected QSOs with $F_{1.4\text{ GHz}} <3$ mJy, the enhancement of rQSOs over cQSOs at 1.4 GHz is a factor of $3.5^{+0.3}_{-0.3}$, which is in agreement with the findings of \citet{Klindt2019FundamentalOrientation}.}. In contrast, the enhancements for the non-compact sources are nearly consistent with unity (with the rQSOs being only slightly enhanced), suggesting that compact sources in any radio survey are responsible for the majority of the observed radio enhancements in rQSOs. The enhancements observed among the compact sources are consistent across all three surveys, within the uncertainties. This aligns with the findings of \citet{Klindt2019FundamentalOrientation} and \citet{Rosario2020FundamentalLoTSS}, who found enhancements of factors of approximately $\approx 3$ among FIRST compact QSOs and LoTSS compact QSOs, respectively, while observing no significant differences in the most extended classifications. We note however that \citet{Rosario2020FundamentalLoTSS} had an additional (LoTSS) morphological classification of "Marginally Extended" sources (extended over $6''-10''$), where a significant enhancement of rQSOs over cQSOs was observed.

\begin{table*}
\centering
\caption{\raggedright The numbers of QSOs in our colour-selected parent samples, and the number (percentage of parent samples) of QSOs in each morphological classification, for each of the three radio surveys used in this work, and the rQSO enhancement in each. We focus only on QSOs that have passed our quality cuts and are defined as bright (a FIRST 1.4 GHz peak flux $>3$ mJy; see Section~\ref{sec:final_sample}). The compact and non-compact classifications are determined from a visual inspection of the images in each survey (see Section~\ref{sec:VI}). Errors are 1$\sigma$ binomial uncertainties.}
\renewcommand{\arraystretch}{1.4} 
\begin{tabular}{cccc}
\hline
\hline
   \multirow{2}{*}{Classification}         & \multicolumn{2}{c}{Number (\% parent sample)} & \multirow{2}{*}{rQSO enhancement} \\
            & rQSOs                  & cQSO                 &                              \\ \hline \hline
\multicolumn{4}{c}{Colour-selected QSO parent samples (in overlapping radio region)}                             \\ \hline
      All      & 10877                  & 51615                & -                            \\  \hline 
\multicolumn{4}{c}{FIRST bright ($F_{1.4\text{ GHz}}>3$mJy)}                                                     \\ \hline
All & 573 ($5.3^{+0.2}_{-0.2}$\%)   &  1278 ($2.48^{+0.07}_{-0.07}$\%) & $2.1^{+0.1}_{-0.1}$  \\
 \hline
\multicolumn{4}{c}{FIRST visual inspection classification}                                               \\ \hline
Compact     & 433 ($4.0^{+0.2}_{-0.2}$\%)                 & 781 ($1.51^{+0.05}_{-0.05}$\%)              &  $2.6^{+0.2}_{-0.2}$                          \\
Non-compact & 140 ($1.3^{+0.1}_{-0.1}$\%)                    & 497 ($0.96^{+0.04}_{-0.04}$\%)                  & $1.3^{+0.2}_{-0.2}$                            \\ 
\hline
\multicolumn{4}{c}{LoTSS visual inspection classification}                                               \\ \hline
Compact     & 366 ($3.4^{+0.2}_{-0.2}$\%)$^\dagger$                   & 585 ($1.13^{+0.05}_{-0.04}$\%)$^\dagger$                 &
$3.0^{+0.3}_{-0.3}$\\
Non-compact & 207 ($1.9^{+0.1}_{-0.1}$\%)                    & 693 ($1.34^{+0.05}_{-0.05}$\%)                  &
$1.4^{+0.2}_{-0.1}$\\ 
 \hline
\multicolumn{4}{c}{VLASS visual inspection classification}                                                  \\ \hline
Compact     & 452 ($4.2^{+0.2}_{-0.2}$\%)$^\ddagger$                    & 873 ($1.69^{+0.06}_{-0.05}$\%)$^\ddagger$                 & 
$2.5^{+0.2}_{-0.2}$\\
Non-compact         & 121 ($1.11^{+0.11}_{-0.09}$\%)                    & 405 ($0.78^{+0.04}_{-0.04}$\%)                 & $1.4^{+0.2}_{-0.2}$                              \\ \hline \hline
\end{tabular}
\label{table:morph_stats}
\begin{tablenotes}
      \small
      \item $\dagger$ Includes three cQSOs and two rQSOs that are undetected in LoTSS, which are given the same classification as in FIRST (compact).
      \item $\ddagger$ Includes seven cQSOs and five rQSOs that are undetected in VLASS, which are given the same classification as in FIRST (compact).
    \end{tablenotes}
\end{table*}

Whilst the enhancements in the compact classifications are consistent across the three frequencies, we find that a slightly higher fraction of the colour-selected parent samples (both rQSOs and cQSOs) are classified as compact in VLASS compared to FIRST. Given that VLASS has the highest resolution ($2\farcs5$) of the three surveys, it is possible that some of the more diffuse radio structure is being resolved out. We also find that a higher fraction (both rQSOs and cQSOs) are classified as compact in FIRST compared to LoTSS. This shows a consistent trend where there is a higher compact fraction toward higher radio frequencies. Given the higher sensitivity of LoTSS (particularly to diffuse emission), and the decade difference in observing frequency between the two surveys (144 MHz vs 1.4 GHz), LoTSS and FIRST are potentially probing different electron populations \citep{Morganti2024WhatSurveys}. Since higher energy (frequency) electrons will radiate away their energy more quickly than lower energy (frequency) electrons, LoTSS can probe relatively older electron populations, a factor that was a key motivation behind our visual inspection.

\subsection{Differences in the low-frequency radio morphologies of rQSOs and cQSOs}
\label{sec:multi_radio}
The previous work of \citet{Klindt2019FundamentalOrientation} performed a visual inspection of FIRST-detected rQSOs and cQSOs and found the FIRST compact sources are driving the enhanced radio detection of rQSOs. With the LoTSS images we have more information on the radio morphologies, so we can investigate differences between rQSOs and cQSOs for a given LoTSS classification. 
 
A particularly interesting result is the number of QSOs that appear compact in the FIRST (and VLASS) images, but are not compact at the lower frequencies (and better sensitivity) of LoTSS. We can see from Table~\ref{table:morph_stats} that there are a fraction of QSOs with differing classifications between the
three frequencies. We find that a significant fraction of FIRST-compact sources are non-compact in LoTSS, making up 16.6$^{+1.9}_{-1.6}$\% of the rQSOs and 26.6$^{+1.6}_{-1.5}$\% of the cQSOs with $F_{1.4\text{ GHz}} >3$ mJy, respectively. More rQSOs remain compact at these lower frequencies than cQSOs, which could give an indication of the relative incidence of relic radio emission or a `restarted' AGN \citep{An2012TheSources,Kharb2016AActivity,Silpa2020ProbingI}. The higher prevalence of extended emission in the cQSOs could suggest they are more likely to have had a previous episode of activity compared to rQSOs, multiple cycles of activity on short time-scales (e.g., \citealp{Nyland2020QuasarsFIRST}), or a longer lasting single period of activity. Alternatively, the extended low frequency emission could suggest that the (possibly ongoing) mechanism responsible for the radio emission was/is able to extend further out of the host galaxy in the cQSOs, and the higher sensitivity of LoTSS to diffuse emission is able to detect this when FIRST and VLASS cannot.

We find that the majority of the FIRST compact QSOs are also compact at the higher resolution and frequency of VLASS, with 416/433 ($96.07\pm0.01$\%) and 749/781 ($95.90\pm0.01$\%) of the FIRST compact rQSOs and cQSOs respectively classified as compact in VLASS. The fraction of VLASS compact classifications is also similar between those that are also LoTSS compact 
or LoTSS non-compact
, i.e., the LoTSS morphology is the main distinguishing factor. The VLASS resolution of 2$\farcs$5 corresponds to projected sizes in the range $8 - 20$ kpc at $z = 0.2 - 2.4$ for our assumed cosmology, suggesting the (probably ongoing rather than relic) radio emission is produced on galactic scales or smaller. This is in broad agreement with the findings of \citet{Rosario2021FundamentalE-MERLIN} but for a much larger sample. 

Given the similar incidence of compact FIRST and VLASS morphologies among rQSOs and cQSOs, but the significant differences in their LoTSS morphologies, we define two morphology categories: 
\begin{itemize}[align=parleft,left=0pt]
    \item \textbf{Truly compact QSOs:} compact in FIRST, VLASS, and LoTSS
    \item \textbf{Fake compact QSOs:} compact in both FIRST and VLASS, but non-compact in LoTSS.
\end{itemize}

We use these definitions in the following analysis, and compare the properties of the truly compact and fake compact rQSOs and cQSOs, since we find the LoTSS data are needed to distinguish sources that appear compact in FIRST but are physically non-compact at lower frequencies.\footnote{We note that the terms truly compact and fake compact are empirical and limited by the sensitivity and spatial resolution of our radio observations. Higher resolution data may of course reveal faint extended radio emission in the truly compact systems and more sensitive low-resolution data may reveal faint extended emission at high frequencies. However, since we do not find a strong 1.4 GHz flux dependence on the fraction of fake compact sources (see Appendix~\ref{sec:app_fake}), we do not expect the fake compact class to be significantly contaminated by truly extended radio sources at the spatial resolution explored in this study.}

\begin{figure*}
\centering
\includegraphics[width=1\textwidth]{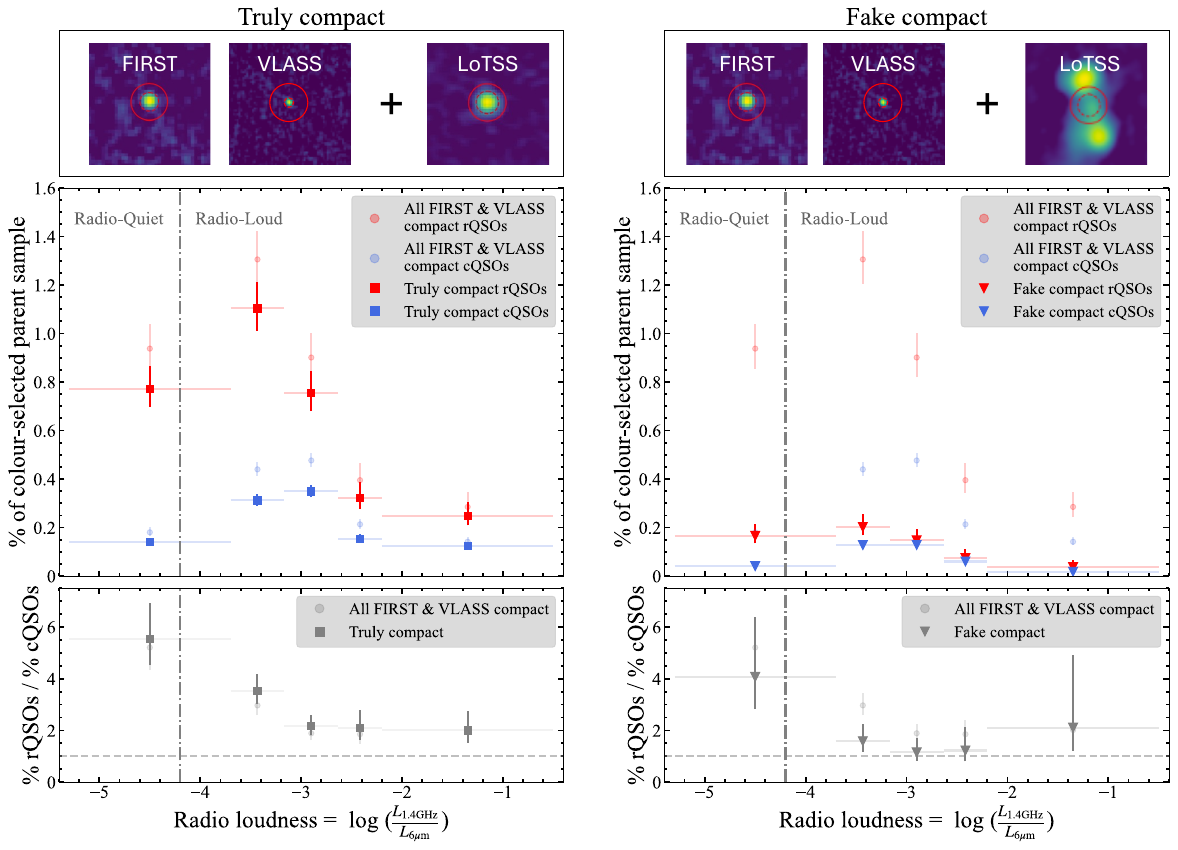}
\caption{The percentage of the colour-selected parent (primary) samples of rQSOs (red markers) and cQSOs
(blue markers) that are bright in FIRST ($F_{1.4\text{ GHz}}>3$ mJy) and are either "truly compact" - compact radio morphologies in FIRST, VLASS and LoTSS (left panel, square markers) or "fake compact" - compact radio morphologies in FIRST and VLASS but non-compact in LoTSS (right panel, triangle markers), as a function of radio loudness. The faded circle markers are the percentages for all FIRST \& VLASS compact QSOs, regardless of the LoTSS morphology (i.e., the sum of the truly and fake compact QSOs), and are identical between both panels. The bottom panels show the respective rQSO enhancements (the fraction of rQSOs in that bin divided by the fraction of cQSOs in that bin) for each morphology classification. The horizontal dashed line indicates an rQSO enhancement of unity (i.e., rQSOs detected at same rate as cQSOs). The enhancements of the FIRST \& VLASS compact rQSOs are mostly driven by the high enhancements of the truly compact rQSOs, whereas the fake compact rQSOs are only slightly enhanced over all but the lowest radio-loudness values, albeit with large uncertainties.}
\label{fig:morphology_fracs}
\end{figure*}

Previous work has found a strong dependence of the FIRST compact rQSO radio enhancement with radio loudness \citep{Klindt2019FundamentalOrientation,Fawcett2021HowProperties}. Figure~\ref{fig:morphology_fracs} shows the fraction of our colour-selected parent samples of rQSOs and cQSOs in bins of radio loudness parameter, $\mathcal{R}$. FIRST compact rQSOs are detected $2.6 \pm 0.2$ times as often as the equivalent cQSOs, in agreement with previous work \citep{Klindt2019FundamentalOrientation}. When separating this into the truly compact QSOs (left panel), and fake compact QSOs (right panel), it is clear that it is the truly compact rQSOs that drive this overall radio enhancement in rQSOs, apart from the most radio-quiet bin, where the red fake compact are also enhanced. The enhancement of truly compact rQSOs to cQSOs (blue square markers) is a factor of $3.0^{+0.3}_{-0.3}$ overall, and they are enhanced across all probed values of radio loudness, from a factor $\approx 2$ for the most radio-loud sources and increasing to a factor $\approx5.5$ in the most radio-quiet bin. 

In contrast, the enhancement within the fake compact classification, of rQSOs to cQSOs, is a factor of $1.6 \pm 0.3$ i.e., the rQSOs are still slightly more likely to be detected than cQSOs with this radio morphology, but it is the truly compact rQSOs (compact in all three surveys) that really drive the overall enhancement. In almost all radio-loudness bins, the rQSO enhancement is consistent with (or only slightly higher than) unity, but increases toward lower radio-loudness values, up to a factor of $\approx 4$ in the lowest radio-loudness bin but with large uncertainty. This suggests that it is predominantly the compact and radio-quiet/intermediate sources that are responsible for the enhanced radio emission in rQSOs, in agreement with \citet{Klindt2019FundamentalOrientation,Fawcett2020FundamentalQuasars,Rosario2020FundamentalLoTSS}.


\subsection{Radio spectral slopes}
\label{sec:slopes}
When radio sources are unresolved and we cannot determine the morphology, radio spectral slope analyses are key to understanding the underlying radio emission mechanism. In the following we focus our analyses mainly on comparing the FIRST$-$VLASS spectral slopes ($\alpha_{1400-3000}$) of truly and fake compact rQSOs and cQSOs, but also briefly consider the LoTSS$-$FIRST spectral slopes ($\alpha_{144-1400}$) in Section~\ref{sec:alphaLF}, although the decadal gap in frequency between LoTSS and FIRST limits what we can infer about the spectral behaviour at low frequencies. Using sources that are classified as compact in both FIRST and VLASS ensures the flux densities used in calculating the radio spectral slopes between FIRST and VLASS are unresolved/compact, since the flux density measurement could include radio emission from multiple different components\footnote{We do not match the resolution of the FIRST ($5''$) and VLASS ($2\farcs5$) radio data used when calculating the radio spectral slope, since it has been shown that the flux biases introduced from the different resolutions only becomes significant for extended sources \citep{Kukreti2023IonisedCycle}. But note all of the of QSOs used in this spectral slope analysis are radio-compact in both FIRST and VLASS by selection.}. Despite the $>10$ year difference in observing epoch between FIRST and VLASS, our results (on average) do not appear to be strongly impacted by variability over this time period (see Appendix~\ref{sec:app_var} for justification).

\begin{figure*}
\centering
\includegraphics[width=1\textwidth]{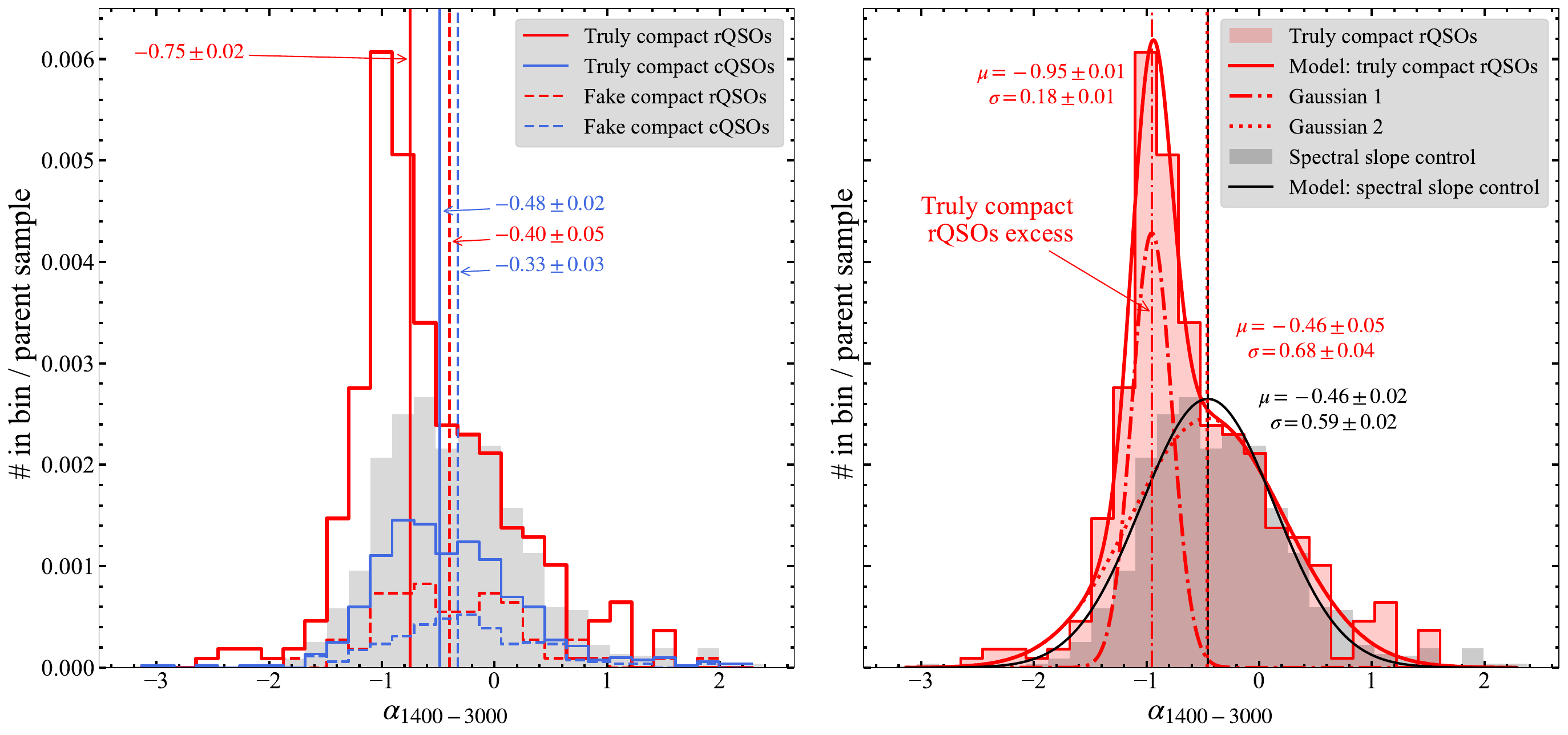}
\caption{\textit{Left panel}: The distributions of $\alpha_{1400-3000}$ for our various colour and radio morphology classifications, normalized by the size of the respective colour-selected parent sample, in order to show the relative detection rates. Red and blue lines indicate the distributions for the rQSOs and cQSOs respectively. The grey histogram is the combined distribution (sum of) all classifications other than the truly compact rQSOs (spectral slope control). The histograms and medians for the truly compact classifications are indicated by the solid lines, and the histograms and medians for the fake compact classifications are given by the dashed lines. \textit{Right panel}: A bimodal distribution fit (red solid line) to the truly compact rQSOs (red histogram), with the two component Gaussians indicated by the dash-dot and dotted red lines. The black line shows the Gaussian fit to the combined distribution (grey histogram). The mean and standard deviation from each fit is indicated on the plot. The truly compact rQSOs have a strong excess of sources with steep ($\alpha\approx-1$) $\alpha_{1400-3000}$ values over all other colour-morphology classes, as indicated by the spectral slope control distribution. }
\label{fig:comb_hists}
\end{figure*}

Here we present our results for the $\alpha_{1400-3000}$ spectral slopes, investigating whether there is evidence for different physical processes in the high frequency slopes ($1.4 - 3$ GHz)\footnote{At the redshift range of our QSOs ($0.2<z<2.4$), the corresponding rest frame spectral slopes range from $1.7-3.6$ GHz up to $4.8-10.2$ GHz. We find no differences in the results when considering the $L_{6\mu\text{m}}-z$ matched samples.} for QSOs with compact or non-compact low-frequency radio morphologies. The left panel of Figure~\ref{fig:comb_hists} shows the distributions of $\alpha_{1400-3000}$ of the rQSOs and cQSOs split into the two morphology classifications: truly compact and fake compact. The distributions are normalised to the size of the respective parent samples, so that the figure is essentially showing detection rate as a function of radio spectral slope.

\begin{table*}
\centering
\caption{\raggedright The median $\alpha_{1400-3000}$ values and $p$-values from two-sided KS tests of the $\alpha_{1400-3000}$ distributions for the various QSO sub-samples. The spectral slope control sub-sample is the combination of all the QSO sub-samples except for the truly compact rQSOs. Bold text indicates $p\leq0.01$ and italic text indicates $0.01<p<0.05$.}
\renewcommand{\arraystretch}{1.2} 
\begin{tabular}{ccccccc}
\hline \hline
 &   & \multicolumn{5}{c}{p-value from two-sided KS test with:}                                                                                                   \\ \cline{3-7} 
QSO sub-sample         & \begin{tabular}[c]{@{}c@{}} Median \\ $\alpha_{1400-3000}$\end{tabular} & \begin{tabular}[c]{@{}c@{}}Truly compact \\ rQSOs\end{tabular} & \begin{tabular}[c]{@{}c@{}}Fake compact \\ rQSOs\end{tabular} & \begin{tabular}[c]{@{}c@{}}Truly compact\\  cQSOs\end{tabular} & \begin{tabular}[c]{@{}c@{}}Fake compact \\ cQSOs\end{tabular} & \begin{tabular}[c]{@{}c@{}}Spectral slope \\ control\end{tabular} \\ \hline
Truly compact rQSOs & $-0.75 \pm0.02 $  & - & \textbf{0.01} & $\mathbf{3 \times 10^{-8}}$  & $\mathbf{3 \times 10^{-9}}$   & $\mathbf{3 \times 10^{-10}}$ \\
Fake compact rQSOs     &$-0.40 \pm0.05$ & $\cdots$ & -  & 0.79  & 0.50 & 0.96 \\
Truly compact cQSOs    & $-0.48 \pm0.02$   & $\cdots$   & $\cdots$   & -    & \textit{0.03}   &  0.91 \\
Fake compact cQSOs     & $-0.33 \pm0.03$  & $\cdots$   & $\cdots$ & $\cdots$   & -  & 0.15    \\
Spectral slope control & $-0.43 \pm0.01$   & $\cdots$   & $\cdots$   & $\cdots$  & $\cdots$  & -    \\ \hline \hline
\end{tabular}
\label{table:p_vals}
\end{table*}

We find that the truly compact rQSOs have a statistically different $\alpha_{1400-3000}$ distribution to all other colour/morphology combinations, which have similar distributions. The truly compact rQSOs are much more likely to be detected with steep radio spectral slopes ($\alpha_{1400-3000} \approx-1$) than any other classification, and whilst they are enhanced compared to the truly compact cQSOs across the full range of spectral slopes (by a factor of $\approx 3$ overall), this enhancement is most pronounced around steep spectral slope values of $-1.2<\alpha_{1400-3000}<-0.8$, by a factor of $\approx 5$. Comparing all combinations of rQSOs vs. cQSOs across both the truly compact and fake compact categories we find the only outlying population are the truly compact rQSOs which are inconsistent with fake compact rQSOs and all FIRST \& VLASS compact cQSOs ($p$-values and medians given in Table~\ref{table:p_vals}). 

Given the similarity in the $\alpha_{1400-3000}$ distributions for all colour-morphology combinations other than the truly compact rQSOs, we construct an overall $\alpha_{1400-3000}$ distribution from all the  FIRST \& VLASS compact cQSOs and the fake compact rQSOs (the grey histogram in both panels of Figure~\ref{fig:comb_hists}) to allow for more detailed analyses with the truly compact rQSOs. We refer to this distribution as the spectral slope control and assume it represents the overall radio spectral slope distribution for normal FIRST \& VLASS compact QSOs with $F_{1.4\text{GHz}} >3$ mJy. The $p$-value from a KS test between this spectral control distribution and the truly compact rQSOs is p $ = 3 \times 10^{-10}$ (also given in Table~\ref{table:p_vals}). Based on this analysis, the truly compact rQSOs appear to be a distinct population, with a predominance of steep $\alpha_{1400-3000}$ values and the highest detection rates (see Section~\ref{sec:morph_results} and Figure~\ref{fig:morphology_fracs}) over any other colour/morphology classification. Comparing the shapes of the spectral slope control distribution and truly compact rQSOs distribution makes it even clearer that the truly compact rQSOs are a distinct population in terms of their $\alpha_{1400-3000}$ values. They appear to contain the full range of spectral slope values found for typical FIRST \& VLASS compact QSOs, but they also host an additional population of sources with steep spectral slopes (right panel, Figure~\ref{fig:comb_hists}). We characterise the differences in the shapes in the following analysis. 

We fit a Gaussian to the spectral slope control distribution, leaving the amplitude, mean ($\mu$), and standard deviation ($\sigma$) as free parameters. The right panel of Figure~\ref{fig:comb_hists} shows the best fit to the data (grey histogram), shown by the solid black line, of a Gaussian with a mean spectral slope of $\mu = -0.46 \pm 0.02$, and a standard deviation of $\sigma = 0.59 \pm 0.02$, where the errors on the parameters are from the covariance matrix. We then fit a bimodal distribution to the spectral slope distribution of the truly compact rQSOs, again leaving the mean, standard deviation, and amplitude of both component Gaussians as free parameters. The resulting best fit model is shown by the solid red line in the right panel of Figure~\ref{fig:comb_hists}, with the two component Gaussians, centred on $\mu = -0.95 \pm 0.01$ and $\mu = -0.46 \pm 0.05$, with standard deviations of $\sigma = 0.18 \pm 0.01$, and $\sigma = 0.68 \pm 0.04$ respectively, shown by the dash-dot and dotted red lines. Without fixing the parameters of the mean and standard deviation we find remarkable agreement between the spectral slope control distribution and the broad component Gaussian of the truly compact rQSOs,  with the means agreeing within the uncertainties, and similar amplitudes and standard deviations. This suggests the truly compact rQSO population is made up of two sub-populations - one composed of rQSOs with "normal" radio spectral slopes found for the overall cQSO (and fake compact rQSO) population, and one with steep slopes centred around $-0.95$\footnote{We checked that the peak in the truly compact rQSO spectral slope distribution is not just due to faint sources being boosted above the detection limit due to Eddington bias. We found that the peak persists even when taking detections with a SNR $>30$ in both FIRST and VLASS. In Appendix~\ref{sec:app_faint} we show that the spectral slopes of the truly compact rQSOs are steeper across the range of FIRST flux densities probed.}.

Ideally, we would want to separate out this distinct excess population of truly compact, steep-slope rQSOs, and study their properties. However there is significant overlap between this sub-population and the typical population, as demonstrated by the overlapping component Gaussian model fits. By integrating the Gaussians we determine that of the truly compact rQSO population, $\sim 31 \%$ are consistent with being in this excess population. To distinguish between the broad Gaussian contribution seen across all of the QSO sub-populations and the excess contribution within the narrow Gaussian seen for the truly compact rQSOs, we estimate the probability of each truly compact rQSO being in this excess population. This probability is given by the height of the narrow Gaussian. Conversely, we also estimate the probability of each compact rQSO being consistent with the normal (non-excess) population. These probabilities allow us to apply weighted statistics and infer the properties of these steep $\alpha_{1400-3000}$ sources in the rest of our analysis.

For comparison to the general population, we estimate the maximum subset of the typical FIRST \& VLASS compact QSOs (i.e., everything except the truly compact rQSOs) that could also be in a similar steep-slope excess sub-population. We apply a bimodal fit to our spectral slope control distribution, this time fixing the mean of the broad Gaussian to be $\mu = -0.46$ (in-line with the best-fit mean from the single Gaussian model), constraining the mean of the narrow Gaussian component to $\alpha <0$, and leaving the other parameters free. We find the best fit to be a narrow Gaussian centred on $\mu = -0.85 \pm 0.06$ with a standard deviation of $\sigma = 0.15 \pm 0.05$, similar to the best-fit narrow Gaussian of the truly compact rQSOs. The best-fit width of the broad Gaussian component remains unchanged with $\sigma = 0.59 \pm 0.02$. As we did for the truly compact rQSOs, we integrate the Gaussians and determine that the fraction of typical FIRST \& VLASS compact QSOs that could be in this excess population is  $\sim 4 \%$, around one-eighth of the fraction of the truly compact rQSOs. This indicates a much lower incidence of steep radio spectral slopes in QSOs with low levels of dust extinction. This is broadly consistent with the result for the truly compact QSOs with $E(B-V)<0.1$ mag (i.e., the least obscured truly compact QSOs) in Section~\ref{sec:radio_dust} (see Figure~\ref{fig:EBV_bins}).

\subsubsection{Comparison to low-frequency spectral slopes}
\label{sec:alphaLF}
The decadal gap in frequency between LoTSS and FIRST limits the diagnostic power of spectral analyses towards low frequencies, which would much more greatly benefit from additional intermediate frequencies such as those utilised in  \citet{Fawcett2025ConnectionQSOs}. However, we can still use the spectral slope between 144 MHz and 1.4 GHz ($\alpha_{144-1400}$) to provide basic constraints on the crude shape of the radio SED, noting that the large gap in frequency coverage could mean we are missing significant spectral complexity (e.g., a turnover in the radio SED). With this limitation in mind, we can still search for differences in the broad radio spectral slopes over 144 - 1400 MHz and investigate whether the steep slopes found for the truly compact rQSOs extend to lower frequencies, which will be more sensitive to the older electron population. \citet{Rosario2020FundamentalLoTSS} previously investigated the $\alpha_{144-1400}$ slopes of red and blue QSOs and found no significant differences between the two populations, even for LoTSS-compact sources, but limited their analyses to just the LoTSS-FIRST spectral slope for the overall rQSO and cQSO population. In our study we focus our analyses on the truly compact sources, which is where significant differences are seen between the FIRST-VLASS spectral slopes of rQSOs and cQSOs, but which also minimises potential complications in the flux determination from multiple extended components.

\begin{figure}
\centering
\includegraphics[width=1\columnwidth]{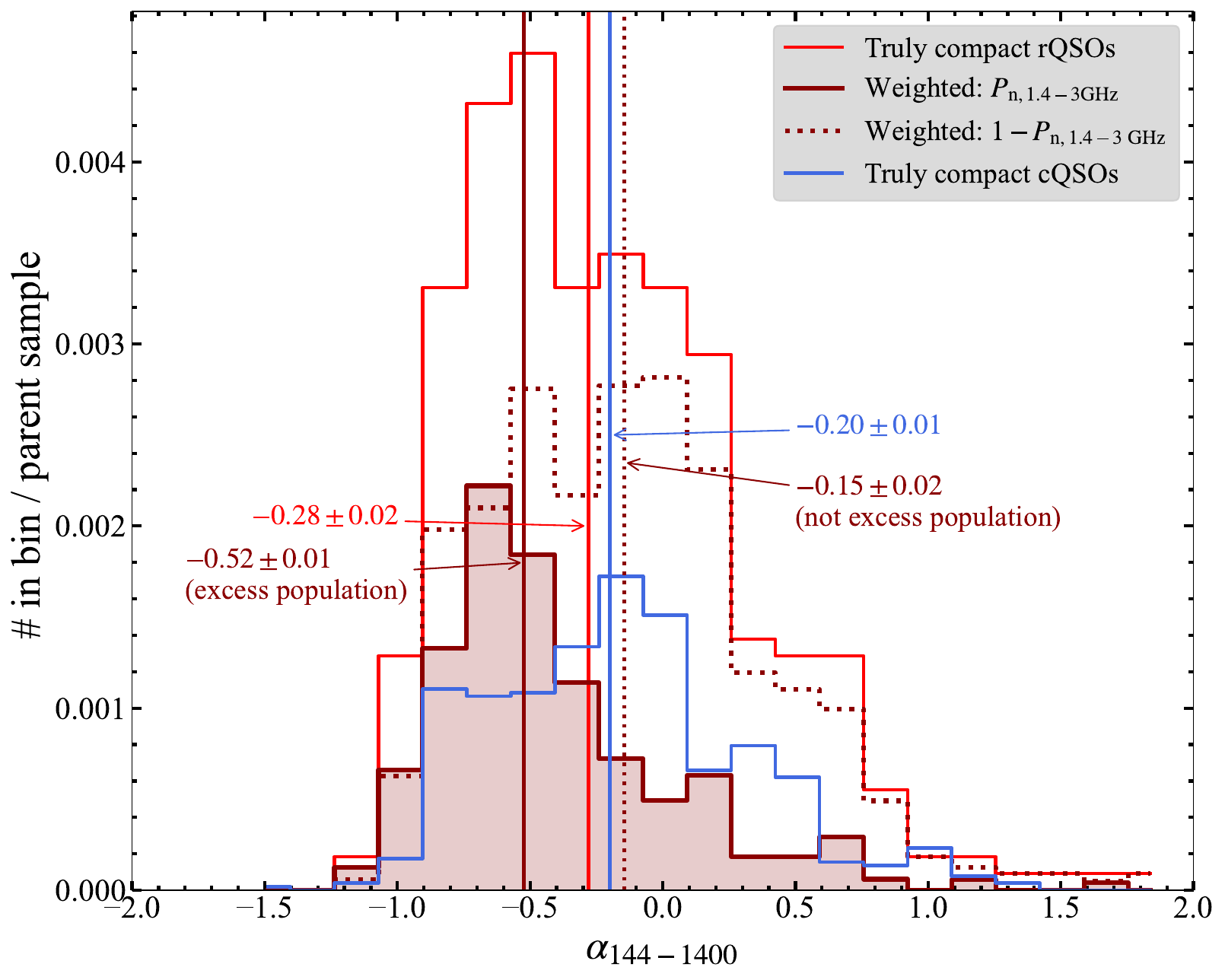}
\caption{The distributions of $\alpha_{144-1400}$ for the truly compact rQSOs and cQSOs. The colour and labeling scheme is the same as in Figure~\ref{fig:comb_hists}, with the addition of weighted distributions to illustrate the spectral slopes of the truly compact rQSOs that likely reside (dark red solid filled histogram) or do not reside (dotted dark red histogram) in the excess population (narrow Gaussian) identified in Figure ~\ref{fig:comb_hists} (right panel). The respective weighted median values are shown with dark red solid/dotted vertical lines and are annotated with the median values. The median $\alpha_{144-1400}$ of the truly compact rQSOs residing in the excess population ($-0.52 \pm 0.01$) is significantly steeper than the truly compact rQSOs not residing in the excess population ($-0.15 \pm 0.02$).}
\label{fig:alphaLF}
\end{figure}

Figure~\ref{fig:alphaLF} shows the distribution of $\alpha_{144-1400}$ for the truly compact rQSOs and cQSOs, normalized to the size of the parent sample (so essentially detection rate and analogous to Figure~\ref{fig:comb_hists}). Interestingly, there is a less clear excess centred around steep $\alpha_{144-1400}$ values, than in the $\alpha_{1400-3000}$ case. The median slopes of the rQSOs and cQSOs are shown by the coloured solid lines, and are relatively close in value, and a KS test between the distributions results in a $p$-value of 0.02, indicating a tentative difference. In dark solid red is the histogram of $\alpha_{144-1400}$ weighted by the probability of an rQSO residing in the excess population with steep $\alpha_{1400-3000}$ spectral slopes (see Figure~\ref{fig:comb_hists}). There is clear evidence that the excess rQSOs with steep $\alpha_{1400-3000}$ spectral slopes also have steeper $\alpha_{144-1400}$ spectral slopes than the cQSOs (weighted median of $-0.52 \pm 0.01$ for the rQSOs compared to a median of $-0.20 \pm 0.01$ for the cQSOs), although the absolute values are not as steep as those found at $\alpha_{1400-3000}$, suggestive of either a steepening at high frequencies or a flattening at low frequencies. A KS-test between the truly compact cQSOs and the truly compact rQSOs weighted toward the $\alpha_{1400-3000}$ excess population, results in a $p$-value of $1.4\times10^{-15}$, indicating significantly different distributions. 

By comparison, the dark red dotted histogram shows the distribution of the truly compact rQSOs weighted in the opposite sense, to those that are a part of the non-excess population in $\alpha_{1400-3000}$. The weighted median of $-0.15 \pm 0.02$ is very flat and close in value to the truly compact cQSOs: indeed, a weighted KS test between these distributions results in a $p$-value of p = 0.3, indicating they are statistically consistent. This corroborates the result from Section~\ref{sec:slopes}, that the excess population of truly compact rQSOs with steep $\alpha_{1400-3000}$ spectral slopes are distinct from the other QSO sub-populations (including the non-excess truly compact rQSOs), which are much more similar in radio spectral properties. For completeness, in Appendix~\ref{sec:app_CC} we also show the radio detection rates of our truly compact rQSOs and cQSOs across the $\alpha_{1400-3000}$--$\alpha_{144-1400}$ radio spectral slope plane. Of course, with such a large gap in frequency coverage we cannot draw any robust conclusions on the overall radio SEDs of the truly compact rQSOs with steep $\alpha_{1400-3000}$ slopes, which would require intermediate-frequency radio observations (e.g., using uGMRT at 400 and 650 MHz, as utilised in \citealp{Fawcett2025ConnectionQSOs}). However, the significantly steeper $\alpha_{144-1400}$ spectral slopes clearly indicate that the differences are not limited to high frequencies.

\subsection{\texorpdfstring{Radio connection with $E(B-V)$}{Radio connection with E(B-V)}}
\label{sec:radio_dust}
\begin{figure}
\centering
\includegraphics[width=1\columnwidth]{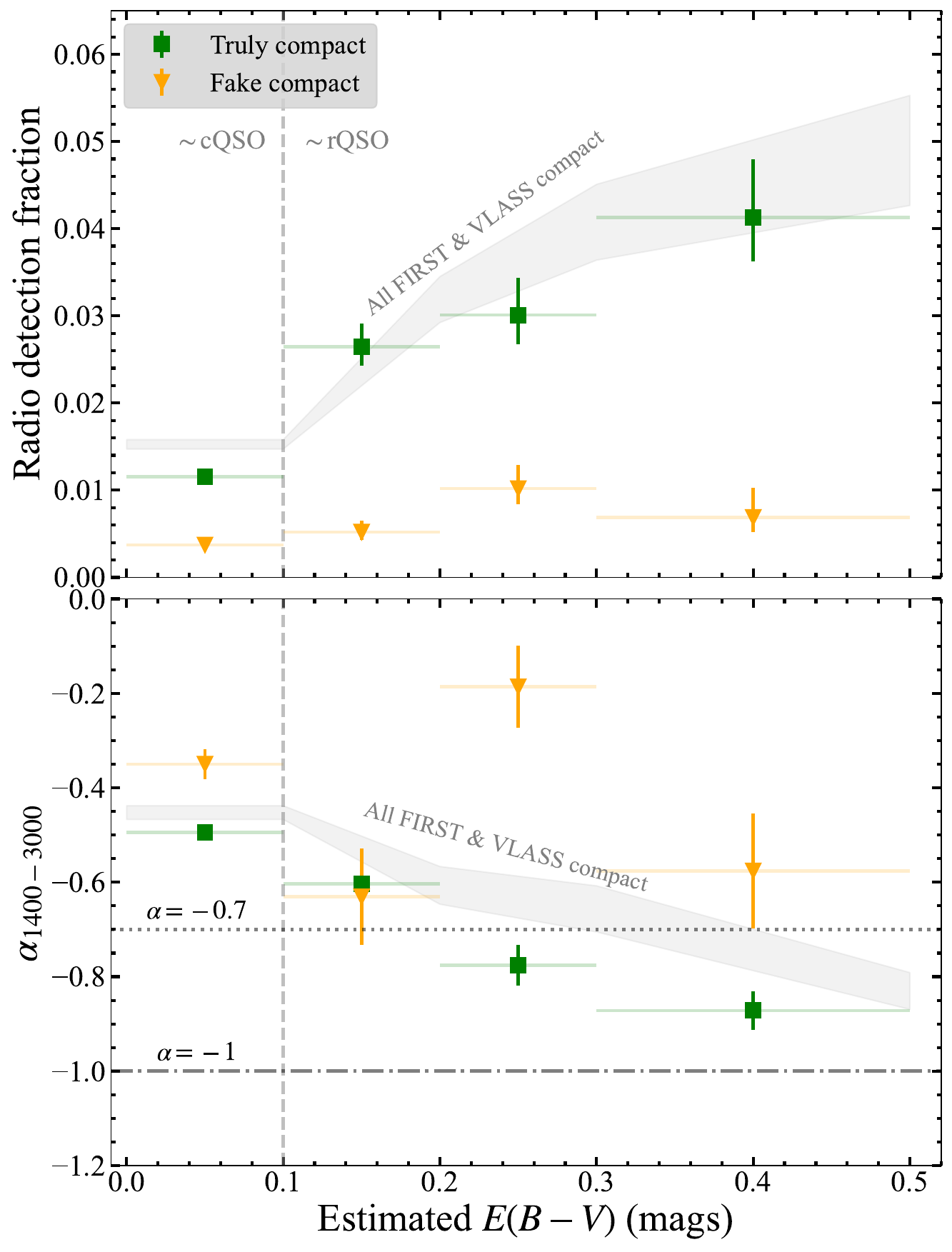}
\caption{\textit{Top panel}: The radio detection fraction of the truly compact QSOs (green square points) and fake compact QSOs (orange triangle points) as a function of estimated $E(B-V)$. The vertical dashed line roughly indicates the separation between a cQSO and an rQSO. The grey region is the radio detection rate for all FIRST \& VLASS compact QSOs. Horizontal error bars represent the bin range, and vertical error bars indicate the 1$\sigma$ binomial uncertainties. \textit{Bottom panel}: The median $\alpha_{1400-3000}$ in bins of $E(B-V)$. The grey region is the median spectral slope for all FIRST \& VLASS compact QSOs. The vertical error bars indicate the median absolute deviation/$\sqrt{\text{bin size}}$. The spectral slopes expected from typical synchrotron emission ($\alpha = -0.7$) and AGN-wind driven shocks ($\alpha = -1$) are indicated.}
\label{fig:EBV_det}
\end{figure}

Previous work has revealed an intrinsic positive correlation between the level of opacity in quasars and the radio detection fraction \citep{Fawcett2023AQSOs,rivera24,Petley2024HowRate}. This is suggestive of a causal connection between the production of the radio emission and opacity, as traced by the inferred column density of dust. In Figure~\ref{fig:EBV_det} (top panel) we show the radio detection rate of our QSOs (combined rQSOs and cQSOs), now in bins of $E(B-V)$ rather than separating into red and control QSO samples. For reference the divide between a rQSO and cQSO is $E(B-V)\approx 0.1$ (and has a small redshift dependence). Similar to previous work we find the radio detection fraction increases with increasing $E(B-V)$, with the grey region showing the detection rate for all of our FIRST \& VLASS compact QSOs, although we note that since our sample is restricted to bright 1.4 GHz flux densities and our QSOs are selected from SDSS, we are restricted to a smaller region of the radio-detection fraction$-E(B-V)$ plane than that probed in the deeper DESI-LoTSS analyses of \citet{Fawcett2023AQSOs}. From Figure~\ref{fig:EBV_det} it is evident that the truly compact QSOs (green points) drive this trend, with no clear trend found for the fake compact QSOs. Therefore, whatever is producing the radio emission is predominantly occurring on host-galaxy scales or smaller, and is intrinsically connected to the level of dust and the lack of extended (potentially relic) low frequency radio emission.

Since the radio emission production is clearly connected to the amount of dust, we can also test whether the radio spectral slope is also correlated with $E(B-V)$. We focus on performing this analysis only for the FIRST$-$VLASS spectral slopes since (1) we want to explore the dust-radio connection on the smallest scales, where by definition all of our QSOs are compact at high frequencies and (2) the decadal frequency gap between LoTSS and FIRST restricts the diagnostic potential of LoTSS-FIRST spectral analyses.

In Figure \ref{fig:EBV_det} we show the median $\alpha_{1400-3000}$ in bins of $E(B-V)$ (lower panel). We observe a trend of steepening radio spectral slope with increasing $E(B-V)$, and as with the radio detection rates, we only find a significant correlation for the truly compact QSOs, with no trend observed for the fake compact QSOs. We check that redshift is not driving this correlation by performing a partial correlation test between $\alpha_{1400-3000}$ and $E(B-V)$ for the truly compact QSOs, whilst holding $z$ constant. This results in a correlation coefficient of $r = -0.12$ with a $p$-value of $p=4\times10^{-4}$, indicating a fairly weak but statistically significant correlation between $\alpha_{1400-3000}$ and $E(B-V)$, independent of redshift.\footnote{\label{foot:EBV_tc} For the truly compact QSOs, we also test whether the $\alpha_{144-1400}$ slopes are correlated with $E(B-V)$ like the higher frequency $\alpha_{1400-3000}$ slopes. A partial correlation between $\alpha_{144-1400}$ and $E(B-V)$ (holding $z$ constant) produces a correlation coefficient of $r = -0.08$ with a $p$-value of $6\times10^{-4}$, indicating a weaker trend than observed for the higher frequency slopes.} For the fake compact sources we find a correlation coefficient of $r = -0.02$ with a $p$-value of $p=0.80$, indicating no correlation.

\begin{figure*}
\centering
\includegraphics[width=1\textwidth]{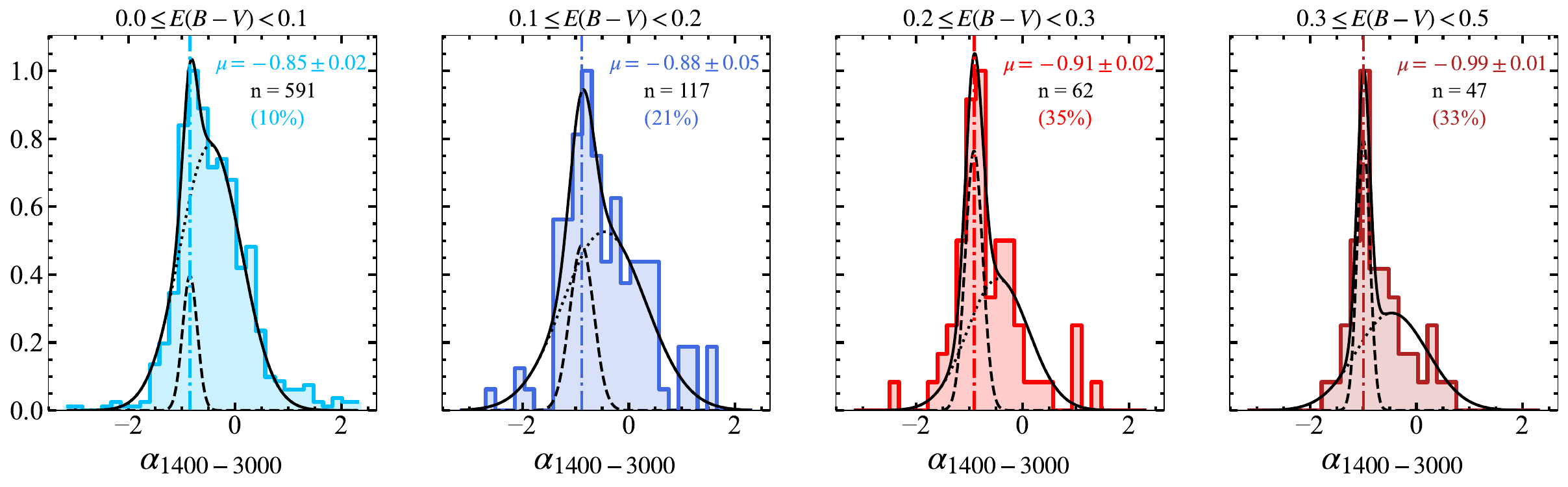}
\caption{The distributions of $\alpha_{1400-3000}$ in four bins of $E(B-V)$, for the truly compact QSOs. In all cases a bimodal distribution model has been fit, with a fixed broad distribution with a mean of $\mu = -0.46$ (following the result for the broad component in Figure~\ref{fig:comb_hists}: spectral slope control) and a free-fitting narrow distribution (excess population). The mean of the narrow distribution is indicated by the dash-dotted line on each panel, along with the number of QSOs in each bin and the corresponding fraction of QSOs within the narrow distribution.}
\label{fig:EBV_bins}
\end{figure*}

We can test whether the spectral slope$-E(B-V)$ trend in the truly compact QSOs is due to a physical relationship between the radio emission and dust, resulting in an overall steepening of the spectral slopes, versus an overall increase in the fraction of sources in the narrow Gaussian toward dustier QSOs. In Figure \ref{fig:EBV_bins} we have split our truly compact QSOs into bins of $E(B-V)$, and fit a bimodal model, made up of one narrow and one broad component, to the $\alpha_{1400-3000}$ distributions in each bin. We fix the mean of the broad Gaussian at $\mu = -0.46$ to be the same as that found for the spectral slope control distribution (see Figure ~\ref{fig:comb_hists}), representative of typical FIRST \& VLASS compact QSOs. The mean of the narrow Gaussian was allowed to vary, but constrained to $\alpha<0$. The standard deviation and amplitude of the broad and narrow Gaussians were all allowed to vary but constrained to be $>0$ to help the fitting. We find in all cases that the best-fitting solution occurs with the narrow Gaussian centred around values of $\alpha \sim-0.85$ to $\alpha \sim -0.99$, with a slight preference for a steeper mean at higher $E(B-V)$. The uncertainties on the mean $\alpha$ (narrow Gaussian fit) are calculated from the covariance matrix and indicated in the figure. The fraction of sources in this narrow Gaussian increases with increasing $E(B-V)$, from $\sim 10\%$ among the least dusty QSOs ($ 0 \leq E(B-V) < 0.1$ mag), up to $\sim 33\% - 35\%$ among the dustiest QSOs in this work ($E(B-V) \geq 0.2 $ mag); see Figure~\ref{fig:EBV_bins}. Thus, although the slopes generally get steeper with $E(B-V)$, the greatest difference is in the fraction of excess sources with steep spectral slopes with increasing $E(B-V)$.
\section{Discussion}
\label{sec:discussion}
In this work, we have used multi-frequency radio data to explore the radio morphologies and radio spectral slopes of red and blue/typical QSOs, to further understand the origin of the enhanced radio detection rates of rQSOs. We have
found evidence for a distinct population of rQSOs, with compact morphologies ($<6''$ at $0.144-3$ GHz, referred to as truly compact) and steep spectral slopes over observed-frame $1.4-3$ GHz. This suggests a distinct physical process is responsible for producing the excess population of steep spectral slope sources, significantly in excess of that found in the general QSO population. The fact that we see a trend for the truly compact QSOs for both the radio detection fractions and steep radio spectral slopes with $E(B-V)$, suggests the mechanism is related to opacity and occurs on approximately host-galaxy scales.

In Section~\ref{sec:steep_discuss} we discuss the physical implications of our empirical results, by comparing the steep radio spectral slopes to other red quasar studies in the literature. We compare our results to the predictions for AGN-driven wind shocks from \citet{Nims2015ObservationalNuclei} in Section~\ref{sec:Nims}, and consider whether this is a feasible mechanism to produce the observed radio luminosities and steep spectral slopes of our truly compact rQSOs. We discuss whether jets or winds (or both) could be the cause of the radio emission (Section~\ref{sec:jet_v_wind}), by comparing to evidence from observations of Compact Steep-Spectrum (CSS) and other similar well-studied radio-source populations in Section~\ref{sec:obs_jet_v_wind}, and comparing to studies of detailed simulations of the radio emission from both jets and winds in Section~\ref{sec:sims}. Finally, in Section~\ref{sec:dusty_blowout}, we discuss our results in the context of a dusty blow-out phase evolutionary model.


\subsection{Steep radio spectral slopes in dust reddened QSOs}
\label{sec:steep_discuss}
What is causing this increasing contribution of excess sources with a steep radio spectral slope (specifically $\alpha_{1400-3000}$) with $E(B-V)$ in QSOs? In an evolutionary model, QSOs in an earlier phase are expected to have higher levels of opacity/dust, and the winds or jets in these systems are expected to interact with this surrounding material, destroying/removing it over time and, hence, decreasing the amount of obscuration. Therefore, an increased amount of dust/opacity would be expected to increase the probability of wind/jet-dust interactions. The observed correlation between both radio detection fraction and radio spectral slopes with $E(B-V)$ in our QSOs suggests that the radio emission is likely a consequence of interactions with the surrounding medium such as shocks, that could be driven by jets or winds.

The peak in the $\alpha_{1400-3000}$ distribution we observe for our excess rQSO population ($\alpha \approx -1$) is steeper than the expected $\alpha \approx-0.7$ from synchrotron emission in an optically thin region (e.g., typical in radio-loud sources with evolved jets), and is much steeper than the other radio/colour classifications (i.e, all FIRST \& VLASS compact cQSOs and the fake compact rQSOs) explored in this paper,  which are generally much flatter (with median spectral slopes of $\alpha \approx -0.5$ to $-0.3$). Steep radio spectral slopes could be due to a number of different processes, possibly indicating either a mechanism with rapid electron cooling or ageing electrons \citep{Kardashev1962NonstationarityEmission}, or an intrinsically steeper injection index \citep{Athreya1998TheGalaxies.} in red QSOs. Star formation is predicted to produce radio emission with a steep ($\alpha<-0.5$) spectral slope \citep{Kimball2011TheAGN, Condon2013AGNQSOs, Rivera2002The2.5}, however we rule out its contribution, since previous work has found red QSOs have star formation rates consistent with typical blue QSOs \citep{Fawcett2020FundamentalQuasars,Rosario2020FundamentalLoTSS,Rivera2021TheReddening,Yue2024AQuasars}. Additionally, as shown in \citet{Fawcett2020FundamentalQuasars}, AGN-related mechanisms dominate the radio emission compared to star-formation above radio loudness values $\mathcal{R}\gtrsim-5$, which is where $>99\%$ of our QSO samples lie (see Figure~\ref{fig:Lradio_L6}).

A number of other studies have also found evidence for steep radio spectral slopes in dust reddened QSOs. For instance, \citet{Glikman2022TheRegime} compared the $1.4-3$ GHz (FIRST$-$VLASS) radio spectral slopes of a sample of MIR-selected red and blue QSOs, and found that the red QSOs exhibited significantly steeper slopes on average, with $\alpha = -0.70 \pm 0.05$ compared to $\alpha = -0.34 \pm 0.04$ for blue QSOs. Similarly, \citet{Hwang2018WindsQuasars} examined a sample of "Extremely Red Quasars" (ERQs), identified based on their extreme optical to mid-infrared ratios, and found that their radio spectral slopes in the frequency range 4.25 GHz to 8.2 GHz, were steeper than those of typical blue QSOs, with a mean spectral slope between $\alpha = -1.0$ and $\alpha = -1.3$, depending on the measurement method. They also found a strong correlation between the radio luminosity and [O~{\sc iii}]$\lambda5007$\AA\ velocity widths, suggesting that wind-driven shocks are the most likely explanation for the radio emission in their sources.

The result of steepening radio spectral slope with increasing $E(B-V)$ was also found in \citet{Fawcett2025ConnectionQSOs}, who used more detailed radio SED fitting with uGMRT data at 400 MHz and 650 MHz, in addition to the three radio surveys utilized in this study. By fitting power-law, broken power-law, and peaked/curved radio SED models, they found that dust reddened QSOs are more likely to have power-law or broken power-law radio SEDs than their control sample of unobscured blue QSOs ($77\pm15$ per cent of red versus $52\pm10$ of blue), and that the spectral slopes of the power-law and broken power-law QSOs became steeper with increasing $E(B-V)$. In this work we are able to firmly establish this trend with a larger sample, and demonstrate that the truly compact QSOs predominantly drive this relation.

The presence of steep radio spectral slopes appears to extend beyond optically selected red QSOs and is also observed in even more heavily obscured systems. For instance, \citet{Patil2022RadioQuasars} analysed radio spectra from 0.1 to 10 GHz for samples of luminous, heavily obscured \textit{WISE}-NVSS selected quasars and found that their high-frequency (generally for $\nu > 1.4$ GHz) spectral indices are steep, with $\alpha \approx -1$. These steep slopes show a weak correlation with the ratio of mid-infrared (MIR) photon energy density to magnetic energy density, suggesting that inverse Compton scattering off an intense MIR photon field (from the AGN) may contribute to the observed spectral steepening in this case. Since we are using SDSS for our optical selection our $E(B-V)$ range is relatively limited, and it would be interesting to investigate whether this trend of an increasing fraction of steep spectral slopes consistently increases toward higher extinction values, for example using the upcoming 4MOST IR AGN survey \citep{Andonie2025AnGoals} which will obtain a large sample ($>100,000$) of obscured quasars.

\subsection{Wind-shock model}
\label{sec:Nims}
\begin{figure}
\centering
\includegraphics[width=1\columnwidth]{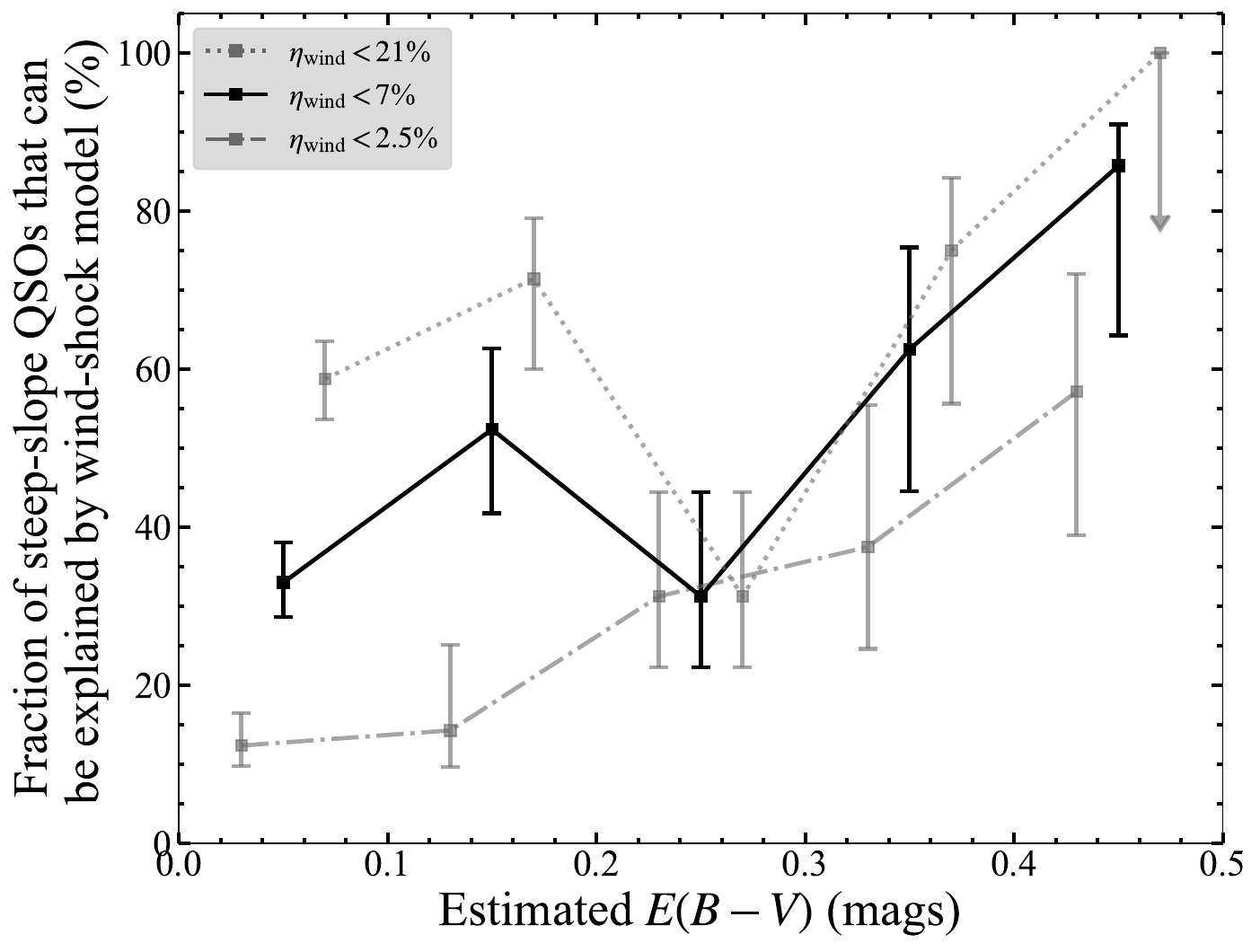}
\caption{The fraction of steep $1.4-3$ GHz spectral slope ($-1.15< \alpha_{\text{1400-3000}} < -0.85$) truly compact QSOs with radio luminosities consistent with the prediction from the \citet{Nims2015ObservationalNuclei} wind-shock model, in bins of $E(B-V)$, for wind efficiency values $\eta_{\text{wind}}<2.5$\%, $<7$\% and $<21$\% (calculated from Equation~\ref{equation: 1}). The number of QSOs that are consistent with this model increases with $E(B-V)$ for any wind efficiency parameter. The 100\% fraction for the $<21$\% wind efficiency model in the highest $E(B-V)$ bin is given as an upper limit. The different wind efficiency points have been offset slightly in the $x$-axis direction for clarity.}
\label{fig:nwind}
\end{figure}

We can perform a basic test of the origin of the radio emission by comparing to predictions from models. \citet{Nims2015ObservationalNuclei} predict that the synchrotron emission from electrons in an ambient ISM accelerated by an AGN wind-driven shock would be of the form: \begin{equation}\label{equation: 1}
    \nu L_\nu \approx10^{-5}\eta_{\text{wind}}L_{\text{AGN}}\left( \frac{L_{\text{kin}}}{0.05L_{\text{AGN}}}\right),
\end{equation}for $\nu \gtrsim \nu_{\text{cool}}$, where $\nu_{\text{cool}}$ is the characteristic cooling frequency, above which the electrons will radiate all the energy provided by the shock acceleration. $L_\text{kin}$ is the luminosity of the wind, $\eta_{\text{wind}}$ is the wind efficiency (the percentage of the shock kinetic energy supplied to the electrons), and $L_{\text{AGN}}$ is the bolometric luminosity of the AGN. They assume that the shock accelerated electrons are produced with a power law distribution, $\dot{n}{(\gamma) \propto \gamma^{-p}}$, with $p=2$ taken since it is consistent with shock acceleration theory and observations of supernova remnants (where $p\approx2-2.5$, e.g., \citealp{Blandford1987ParticleOrigin}). This assumption is also appropriate for AGN-driven winds, where electrons are expected to be accelerated by first-order Fermi processes at non-relativistic or mildly relativistic shocks, similar to those in supernova remnants. As discussed by \citet{Kirk1987OnFronts} and \citet{Athreya1998TheGalaxies.}, the slope of the electron injection spectrum is sensitive to the upstream velocity of the shock, with lower velocities producing steeper electron spectra than the flatter indices typically associated with relativistic jet shocks. There are no strong differences expected between AGN-driven and supernova shocks that would significantly alter the average acceleration behaviour, making the assumed injection index of $p\sim2-2.5$ reasonable for AGN winds.

Under this assumption of $p=2$, Equation \ref{equation: 1} predicts that $\nu L_{\nu}$ is $\sim$ constant above the cooling frequency, or that the radio luminosity, $L_\nu$ is inversely proportional to the frequency: $L_{\nu} \propto \nu^{-1} $, corresponding to a spectral slope of $\sim -1$. This is in excellent agreement with the peak of the spectral slope distribution for our truly compact rQSOs of $\alpha_{1400-3000} \sim -0.95$, and the peak in our highest $E(B-V)$ bin of $\alpha_{1400-3000} \sim -0.99$ (Figure \ref{fig:EBV_bins}). We can also calculate the wind efficiency values, $\eta_{\text{wind}}$, required to reproduce the observed 1.4 GHz radio luminosities of our QSO samples, assuming that the radio luminosity is produced purely due to the AGN-driven wind shock (following Equation \ref{equation: 1}). We assume that the kinetic energy of the wind, $L_{\text{kin}}$, is 5\% of the AGN bolometric luminosity, $L_{\text{AGN}}$, which we take to be $L_{\text{AGN}} = BC_{6\mu\text{m}} \times L_{6\mu\text{m}}$\footnote{We also calculated $L_{\text{AGN}}$ using the optical luminosities from \citet{Rakshit2020SpectralCatalog} and correcting for the dust extinction $E(B-V)$. The wind efficiency values do not change significantly compared to using $L_{6\mu\text{m}}$}, with $BC_{6\mu\text{m}} = 8$, following \citet{Fawcett2025ConnectionQSOs} and \citet{Richards2006TheThree}. We calculate the wind efficiency for each QSO using its observed rest-frame $L_{1.4\text{ GHz}}$ and $L_{6\mu\text{m}}$ and we find the required median wind efficiency is 6\% (inter-quartile range [IQR] $3-21$\%) for the truly compact rQSOs, and 11\% (IQR $4-30$\%) for the truly compact cQSOs. The typical wind efficiency values from observations of ionised outflows in luminous AGN are $\lesssim 5-7\%$ (e.g., \citealp{Harrison2012EnergeticActivity,Liu2013ObservationsNebulae,Sun2017SizesImages,Harrison2018AGNOn}), with $\lesssim$5\% typically assumed in analytical models and simulations (e.g., \citealp{Ward2024AGN-driRelations}), which is in good agreement with the median wind efficiencies of our QSOs. Therefore, the wind-shock model is a reasonable model to investigate for our QSOs. There are however large uncertainties on wind efficiency measurements and the bolometric correction \citep{Harrison2018AGNOn}.

Restricting this analysis to only truly compact QSOs with steep $1.4-3$ GHz spectral slopes ($-1.15 < \alpha_{\text{1400-3000}} < -0.85$), i.e., within a tolerance of $\Delta\alpha = 0.15$ (the median uncertainty on the spectral slope measurement) from $\alpha_{\text{1400-3000}} = -1$, we find a high fraction of these QSOs can be explained by the wind-shock model with reasonable wind-efficiency values. In Figure \ref{fig:nwind} we show the fraction of these steep-slope truly compact QSOs with calculated wind efficiency values up to $\eta_{\text{wind}}<$2.5\%, $<$7\% and $<$21\%, in bins of $E(B-V)$. We assume that a shock could always produce a lower $\eta_\text{wind}$, and so anything below the wind efficiency upper limit is considered consistent with the wind-shock model. We take $\eta_{\text{wind}}\approx7$\% as the fiducial value (the broad range in $\eta_{\text{wind}}$ from $\approx1/3 \times$ to $\approx3\times$ the fiducial value is to account for the large uncertainties on the wind efficiency). 

Toward higher values of $E(B-V)$, the majority of the observed radio luminosities in this sample can be explained by the wind-shock model with values $\eta_{\text{wind}} <7\%$ or $<21\%$, with $86_{-21}^{+5}\%$ and up to 100\% being consistent with the model, respectively, in the highest $E(B-V)$ bin. Even taking a conservative value of $\eta_{\text{wind}} <2.5\%$, $57_{-18}^{+15}\%$ of the QSOs in the highest $E(B-V)$ bin are consistent with the model. We note there are large uncertainties and assumptions on the parameters of the wind-shock model, but a significant fraction, particularly among the dustiest QSOs, can be explained by this model. Furthermore, given that we see an increase in the fraction of QSOs consistent with the model with increasing $E(B-V)$, this suggests there is a link between the dust extinction and wind-driven shocks. However, it is also possible that other physical processes may also be contributing (e.g., low-powered jets or shocks due to low-powered jets). Since redder QSOs are more likely to be radio-quiet/intermediate than their blue/control counterparts (see Figure \ref{fig:morphology_fracs}, and \citealp{Klindt2019FundamentalOrientation,Fawcett2020FundamentalQuasars,Rosario2020FundamentalLoTSS}), it is perhaps not surprising that a higher fraction of them are consistent with the wind-shock model, since radio-loudness is essentially what Equation \ref{equation: 1} traces - the ratio of radio luminosity, $L_{\nu}$, to $L_{\text{AGN}}$, traced by $L_{6\mu\text{m}}$. The $F_{1.4\text{GHz}} >3$ mJy cut we applied to our initial sample means we are missing the most radio-quiet QSOs (see Figure \ref{fig:Lradio_L6}), and it would be interesting to investigate how many of these show the steep slopes and radio luminosities required to be consistent with the wind-shock, to see if this potential mechanism extends down to the faintest QSOs. 

The above prediction of the radio luminosity and slopes from the \citet{Nims2015ObservationalNuclei} wind shock model only applies to (rest-frame) frequencies $\nu > \nu_{\text{cool}}$, which depends strongly on unknown parameters of the velocity of the shock, $v_s$, the magnetic field strength, $B$, and the radius of the forward shock, $R_s$, as \begin{equation}
    \nu_{\text{cool}} \approx 10 \text{ MHz} \left(\frac{v_s}{1000\text{ km s}^{-1}} \right)^2 \left(\frac{3 \text{ mG}}{B} \right)^3\left(\frac{100 \text{pc}}{R_s}\right)^2.
\end{equation}
For $\nu<\nu_{\text{cool}}$, the predicted emission is $\nu L_{\nu} \propto \nu^{1/2}$, or a spectral slope of $\alpha \approx -0.5$ (approximately the weighted $\alpha_{144-1400}$ median for the truly compact rQSOs). Assuming the fiducial values of magnetic field, $B= 3$ mG, and shock radius, $R_s = 100$ pc, the shock velocities required for $\nu_{\text{cool}} > \nu_{\text{rest}} = 173 - 490$ MHz (i.e., $\nu_{\text{obs}}>144$ MHz at the $z$ range of our sources, which would put the flatter slope in the $144-1400$ MHz spectral slope range), are in the range $v_s > (4.2 - 7.0) \times 1000$ km s$^{-1}$. This is generally higher than expected for host-galaxy scale outflows in quasars (e.g., \citealp{Harrison2012EnergeticActivity, Faucher-Giguere2012TheNuclei}), however this depends on a lot of unknowns, with a particularly strong dependence on the $B$-field. Therefore, it could be possible that for some QSOs with fast outflows, the flattening we observe in the $144-1400$ MHz spectral slope is due to this effect, but we would require radio data in this range (for example with uGMRT), to reliably characterise the radio SEDs and determine the nature of the flattening to low frequencies.

\subsection{Steep slopes from jets or winds?}
\label{sec:jet_v_wind}
\subsubsection{Observational evidence}
\label{sec:obs_jet_v_wind}
Previous studies have proposed that the radio emission observed in red QSOs may be powered by low-luminosity jets and/or AGN-driven winds (e.g., \citealp{Klindt2019FundamentalOrientation,Fawcett2020FundamentalQuasars,Rosario2020FundamentalLoTSS,Rosario2021FundamentalE-MERLIN}). As we have shown, the radio spectral slopes and luminosities we find for a large fraction of our truly compact rQSOs are consistent with the above \citet{Nims2015ObservationalNuclei} AGN-driven wind model, but the possibility that these systems host compact or low-power jets cannot be excluded. Evidence also suggests (e.g., \citealp{Tombesi2014UltrafastNuclei,Mehdipour2019RelationAGN}) that both jets and AGN-driven winds can occur within the same source, and so jets or jet-induced shocks could contribute to, or dominate, the observed radio properties. If jets are present in our truly compact (r)QSOs, they must be relatively compact and low- to moderately powered (in most cases), given the unresolved radio emission in VLASS (corresponding to scales of $< 8-20$ kpc depending on the redshift), and the low- to intermediate- radio loudness of the rQSOs. Although jets were initially thought to occur only in radio-loud quasars (e.g., \citealp{Kellermann1989VLASurvey,Urry1995UnifiedNuclei}), recent studies have revealed that even radio-quiet or intermediate systems can host compact radio jets (e.g., \citealp{Blundell2018TheQuasars,Jarvis2021TheGas,Njeri2025TheE-MERLIN}), and can significantly impact the ISM on sub-kpc scales, suggesting that small-scale jet-ISM interactions may be more common than previously thought. 

Without high resolution radio observations, it is challenging to distinguish compact jets from winds. However, radio SEDs may provide a unique signature of jets, and a curved/peaked radio SED is often used to identify Gigahertz-peaked spectrum (GPS) and compact steep spectrum (CSS) sources. These are both classes of compact radio sources that exhibit characteristic spectral turnovers at low frequencies due to synchrotron self-absorption. GPS sources are typically compact ($<$1 kpc) and peak at $\sim$1 GHz, and are thought to be the progenitors to CSS sources, which have larger sizes (up to $\sim$20 kpc) and turnover at lower frequencies ($\sim$100 MHz), with steep spectra ($\alpha<-0.5$) at higher frequencies \citep{ODea1998TheSources}. These spectral shapes reflect the evolution of the source as the emitting region expands and the turnover shifts to lower frequencies \citep{Bicknell2018RelativisticGalaxies}. Both GPS and CSS sources are commonly interpreted as young AGN, representing early stages in the life cycle of radio galaxies, and have been found to have broad [O~{\sc iii}]$\lambda5007$\AA\ outflows and disturbed gas kinematics \citep{Kukreti2023IonisedCycle,Kukreti2024ConnectingAGN}. In this scenario, GPS sources evolve into CSS sources and eventually into larger-scale radio galaxies, where the radio jets have "burst" out the host galaxy \citep{An2012TheSources}. Alternatively, their compactness may result not from their youth but from confinement by a dense interstellar medium that prevents the jets from expanding, i.e., "frustrated jets" \citep{ODea1998TheSources}.  

This scenario could explain both the steep $1.4-3$ GHz slopes and the flatter $144-1400$ MHz spectral slopes we observe in both our rQSO and cQSO samples, where either the radio SED is peaking within the $144-1400$ MHz frequency range, or is beginning to turn over toward frequencies below 144 MHz. In our sample, the current resolution puts upper limits on the physical size of $<8-20$ kpc, broadly consistent with CSS scales or potentially smaller. The radio luminosities of CSS sources are in the range $\sim 10^{25}-10^{29}$ WHz$^{-1}$ (see Figure 7 of \citealp{Jarvis2019PrevalenceQuasars}), putting our steep-spectrum QSOs, with radio luminosities $\sim 10^{24}-10^{28}$ WHz$^{-1}$ (this covers the same range of radio luminosities as for all the radio-detected QSOs; see Figure~\ref{fig:Lradio_L6}), at the lower end of this range. To confidently determine whether our steep-slope truly compact rQSOs are jet-dominated CSS/GPS sources would require more data in the $144-1400$ MHz frequency range, similar to \citet{Fawcett2025ConnectionQSOs}, but focused on those with steep radio spectral slopes of $\alpha\sim-1$, to see if they show evidence for a turnover in their spectrum, or are consistent with the wind-shock model down to lower radio frequencies. Given that jet-ISM interactions are also known to induce/produce shocks in the ISM \citep{Morganti2014ExtragalacticGas}, it is possible they could be producing the steep radio slopes we observe through a similar shock mechanism as an AGN-driven wind (see Section~\ref{sec:Nims}). Below we compare to simulations of jets and winds that predict the observed radio spectral slopes.

\subsubsection{Comparison to simulations}
\label{sec:sims}
Disentangling the contributions of winds and jets in radio sources remains challenging, particularly on small scales where the emission is unresolved. Recent studies (e.g., \citealp{Meenakshi2023AFields,Meenakshi2024APolarization,Yamada2024DecipheringNuclei,Xia2025RadioObservations}) have aimed to predict the observable signatures of jet- and wind-driven radio emission using numerical simulations. For example, \citet{Meenakshi2024APolarization} performed 3D relativistic magnetohydrodynamic (RMHD) simulations to model both jets and wide-angle AGN-driven winds interacting with an ambient ISM, spanning a range of jet/wind powers and densities. Their results indicate that jets typically produce narrow, collimated spines with compact hotspots, while winds tend to create broader emission structures, often enclosed by bright terminal arcs; however, they note that the morphology of the radio emission in the wind scenarios is sensitive to the initial density ratio between the wind and the ambient medium at the radius of injection, denoted by $\eta_w$ in their paper. Their analysis shows that the ability to distinguish between radio emission from QSO-driven winds and jets is strongly resolution-dependent. At lower resolutions, the resulting morphologies of winds and jets can appear indistinguishable. Very high spatial resolution observations may therefore be able to distinguish between the intricate structures associated with collimated jets and wide-angled winds. However, additional diagnostics, such as the spectral index, can provide useful complementary information. In this context, Meenakshi et al. (\textit{in prep.}) have examined the spatial variation of spectral slopes across different frequency ranges for compact jet and wind ($\sim4$ kpc) simulations with powers of  $10^{43}-10^{44}$ erg s$^{-1}$. They also computed integrated, flux-weighted spectral slopes between 1.4 and 3 GHz, and the following trends are seen: 

\begin{itemize}[align=parleft,left=0pt]
\item Stable higher-power jets ($10^{44}$ erg s$^{-1}$) show relatively flat spectra with spectral slopes near $\alpha=-0.6$, and the lower-power, kink-unstable\footnote{A current-driven instability in magnetically dominated jets caused by a strong toroidal magnetic field (e.g., \citealp{unstablejets1,unstablejets2}).} jets show steeper slopes of about $\alpha=-0.9$.
\item Light winds ($\eta_w = 0.04$) of all powers generally exhibit steep spectral slopes around $\alpha = -1$.
\item Dense winds ($\eta_w=4$) show intermediate values of about $\alpha=-0.7$.
\end{itemize}

These variations in spectral slopes arise from differences in the emission contributions and electron evolution in jet versus wind scenarios. In jets, electrons experience multiple shock encounters along the spine before reaching the jet head. As a result, the projected maps typically show flatter spectral indices near the head, where electrons retain higher energies and suffer fewer radiative losses, and steeper indices farther away toward the core. Consequently, the jet head and its nearby regions appear brighter in the stable higher-power jet, leading to a mean spectral slope of $\alpha =-0.6$. In contrast, low-power, kink-unstable jets evolve less smoothly; the jet's  head is weak, and the emission in the cocoon becomes dominated by the aged electrons, leading to a steeper spectrum at low frequencies. However, the mean value measured from regions near the head ($\sim2$ kpc towards the core) gives a shallower value of $\alpha =-0.7$. 

In wind-driven cases, the termination shock can produce relatively flat local spectra. However, in light winds, the extended regions above the shock are dominated by cooling electrons that flow outward beyond the termination shock. These regions exhibit steep spectral slopes and contribute more significantly to the total radio flux, resulting in a steeper integrated spectral slope ($\alpha\approx -1$). In dense winds, the termination shock is stronger and more extended than in light winds, but the resulting cocoon is less elongated in the forward direction. Consequently, a comparatively flatter spectral slope ($\alpha \approx -0.7$) is seen in this case. We note that these simulations do not include a multi-phase ISM, which would further restrict the propagation of light winds and could lead to steeper spectral slopes.

Overall, these simulations reveal that the evolution of the wind, and hence of the electrons transported along with the wind, depends sensitively on both the intrinsic wind properties and the characteristics of the surrounding medium. In particular, the spectral slopes in wind scenarios are highly responsive to the density contrast between the wind and the ambient ISM, parameters that are observationally difficult to constrain. We also note that the spectral slopes from these simulations are calculated at rest-frame $1.4-3$ GHz, whereas our spectral slopes are calculated from observed-frame $1.4-3$ GHz fluxes. Despite these limitations, it is interesting that light winds, or equivalently, winds expanding into denser ISM (and potentially therefore dusty) environments tend to produce steeper integrated spectral slopes. Recent studies (e.g., \citealp{Molyneux2025Evidence2}, and references therein) have shown that dust-reddened QSOs generally exhibit higher gas fractions than unobscured QSOs, with a modest trend toward increasing gas content in more dust-obscured systems (e.g. HotDOGs; \citealp{Sun2024PhysicalALMA}). Therefore, the light-wind model may plausibly explain the observed higher fraction of QSOs with steep radio spectral slopes at greater levels of dust extinction. However, we can not rule out that low-powered jets (which in these simulations can also produce up to $\alpha = -0.9$ for some lower-power kink-unstable scenarios) embedded within dusty QSOs could likewise produce the observed steep spectral slopes.

\subsection{Truly compact red QSOs as a dusty blow out phase}
\label{sec:dusty_blowout}
As discussed in Section~\ref{sec:radio_dust}, we only observe a trend of increasing radio detection fraction and steeper spectral slope with $E(B-V)$ among the truly compact QSOs, and not for the fake compact QSOs, for which the radio-detection fraction is independent of $E(B-V)$. This indicates the mechanism producing the excess radio emission in red QSOs is linked to whether or not the QSO has extended low-frequency emission.  The lack of extended low-frequency emission in the truly compact QSOs could suggest a lack of a previous recent episode of AGN radio activity, and it is these QSOs that show the most distinct radio properties and correlations between the dust and radio. 

In the dusty blow-out phase scenario, red QSOs are expected to host winds/jets that shock or interact with the surrounding ISM and expel or destroy the obscuring dust. The excess of steep radio spectral slopes we observe in some of our truly compact rQSOs is consistent with the prediction from an AGN-driven wind shock (\citealp{Nims2015ObservationalNuclei}; see Section~\ref{sec:Nims}). The clear trend we observe in radio detection fraction and spectral slopes with $E(B-V)$ strongly suggests that the radio emission in red QSOs is due to an interaction with the surrounding dusty medium. For any given (unresolved) QSO, the observed spectral slope will be a superposition of multiple radio-producing mechanisms within the QSO. The correlation we observe between steepening radio spectral slopes and $E(B-V)$ in our QSOs is therefore likely due to an increasing proportion of shock-dominated systems producing a spectral slope of $\alpha \sim-1$ as the dust (and most likely gas) density increases, hence steepening the median spectral slope and increasing the radio detection fraction. 

Recent studies have shown that red QSOs exhibit stronger C~{\sc iv}~$\lambda1549$\AA\ and [O~{\sc iii}]$\lambda5007$\AA\ outflows than typical blue QSOs, as well as a hot dust excess \citep{Rivera2021TheReddening,rivera24}, suggesting that AGN-driven winds are shocking and heating the surrounding dust. The red QSOs with the most prominent high-velocity wind components also tend to show the strongest MIR excesses, with the strength of this feature increasing with increasing $E(B-V)$, indicating a causal link between the reddening and hot dust, potentially distributed within AGN-driven winds. Furthermore, \citet{rivera24} find that red QSOs at intermediate radio-loudness exhibit significantly larger [O~{\sc iii}]$\lambda5007$\AA\ wind velocities, while no such dependence is seen in the control QSO sample, further supporting the idea that the presence of moderate radio emission in red QSOs may be tied to enhanced wind (or jet) activity. However, whether the outflows are triggered strictly by winds or jets is unclear. For example \citet{Molyneux2019ExtremeGalaxies} found that there is a link between radio-compactness and fast ionised [O~{\sc iii}]$\lambda5007$\AA\ outflows in spectroscopically identified AGN  - with the most compact radio sources having the most extreme outflow kinematics. High resolution follow-up observations for a subset of their sample showed the presence of radio jets and lobes, indicating the outflows could be triggered by young or compact radio jets.

However, there is evidence from known wind-dominated sources that winds are capable of producing excess radio emission. Broad absorption line QSOs  (BALQSOs; \citealp{Foltz1987The1303+308,Weymann1991ComparisonsObjects}) which host strong AGN-driven winds, are more likely to be radio detected than non-BALQSOs in sensitive low-frequency LOFAR observations (e.g., \citealp{Morabito2019TheSurvey,Petley2022ConnectingQuasars}). Additionally, \citet{Petley2022ConnectingQuasars} found that radio-detected BALQSOs have different absorption profiles to non-radio-detected BALQSOs, with increased reddening and broader C~{\sc iv}~$\lambda1549$\AA\ absorption features, which they speculate is due to a connection between the radio emission and the velocity of the outflow, potentially due to wind-driven shocks. Interestingly, BALQSOs are more likely to be dust-reddened compared to non-BALQSOs \citep{Sprayberry1992ExtinctionObjects,Richards2003RedSurvey,Urrutia2009TheQuasars}, suggesting a causal connection between AGN-driven winds and the presence of dust. \citet{Petley2024HowRate} found that in BALQSOs, reddening has a significant impact on the radio detection rate, although there were still differences between the BALQSOs and non-BALQSOs after controlling for colour. However, the exact relationship between the dust, winds and radio emission is unclear, with some work suggesting that windy QSOs are more likely to be dusty - as it has been predicted that outflows can be enhanced by dust (e.g., \citealp{Costa2018DrivingSimulations,Costa2018QuenchingRadiation})
or even provide favourable conditions for forming dust (e.g., \citealp{Sarangi2019DustWinds}).

Exploring whether the trends in spectral slope and radio detection rate extends to even dustier QSOs could provide more evidence for the dusty blow-out phase scenario. It is important to consider how ubiquitous this radio-opacity interaction is within the red QSO population. While it appears to become more prevalent at higher values of $E(B-V)$, we have only considered QSOs that are relatively bright ($F_{1.4\text{GHz}} > 3$~mJy), and it would be valuable exploring whether the same trend is found to much fainter radio fluxes, where the majority of the radio-emitting QSO population are detected. We note that, given the strong radio-detection fraction--$E(B-V)$ connection, which persists down to faint radio fluxes (e.g., \citealp{Fawcett2023AQSOs} finds a radio-detection fraction of up to 40\% with sensitive LoTSS low-frequency data), it appears likely that the excess of steep radio spectral slopes in truly compact rQSOs also persists to faint radio fluxes (see Appendix~\ref{sec:app_faint}). Our analysis focuses on the $\sim$5\% of rQSOs with necessary radio detections, and it will be important to test with future, deeper surveys whether lower-power radio systems display the same behaviour or whether the result is specific to the intermediate-power population probed here. Upcoming facilities such as the Square Kilometre Array \citep{SKA}, and upgrades such as the next generation VLA \citep{ngVLA} should provide the sensitivity required to push to significantly lower radio flux densities over wide areas. 

\section{Conclusions}
In this work, we have used samples of optically selected quasars ($0.2 < z< 2.4$) from SDSS with mid-infrared counterparts and radio counterparts in the LoTSS (144 MHz), FIRST (1.4 GHz), and VLASS (3 GHz) surveys to investigate differences in the multi-frequency radio properties of red (rQSOs) and control, i.e., typical blue QSOs (cQSOs). From a comparison of the radio morphologies and radio spectral shapes of red and blue QSOs, we have found:

\begin{itemize}[align=parleft,left=0pt]
\item \textbf{Not all high frequency radio compact QSOs are compact at low frequency}: A non-negligible fraction of FIRST-compact sources are extended in lower frequency LoTSS imaging (see Figure \ref{fig:morphology_fracs} \& Section~\ref{sec:multi_radio}) at the same approximate spatial resolution: $\sim 17\%$ of FIRST-compact rQSOs and $\sim 27\%$ of FIRST-compact cQSOs are extended at the lower frequency of LoTSS, potentially indicating a higher incidence of previous radio activity in cQSOs.
\item \textbf{The excess radio emission from truly compact rQSOs is likely due to shocks:} The truly compact rQSOs are made up of two sub-populations, one of which shows $\alpha_{1400-3000}$ values typical of FIRST compact QSOs, and the other of which shows the strongest radio detection excess around steep $\alpha_{1400-3000}$ values of $\sim-1$ (see Figure~\ref{fig:comb_hists} \& Section~\ref{sec:slopes}). This excess sub-population shows $\alpha_{1400-3000}$ values and radio luminosities broadly consistent with the prediction from an AGN-driven wind shock model (see Section~\ref{sec:Nims}). The excess sub-population also shows steeper $\alpha_{144-1400}$ spectral slopes than the non-excess sub-population, but they are flatter (on average) than the $\alpha_{1400-3000}$ slopes and this requires intermediate frequency radio observations to determine the origin.
\item \textbf{The contribution from the steep $\alpha_{1400-3000}$ population increases with dust extinction:} The increasing contribution of steep radio spectral slopes with $E(B-V)$ suggests dustier systems are more likely to be shock-dominated (see Figures~\ref{fig:EBV_det}-~\ref{fig:EBV_bins} \& Section~\ref{sec:radio_dust}), and the connection to $E(B-V)$ suggests a causal connection to opacity. We only observe this increasing contribution in the truly compact QSOs and not the fake compact QSOs, suggesting the excess radio emission is due to opacity on host galaxy scales or smaller.
\item \textbf{The radio--dust connection is only found for truly compact QSOs}: Both the radio detection fraction and radio spectral slopes are correlated with $E(B-V)$ for the truly compact QSOs, but not in QSOs with extended low frequency emission (see Figure~\ref{fig:EBV_det} \& Section~\ref{sec:radio_dust}). If the presence of extended low-frequency emission is indicative of either a more evolved QSO or a system which has had a recent earlier QSO episode, the correlation of the radio properties with dust found for only the truly compact QSOs suggests there is something unique about this potentially early (or first) phase of QSO activity. 
\end{itemize}

Overall, our results confirm previous work showing that the excess radio emission in red QSOs is likely due to an interaction of a wind or a compact jet with a surrounding dusty medium, and builds on this result to demonstrate that the dominant cause of this is a specific sub-population of red QSOs with compact radio morphologies at low and high frequency and steep radio spectral slopes. The intrinsic connection between the radio properties (detection rate and spectral slopes) and the amount of dust, found both in this work and previous works \citep{Fawcett2023AQSOs,Petley2024HowRate,rivera24}, shows that this interaction becomes more common as the population of QSOs becomes dustier. This is consistent with red QSOs being in a younger evolutionary phase, where this interaction decreases the amount of dust obscuration over time.

To further constrain if an AGN-driven wind or a jet is driving the excess radio emission in red QSOs, in a forthcoming study we will investigate the optical-MIR properties of our various colour/radio-morphology classifications. We will look for signatures of outflows (e.g., in C~{\sc
iv} $\lambda1549$\AA\ and [O~{\sc iii}]$\lambda5007$\AA) in the optical spectra or search for evidence of dust heating (MIR) from this interaction, and investigate if there is a link to the steep radio spectral slopes. We will also investigate if there are differences in the accretion properties between truly and fake compact rQSOs and cQSOs. Whilst previous work has found no strong difference in the accretion properties of red and blue QSOs \citep{Rivera2021TheReddening,Fawcett2022FundamentalX-shooter}, we can now hone in on the population of red QSOs (steep radio spectral slopes, compact at all radio frequencies), where the radio excess is most strongly occurring. Upcoming studies such as Sweijen et al. (\textit{in prep.}) investigate the LOFAR morphologies of red and blue QSOs across multiple resolutions (0.3 to 6 arcsec) and will be useful for constraining the scales and mechanism of the radio excess in red QSOs. Future mid-frequency observations (for example with uGMRT) are essential to bridge the decadal gap in frequency between LoTSS and FIRST to allow for more complete radio SEDs and to better understand the origin of the differences in the radio emission between truly and fake compact rQSOs and cQSOs.

\section*{Acknowledgements}
We would like to thank the anonymous referee for their constructive comments that improved the clarity of this work. The authors would like to thank Chris Done, Alastair Edge, Dipanjan Mukherjee, David Rosario and Frits Sweijen for their valuable discussions. CLS acknowledges support from the UK Science and Technology Facilities Council studentship under the grant ST/Y509346/1. DMA and CLG acknowledge support from the Science and Technology Facilities Council (grant codes: ST/T000244/1 and ST/X001075/1). CLG, VAF and CMH acknowledge support from United Kingdom Research and Innovation (grant code: MR/V022830/1). LKM is grateful for support from a UKRI Future Leaders Fellowship [MR/Y020405/1] and from STFC via LOFAR-UK [ST/V002406/1]. MM acknowledges support by the European Research Council under ERC-AdG grant PICOGAL-101019746, and support by the DFG Research Unit FOR-5195. RCH acknowledges support from NASA via grant numbers 80NSSC22K0862 and 80NSSC23K0485.

Funding for the Sloan Digital Sky Survey IV has been provided by the Alfred P. Sloan Foundation, the U.S. Department of Energy Office of Science, and the Participating Institutions. SDSS-IV acknowledges support and resources from the Center for High Performance Computing  at the University of Utah. The SDSS website is www.sdss4.org. SDSS-IV is managed by the Astrophysical Research Consortium for the Participating Institutions of the SDSS Collaboration including the Brazilian Participation Group, the Carnegie Institution for Science, Carnegie Mellon University, Center for Astrophysics | Harvard \& Smithsonian, the Chilean Participation Group, the French Participation Group, Instituto de Astrof\'isica de Canarias, The Johns Hopkins University, Kavli Institute for the Physics and Mathematics of the Universe (IPMU) / University of Tokyo, the Korean Participation Group, Lawrence Berkeley National Laboratory, Leibniz Institut f\"ur Astrophysik Potsdam (AIP),  Max-Planck-Institut f\"ur Astronomie (MPIA Heidelberg), Max-Planck Institut f\"ur Astrophysik (MPA Garching), Max-Planck-Institut f\"ur Extraterrestrische Physik (MPE), National Astronomical Observatories of China, New Mexico State University, New York University, University of Notre Dame, Observat\'ario Nacional / MCTI, The Ohio State University, Pennsylvania State University, Shanghai Astronomical Observatory, United Kingdom Participation Group, Universidad Nacional Aut\'onoma de M\'exico, University of Arizona, University of Colorado Boulder, University of Oxford, University of Portsmouth, University of Utah, University of Virginia, University of Washington, University of Wisconsin, Vanderbilt University, and Yale University.

This publication makes use of data products from the Wide-field Infrared Survey Explorer, which is a joint project of the University of California, Los Angeles, and the Jet Propulsion Laboratory/California Institute of Technology, funded by the National Aeronautics and Space Administration.

The National Radio Astronomy Observatory and Green Bank Observatory are facilities of the U.S. National Science Foundation operated under cooperative agreement by Associated Universities, Inc.

LOFAR data products were provided by the LOFAR Surveys Key Science project (LSKSP; https://lofar-surveys.org/) and were derived from observations with the International LOFAR Telescope (ILT). LOFAR (van Haarlem et al. 2013) is the Low Frequency Array designed and constructed by ASTRON. It has observing, data processing, and data storage facilities in several countries, which are owned by various parties (each with their own funding sources), and which are collectively operated by the ILT foundation under a joint scientific policy. The efforts of the LSKSP have benefited from funding from the European Research Council, NOVA, NWO, CNRS-INSU, the SURF Co-operative, the UK Science and Technology Funding Council and the Jülich Supercomputing Centre.

This research used data obtained with the Dark Energy Spectroscopic Instrument (DESI). DESI construction and operations is managed by the Lawrence Berkeley National Laboratory. This material is based upon work supported by the U.S. Department of Energy, Office of Science, Office of High-Energy Physics, under Contract No. DE–AC02–05CH11231, and by the National Energy Research Scientific Computing Center, a DOE Office of Science User Facility under the same contract. Additional support for DESI was provided by the U.S. National Science Foundation (NSF), Division of Astronomical Sciences under Contract No. AST-0950945 to the NSF’s National Optical-Infrared Astronomy Research Laboratory; the Science and Technology Facilities Council of the United Kingdom; the Gordon and Betty Moore Foundation; the Heising-Simons Foundation; the French Alternative Energies and Atomic Energy Commission (CEA); the National Council of Humanities, Science and Technology of Mexico (CONAHCYT); the Ministry of Science and Innovation of Spain (MICINN), and by the DESI Member Institutions: www.desi.lbl.gov/collaborating-institutions. The DESI collaboration is honored to be permitted to conduct scientific research on I’oligam Du’ag (Kitt Peak), a mountain with particular significance to the Tohono O’odham Nation. Any opinions, findings, and conclusions or recommendations expressed in this material are those of the author(s) and do not necessarily reflect the views of the U.S. National Science Foundation, the U.S. Department of Energy, or any of the listed funding agencies.

\section*{Data Availability}
The data underlying this article were derived from sources in the public domain. Optical spectroscopic and photometric data were obtained from the SDSS (see Section~\ref{sec:optical}). Imaging and photometric data were taken from the DESI Legacy Imaging Surveys and WISE (see Sections~\ref{sec:optical} \& \ref{sec:EBV}). Radio data were retrieved from the archival radio surveys FIRST, VLASS and LoTSS (see Section~\ref{sec:radio_data}).


\bibliographystyle{mnras}
\bibliography{references} 

@article{Kormendy2013CoevolutionGalaxies,
    title = {{Coevolution (or not) of supermassive black holes and host galaxies}},
    year = {2013},
    journal = {Annu. Rev. Astron. Astrophys.},
    author = {Kormendy, John and Ho, Luis C.},
    month = {8},
    pages = {511--653},
    volume = {51},
    doi = {10.1146/annurev-astro-082708-101811},
    issn = {00664146},
    arxivId = {1304.7762}
}

@article{Alexander2012WhatHoles,
    title = {{What drives the growth of black holes?}},
    year = {2012},
    journal = {\nar},
    author = {Alexander, D. M. and Hickox, R. C.},
    number = {4},
    month = {6},
    pages = {93--121},
    volume = {56},
    publisher = {North-Holland},
    doi = {10.1016/J.NEWAR.2011.11.003},
    issn = {1387-6473},
    arxivId = {1112.1949}
}

@ARTICLE{Hickox2018ObscuredNuclei,
       author = {{Hickox}, Ryan C. and {Alexander}, David M.},
        title = "{Obscured Active Galactic Nuclei}",
      journal = {\araa},
         year = 2018,
        month = sep,
       volume = {56},
        pages = {625-671},
          doi = {10.1146/annurev-astro-081817-051803},
archivePrefix = {arXiv},
       eprint = {1806.04680},
 primaryClass = {astro-ph.GA},
}

@article{Fabian2012ObservationalFeedback,
    author = "Fabian, A. C.",
    title = "{Observational Evidence of AGN Feedback}",
    eprint = "1204.4114",
    archivePrefix = "arXiv",
    primaryClass = "astro-ph.CO",
    doi = "10.1146/annurev-astro-081811-125521",
    journal = "Ann. Rev. Astron. Astrophys.",
    volume = "50",
    pages = "455--489",
    year = "2012"
}

@article{Harrison2024ObservationalInterpretation,
    title = {{Observational Tests of Active Galactic Nuclei Feedback: An Overview of Approaches and Interpretation}},
    year = {2024},
    journal = {Galaxies},
    author = {Harrison, Chris M. and Ramos Almeida, Cristina},
    number = {2},
    month = {4},
    volume = {12},
    publisher = {Multidisciplinary Digital Publishing Institute (MDPI)},
    url = {https://arxiv.org/abs/2404.08050v1},
    doi = {10.3390/galaxies12020017},
    issn = {20754434},
    arxivId = {2404.08050}
}

@article{Hopkins2006TheUniverse,
    title = {{The Relation between Quasar and Merging Galaxy Luminosity Functions and the Merger-driven Star Formation History of the Universe}},
    year = {2006},
    journal = {\apj},
    author = {Hopkins, Philip F. and Somerville, Rachel S. and Hernquist, Lars and Cox, Thomas J. and Robertson, Brant and Li, Yuexing},
    number = {2},
    month = {12},
    pages = {864},
    volume = {652},
    publisher = {American Astronomical Society},
    url = {https://ui.adsabs.harvard.edu/abs/2006ApJ...652..864H/abstract},
    doi = {10.1086/508503},
    issn = {0004-637X}
}

@article{Fabian1999TheHoles,
    title = {{The obscured growth of massive black holes}},
    year = {1999},
    journal = {\mnras},
    author = {Fabian, A. C.},
    number = {4},
    month = {10},
    pages = {L39-L43},
    volume = {308},
    publisher = {Blackwell Publishing Ltd},
    url = {https://ui.adsabs.harvard.edu/abs/1999MNRAS.308L..39F/abstract},
    doi = {10.1046/j.1365-8711.1999.03017.x},
    issn = {00358711},
    arxivId = {astro-ph/9908064}
}

@article{Antonucci1993UnifiedQuasars.,
    title = {{Unified models for active galactic nuclei and quasars.}},
    year = {1993},
    journal = {\araa},
    author = {Antonucci, Robert},
    number = {1},
    pages = {473},
    volume = {31},
    publisher = {Annual Reviews Inc.},
    url = {https://ui.adsabs.harvard.edu/abs/1993ARA%26A..31..473A/abstract},
    doi = {10.1146/ANNUREV.AA.31.090193.002353},
    issn = {0066-4146}
}

@article{Urry1995UnifiedNuclei,
    title = {{Unified Schemes for Radio-Loud Active Galactic Nuclei}},
    year = {1995},
    journal = {\pasp},
    author = {Urry, C. Megan and Padovani, Paolo},
    number = {715},
    month = {9},
    pages = {803},
    volume = {107},
    publisher = {IOP Publishing},
    url = {https://ui.adsabs.harvard.edu/abs/1995PASP..107..803U/abstract},
    doi = {10.1086/133630},
    issn = {0004-6280},
    arxivId = {astro-ph/9506063}
}

@article{Netzer2015RevisitingNuclei,
    title = {{Revisiting the unified model of active galactic nuclei}},
    year = {2015},
    journal = {\araa},
    author = {Netzer, Hagai},
    number = {1},
    month = {8},
    pages = {365--408},
    volume = {53},
    publisher = {Annual Reviews Inc.},
    url = {https://ui.adsabs.harvard.edu/abs/2015ARA&A..53..365N/abstract},
    doi = {10.1146/annurev-astro-082214-122302},
    issn = {00664146},
    arxivId = {1505.00811}
}

@article{Fawcett2022FundamentalX-shooter,
    title = {{Fundamental differences in the properties of red and blue quasars: Measuring the reddening and accretion properties with X-shooter}},
    year = {2022},
    journal = {\mnras},
    author = {Fawcett, V. A. and Alexander, D. M. and Rosario, D. J. and Klindt, L. and Lusso, E. and Morabito, L. K. and Calistro Rivera, G.},
    number = {1},
    month = {6},
    pages = {1254--1274},
    volume = {513},
    publisher = {Oxford University Press},
    url = {https://ui.adsabs.harvard.edu/abs/2022MNRAS.513.1254F/abstract},
    doi = {10.1093/mnras/stac945},
    issn = {13652966},
    arxivId = {2201.04139}
}

@article{Klindt2019FundamentalOrientation,
    title = {{Fundamental differences in the radio properties of red and blue quasars: evolution strongly favoured over orientation}},
    year = {2019},
    journal = {MNRAS},
    author = {Klindt, L and Alexander, D M and Rosario, D J and Lusso, E and Fotopoulou, S},
    pages = {3109--3128},
    volume = {488},
    url = {https://academic.oup.com/mnras/article/488/3/3109/5526240},
    doi = {10.1093/mnras/stz1771}
}

@article{Hopkins2008AActivity,
    title = {{A Cosmological Framework for the Co‐evolution of Quasars, Supermassive Black Holes, and Elliptical Galaxies. I. Galaxy Mergers and Quasar Activity}},
    year = {2008},
    journal = {\apjs},
    author = {Hopkins, Philip F. and Hernquist, Lars and Cox, Thomas J. and Kere{\v{s}}, Dušan},
    number = {2},
    month = {4},
    pages = {356--389},
    volume = {175},
    publisher = {American Astronomical Society},
    url = {https://ui.adsabs.harvard.edu/abs/2008ApJS..175..356H/abstract},
    doi = {10.1086/524362},
    issn = {0067-0049},
    arxivId = {0706.1243}
}

@article{Sanders1988UltraluminousQuasars,
    title = {{Ultraluminous Infrared Galaxies and the Origin of Quasars}},
    year = {1988},
    journal = {ApJ},
    author = {Sanders, D. B. and Soifer, B. T. and Elias, J. H. and Madore, B. F. and Matthews, K. and Neugebauer, G. and Scoville, N. Z. and Sanders, D. B. and Soifer, B. T. and Elias, J. H. and Madore, B. F. and Matthews, K. and Neugebauer, G. and Scoville, N. Z.},
    month = {2},
    pages = {74},
    volume = {325},
    publisher = {American Astronomical Society},
    url = {https://ui.adsabs.harvard.edu/abs/1988ApJ...325...74S/abstract},
    doi = {10.1086/165983},
    issn = {0004-637X}
}

@ARTICLE{Yue2024AQuasars,
       author = {{Yue}, B.-H. and {Best}, P.~N. and {Duncan}, K.~J. and {Calistro-Rivera}, G. and {Morabito}, L.~K. and {Petley}, J.~W. and {Prandoni}, I. and {R{\"o}ttgering}, H.~J.~A. and {Smith}, D.~J.~B.},
        title = "{A novel Bayesian approach for decomposing the radio emission of quasars: I. Modelling the radio excess in red quasars}",
      journal = {\mnras},
         year = 2024,
        month = apr,
       volume = {529},
       number = {4},
        pages = {3939-3957},
          doi = {10.1093/mnras/stae725},
archivePrefix = {arXiv},
       eprint = {2403.07074},
 primaryClass = {astro-ph.GA}
}

@ARTICLE{Petley2024HowRate,
       author = {{Petley}, James W. and {Morabito}, Leah K. and {Rankine}, Amy L. and {Richards}, Gordon T. and {Thomas}, Nicole L. and {Alexander}, David M. and {Fawcett}, Victoria A. and {Calistro Rivera}, Gabriela and {Prandoni}, Isabella and {Best}, Philip N. and {Kolwa}, Sthabile},
        title = "{How does the radio enhancement of broad absorption line quasars relate to colour and accretion rate?}",
      journal = {\mnras},
         year = 2024,
        month = apr,
       volume = {529},
       number = {3},
        pages = {1995-2007},
          doi = {10.1093/mnras/stae626},
archivePrefix = {arXiv},
       eprint = {2402.18623},
 primaryClass = {astro-ph.GA}
}

@ARTICLE{Fawcett2023AQSOs,
       author = {{Fawcett}, V.~A. and {Alexander}, D.~M. and {Brodzeller}, A. and {Edge}, A.~C. and {Rosario}, D.~J. and {Myers}, A.~D. and {Aguilar}, J. and {Ahlen}, S. and {Alfarsy}, R. and {Brooks}, D. and {Canning}, R. and {Circosta}, C. and {Dawson}, K. and {de la Macorra}, A. and {Doel}, P. and {Fanning}, K. and {Font-Ribera}, A. and {Forero-Romero}, J.~E. and {Gontcho A Gontcho}, S. and {Guy}, J. and {Harrison}, C.~M. and {Honscheid}, K. and {Juneau}, S. and {Kehoe}, R. and {Kisner}, T. and {Kremin}, A. and {Landriau}, M. and {Manera}, M. and {Meisner}, A.~M. and {Miquel}, R. and {Moustakas}, J. and {Nie}, J. and {Percival}, W.~J. and {Poppett}, C. and {Pucha}, R. and {Rossi}, G. and {Schlegel}, D. and {Siudek}, M. and {Tarl{\'e}}, G. and {Weaver}, B.~A. and {Zhou}, Z. and {Zou}, H.},
        title = "{A striking relationship between dust extinction and radio detection in DESI QSOs: evidence for a dusty blow-out phase in red QSOs}",
      journal = {\mnras},
         year = 2023,
        month = nov,
       volume = {525},
       number = {4},
        pages = {5575-5596},
          doi = {10.1093/mnras/stad2603},
archivePrefix = {arXiv},
       eprint = {2308.14790},
 primaryClass = {astro-ph.GA},
}

@article{Fawcett2020FundamentalQuasars,
    title = {{Fundamental differences in the radio properties of red and blue quasars: Enhanced compact AGN emission in red quasars}},
    year = {2020},
    journal = {\mnras},
    author = {Fawcett, V. A. and Alexander, D. M. and Rosario, D. J. and Klindt, L. and Fotopoulou, S. and Lusso, E. and Morabito, L. K. and Calistro Rivera, G.},
    number = {4},
    month = {6},
    pages = {4802--4818},
    volume = {494},
    publisher = {Oxford University Press},
    url = {https://ui.adsabs.harvard.edu/abs/2020MNRAS.494.4802F/abstract},
    doi = {10.1093/mnras/staa954},
    issn = {13652966},
    arxivId = {2004.01197}
}

@article{Rosario2020FundamentalLoTSS,
    title = {{Fundamental differences in the radio properties of red and blue quasars: Insight from the LOFAR Two-metre Sky Survey (LoTSS)}},
    year = {2020},
    journal = {\mnras},
    author = {Rosario, D. J. and Fawcett, V. A. and Klindt, L. and Alexander, D. M. and Morabito, L. K. and Fotopoulou, S. and Lusso, E. and Calistro Rivera, G.},
    number = {3},
    pages = {3061--3079},
    volume = {494},
    publisher = {Oxford University Press},
    url = {https://ui.adsabs.harvard.edu/abs/2020MNRAS.494.3061R/abstract},
    doi = {10.1093/MNRAS/STAA866},
    issn = {13652966},
    arxivId = {2004.01196}
}

@article{Rosario2021FundamentalE-MERLIN,
    title = {{Fundamental differences in the radio properties of red and blue quasars: Kiloparsec-scale structures revealed by e-MERLIN}},
    year = {2021},
    journal = {\mnras},
    author = {Rosario, D. J. and Alexander, D. M. and Moldon, J. and Klindt, L. and Thomson, A. P. and Morabito, L. and Fawcett, V. A. and Harrison, C. M.},
    number = {4},
    month = {8},
    pages = {5283--5300},
    volume = {505},
    publisher = {Oxford University Press},
    url = {https://ui.adsabs.harvard.edu/abs/2021MNRAS.505.5283R/abstract},
    doi = {10.1093/mnras/stab1653},
    issn = {13652966},
    arxivId = {2106.02646}
}

@article{Rivera2021TheReddening,
    title = {{The multiwavelength properties of red QSOs: Evidence for dusty winds as the origin of QSO reddening}},
    year = {2021},
    journal = {A{\&}A},
    author = {Calistro Rivera, G and Alexander, D M and Rosario, D J and Harrison, C M and Stalevski, M and Rakshit, S and Fawcett, V A and Morabito, L K and Klindt, L and Best, P N and Bonato, M and Bowler, R A A and Costa, T and Kondapally, R},
    pages = {102},
    volume = {649},
    url = {https://doi.org/10.1051/0004-6361/202040214},
    doi = {10.1051/0004-6361/202040214}
}

@article{Panessa2019TheNuclei,
    title = {{The origin of radio emission from radio-quiet active galactic nuclei}},
    year = {2019},
    journal = {NatAs},
    author = {Panessa, Francesca and Baldi, Ranieri Diego and Laor, Ari and Padovani, Paolo and Behar, Ehud and McHardy, Ian and Panessa, Francesca and Baldi, Ranieri Diego and Laor, Ari and Padovani, Paolo and Behar, Ehud and McHardy, Ian},
    number = {5},
    month = {5},
    pages = {387--396},
    volume = {3},
    publisher = {Nature Publishing Group},
    url = {https://ui.adsabs.harvard.edu/abs/2019NatAs...3..387P/abstract},
    doi = {10.1038/S41550-019-0765-4},
    issn = {2397-3366},
    arxivId = {arXiv:1902.05917}
}

@article{Laor2019WhatQuasars,
    title = {{What drives the radio slopes in radio-quiet quasars?}},
    year = {2019},
    journal = {MNRAS},
    author = {Laor, Ari and Baldi, Ranieri D and Behar, Ehud},
    pages = {5513--5523},
    volume = {482},
    url = {https://academic.oup.com/mnras/article/482/4/5513/5185104},
    doi = {10.1093/mnras/sty3098}
}

@article{Nims2015ObservationalNuclei,
    title = {{Observational signatures of galactic winds powered by active galactic nuclei}},
    year = {2015},
    journal = {\mnras},
    author = {Nims, Jesse and Quataert, Eliot and Faucher-Gigu{\'{e}}re, Claude André},
    number = {4},
    month = {3},
    pages = {3612--3622},
    volume = {447},
    publisher = {Oxford Academic},
    url = {https://dx.doi.org/10.1093/mnras/stu2648},
    doi = {10.1093/MNRAS/STU2648},
    issn = {0035-8711},
    arxivId = {1408.5141}
}

@article{Fawcett2025ConnectionQSOs,
    title = {{Connection between steep radio spectral slopes and dust extinction in QSOs: evidence for outflow-driven shocks in dusty QSOs}},
    year = {2025},
    journal = {MNRAS},
    author = {Fawcett, V. A. and Harrison, C. M. and Alexander, D. M. and Morabito, L. K. and Kharb, P. and Rosario, D. J. and Baghel, Janhavi and Ghosh, Salmoli and Sasikumar, Silpa and Petley, J. and Sargent, C. and CalistroRivera, G.},
    number = {2},
    month = {2},
    pages = {2003--2023},
    volume = {537},
    publisher = {Oxford University Press},
    url = {https://ui.adsabs.harvard.edu/abs/2025MNRAS.537.2003F/abstract},
    doi = {10.1093/MNRAS/STAF105},
    issn = {0035-8711},
    arxivId = {arXiv:2501.10501}
}

@article{Morganti2024WhatSurveys,
    title = {{What Have We Learned about the Life Cycle of Radio Galaxies from New Radio Surveys}},
    year = {2024},
    journal = {Galaxies},
    author = {Morganti, Raffaella},
    volume = {12},
    number = {2},
    url = {https://doi.org/10.3390/galaxies12020011},
    doi = {10.3390/galaxies12020011},
    arxivId = {2403.13329v1}
}

@article{Paris2018TheRelease,
    title = {{The sloan digital sky survey quasar catalog: Fourteenth data release}},
    year = {2018},
    journal = {\aap},
    author = {P{\^{a}}ris, Isabelle and Petitjean, Patrick and Aubourg, Éric and Myers, Adam D. and Streblyanska, Alina and Lyke, Brad W. and Anderson, Scott F. and Armengaud, Éric and Bautista, Julian and Blanton, Michael R. and Blomqvist, Michael and Brinkmann, Jonathan and Brownstein, Joel R. and Brandt, William Nielsen and Burtin, Étienne and Dawson, Kyle and De La Torre, Sylvain and Georgakakis, Antonis and Gil-Mar{\'{i}}n, Héctor and Green, Paul J. and Hall, Patrick B. and Kneib, Jean Paul and LaMassa, Stephanie M. and Le Goff, Jean Marc and MacLeod, Chelsea and Mariappan, Vivek and McGreer, Ian D. and Merloni, Andrea and Noterdaeme, Pasquier and Palanque-Delabrouille, Nathalie and Percival, Will J. and Ross, Ashley J. and Rossi, Graziano and Schneider, Donald P. and Seo, Hee Jong and Tojeiro, Rita and Weaver, Benjamin A. and Weijmans, Anne Marie and Y{\`{e}}che, Christophe and Zarrouk, Pauline and Zhao, Gong Bo},
    month = {5},
    pages = {A51},
    volume = {613},
    publisher = {EDP Sciences},
    url = {https://ui.adsabs.harvard.edu/abs/2018A%26A...613A..51P/abstract},
    doi = {10.1051/0004-6361/201732445},
    issn = {14320746},
    arxivId = {1712.05029}
}

@article{Aghanim2020PlanckParameters,
    title = {{Planck 2018 results - VI. Cosmological parameters}},
    year = {2020},
    journal = {\aap},
    author = {Aghanim, N. and Akrami, Y. and Ashdown, M. and Aumont, J. and Baccigalupi, C. and Ballardini, M. and Banday, A. J. and Barreiro, R. B. and Bartolo, N. and Basak, S. and Battye, R. and Benabed, K. and Bernard, J. P. and Bersanelli, M. and Bielewicz, P. and Bock, J. J. and Bond, J. R. and Borrill, J. and Bouchet, F. R. and Boulanger, F. and Bucher, M. and Burigana, C. and Butler, R. C. and Calabrese, E. and Cardoso, J. F. and Carron, J. and Challinor, A. and Chiang, H. C. and Chluba, J. and Colombo, L. P.L. and Combet, C. and Contreras, D. and Crill, B. P. and Cuttaia, F. and De Bernardis, P. and De Zotti, G. and Delabrouille, J. and Delouis, J. M. and Di Valentino, E. and Diego, J. M. and Dor{\'{e}}, O. and Douspis, M. and Ducout, A. and Dupac, X. and Dusini, S. and Efstathiou, G. and Elsner, F. and En{\ss}lin, T. A. and Eriksen, H. K. and Fantaye, Y. and Farhang, M. and Fergusson, J. and Fernandez-Cobos, R. and Finelli, F. and Forastieri, F. and Frailis, M. and Fraisse, A. A. and Franceschi, E. and Frolov, A. and Galeotta, S. and Galli, S. and Ganga, K. and G{\'{e}}nova-Santos, R. T. and Gerbino, M. and Ghosh, T. and Gonz{\'{a}}lez-Nuevo, J. and G{\'{o}}rski, K. M. and Gratton, S. and Gruppuso, A. and Gudmundsson, J. E. and Hamann, J. and Handley, W. and Hansen, F. K. and Herranz, D. and Hildebrandt, S. R. and Hivon, E. and Huang, Z. and Jaffe, A. H. and Jones, W. C. and Karakci, A. and Keih{\"{a}}nen, E. and Keskitalo, R. and Kiiveri, K. and Kim, J. and Kisner, T. S. and Knox, L. and Krachmalnicoff, N. and Kunz, M. and Kurki-Suonio, H. and Lagache, G. and Lamarre, J. M. and Lasenby, A. and Lattanzi, M. and Lawrence, C. R. and Le Jeune, M. and Lemos, P. and Lesgourgues, J. and Levrier, F. and Lewis, A. and Liguori, M. and Lilje, P. B. and Lilley, M. and Lindholm, V. and L{\'{o}}pez-Caniego, M. and Lubin, P. M. and Ma, Y. Z. and Maci{\'{a}}s-P{\'{e}}rez, J. F. and Maggio, G. and Maino, D. and Mandolesi, N. and Mangilli, A. and Marcos-Caballero, A. and Maris, M. and Martin, P. G. and Martinelli, M. and Mart{\'{i}}nez-Gonz{\'{a}}lez, E. and Matarrese, S. and Mauri, N. and McEwen, J. D. and Meinhold, P. R. and Melchiorri, A. and Mennella, A. and Migliaccio, M. and Millea, M. and Mitra, S. and Miville-Desch{\^{e}}nes, M. A. and Molinari, D. and Montier, L. and Morgante, G. and Moss, A. and Natoli, P. and N{\o}rgaard-Nielsen, H. U. and Pagano, L. and Paoletti, D. and Partridge, B. and Patanchon, G. and Peiris, H. V. and Perrotta, F. and Pettorino, V. and Piacentini, F. and Polastri, L. and Polenta, G. and Puget, J. L. and Rachen, J. P. and Reinecke, M. and Remazeilles, M. and Renzi, A. and Rocha, G. and Rosset, C. and Roudier, G. and Rubi{\~{n}}o-Mart{\'{i}}n, J. A. and Ruiz-Granados, B. and Salvati, L. and Sandri, M. and Savelainen, M. and Scott, D. and Shellard, E. P.S. and Sirignano, C. and Sirri, G. and Spencer, L. D. and Sunyaev, R. and Suur-Uski, A. S. and Tauber, J. A. and Tavagnacco, D. and Tenti, M. and Toffolatti, L. and Tomasi, M. and Trombetti, T. and Valenziano, L. and Valiviita, J. and Van Tent, B. and Vibert, L. and Vielva, P. and Villa, F. and Vittorio, N. and Wandelt, B. D. and Wehus, I. K. and White, M. and White, S. D.M. and Zacchei, A. and Zonca, A.},
    month = {9},
    pages = {A6},
    volume = {641},
    publisher = {EDP Sciences},
    url = {https://www.aanda.org/articles/aa/full_html/2020/09/aa33910-18/aa33910-18.html https://www.aanda.org/articles/aa/abs/2020/09/aa33910-18/aa33910-18.html},
    doi = {10.1051/0004-6361/201833910},
    issn = {0004-6361},
    arxivId = {1807.06209}
}

@article{Cameron2011OnApproach,
    title = {{On the estimation of confidence intervals for binomial population proportions in astronomy: The simplicity and superiority of the Bayesian approach}},
    year = {2011},
    journal = {\pasa},
    author = {Cameron, Ewan},
    number = {2},
    pages = {128--139},
    volume = {28},
    url = {https://ui.adsabs.harvard.edu/abs/2011PASA...28..128C/abstract},
    doi = {10.1071/AS10046},
    issn = {13233580},
    arxivId = {1012.0566}
}

@article{Wright2010ThePerformance,
    title = {{The Wide-field Infrared Survey Explorer (wise): Mission description and initial on-orbit performance}},
    year = {2010},
    journal = {\aj},
    author = {Wright, Edward L. and Eisenhardt, Peter R.M. and Mainzer, Amy K. and Ressler, Michael E. and Cutri, Roc M. and Jarrett, Thomas and Kirkpatrick, J. Davy and Padgett, Deborah and McMillan, Robert S. and Skrutskie, Michael and Stanford, S. A. and Cohen, Martin and Walker, Russell G. and Mather, John C. and Leisawitz, David and Gautier, Thomas N. and McLean, Ian and Benford, Dominic and Lonsdale, Carol J. and Blain, Andrew and Mendez, Bryan and Irace, William R. and Duval, Valerie and Liu, Fengchuan and Royer, Don and Heinrichsen, Ingolf and Howard, Joan and Shannon, Mark and Kendall, Martha and Walsh, Amy L. and Larsen, Mark and Cardon, Joel G. and Schick, Scott and Schwalm, Mark and Abid, Mohamed and Fabinsky, Beth and Naes, Larry and Tsai, Chao Wei},
    number = {6},
    month = {12},
    pages = {1868--1881},
    volume = {140},
    url = {https://ui.adsabs.harvard.edu/abs/2010AJ....140.1868W/abstract},
    doi = {10.1088/0004-6256/140/6/1868},
    issn = {00046256},
    arxivId = {1008.0031}
}

@ARTICLE{rivera24,
       author = {{Calistro Rivera}, G. and {Alexander}, D.~M. and {Harrison}, C.~M. and {Fawcett}, V.~A. and {Best}, P.~N. and {Williams}, W.~L. and {Hardcastle}, M.~J. and {Rosario}, D.~J. and {Smith}, D.~J.~B. and {Arnaudova}, M.~I. and {Escott}, E. and {G{\"u}rkan}, G. and {Kondapally}, R. and {Miley}, G. and {Morabito}, L.~K. and {Petley}, J. and {Prandoni}, I. and {R{\"o}ttgering}, H.~J.~A. and {Yue}, B. -H.},
        title = "{Ubiquitous radio emission in quasars: Predominant AGN origin and a connection to jets, dust, and winds}",
      journal = {\aap},
         year = 2024,
        month = nov,
       volume = {691},
          eid = {A191},
        pages = {A191},
          doi = {10.1051/0004-6361/202348982},
archivePrefix = {arXiv},
       eprint = {2312.10177},
 primaryClass = {astro-ph.GA},
}

@article{Lyke2020TheRelease,
    title = {{The Sloan Digital Sky Survey Quasar Catalog: Sixteenth Data Release}},
    year = {2020},
    journal = {\apjs},
    author = {Lyke, Brad W. and Higley, Alexandra N. and McLane, J. N. and Schurhammer, Danielle P. and Myers, Adam D. and Ross, Ashley J. and Dawson, Kyle and Chabanier, Solène and Martini, Paul and Busca, Nicolás G. and Mas des Bourboux, Hélion du and Salvato, Mara and Streblyanska, Alina and Zarrouk, Pauline and Burtin, Etienne and Anderson, Scott F. and Bautista, Julian and Bizyaev, Dmitry and Brandt, W. N. and Brinkmann, Jonathan and Brownstein, Joel R. and Comparat, Johan and Green, Paul and Macorra, Axel de la and Guti{\'{e}}rrez, Andrea Muñoz and Hou, Jiamin and Newman, Jeffrey A. and Palanque-Delabrouille, Nathalie and P{\^{a}}ris, Isabelle and Percival, Will J. and Petitjean, Patrick and Rich, James and Rossi, Graziano and Schneider, Donald P. and Smith, Alexander and Vivek, M. and Weaver, Benjamin Alan},
    number = {1},
    month = {9},
    pages = {8},
    volume = {250},
    publisher = {American Astronomical Society},
    url = {https://ui.adsabs.harvard.edu/abs/2020ApJS..250....8L/abstract},
    doi = {10.3847/1538-4365/aba623},
    issn = {0067-0049},
    arxivId = {2007.09001}
}

@article{FIRST1,
	author = {{Becker}, Robert H. and {White}, Richard L. and {Helfand}, David J.},
	doi = {10.1086/176166},
	journal = {ApJ},
	month = {9},
	pages = {559},
	title = {{The FIRST Survey: Faint Images of the Radio Sky at Twenty Centimeters}},
	volume = {450},
	year = 1995}

@article{Lacy_2020,
	author = {M. Lacy and S. A. Baum and C. J. Chandler and S. Chatterjee and T. E. Clarke and S. Deustua and J. English and J. Farnes and B. M. Gaensler and N. Gugliucci and G. Hallinan and B. R. Kent and A. Kimball and C. J. Law and T. J. W. Lazio and J. Marvil and S. A. Mao and D. Medlin and K. Mooley and E. J. Murphy and S. Myers and R. Osten and G. T. Richards and E. Rosolowsky and L. Rudnick and F. Schinzel and G. R. Sivakoff and L. O. Sjouwerman and R. Taylor and R. L. White and J. Wrobel and H. Andernach and A. J. Beasley and E. Berger and S. Bhatnager and M. Birkinshaw and G. C. Bower and W. N. Brandt and S. Brown and S. Burke-Spolaor and B. J. Butler and J. Comerford and P. B. Demorest and H. Fu and S. Giacintucci and K. Golap and T. G{\"u}th and C. A. Hales and R. Hiriart and J. Hodge and A. Horesh and {\v Z}. Ivezi{\'c} and M. J. Jarvis and A. Kamble and N. Kassim and X. Liu and L. Loinard and D. K. Lyons and J. Masters and M. Mezcua and G. A. Moellenbrock and T. Mroczkowski and K. Nyland and C. P. O'Dea and S. P. O'Sullivan and W. M. Peters and K. Radford and U. Rao and J. Robnett and J. Salcido and Y. Shen and A. Sobotka and S. Witz and M. Vaccari and R. J. van Weeren and A. Vargas and P. K. G. Williams and I. Yoon},
	doi = {10.1088/1538-3873/ab63eb},
	journal = {\pasp},
	month = {jan},
	number = {1009},
	pages = {035001},
	publisher = {The Astronomical Society of the Pacific},
	title = {The Karl G. Jansky Very Large Array Sky Survey (VLASS). Science Case and Survey Design},
	url = {https://dx.doi.org/10.1088/1538-3873/ab63eb},
	volume = {132},
	year = {2020}}

@ARTICLE{vanHaarlem2013LOFAR:ARray,
       author = {{van Haarlem}, M.~P. and {Wise}, M.~W. and {Gunst}, A.~W. and {Heald}, G. and {McKean}, J.~P. and {Hessels}, J.~W.~T. and {de Bruyn}, A.~G. and {Nijboer}, R. and {Swinbank}, J. and {Fallows}, R. and {Brentjens}, M. and {Nelles}, A. and {Beck}, R. and {Falcke}, H. and {Fender}, R. and {H{\"o}randel}, J. and {Koopmans}, L.~V.~E. and {Mann}, G. and {Miley}, G. and {R{\"o}ttgering}, H. and {Stappers}, B.~W. and {Wijers}, R.~A.~M.~J. and {Zaroubi}, S. and {van den Akker}, M. and {Alexov}, A. and {Anderson}, J. and {Anderson}, K. and {van Ardenne}, A. and {Arts}, M. and {Asgekar}, A. and {Avruch}, I.~M. and {Batejat}, F. and {B{\"a}hren}, L. and {Bell}, M.~E. and {Bell}, M.~R. and {van Bemmel}, I. and {Bennema}, P. and {Bentum}, M.~J. and {Bernardi}, G. and {Best}, P. and {B{\^\i}rzan}, L. and {Bonafede}, A. and {Boonstra}, A.-J. and {Braun}, R. and {Bregman}, J. and {Breitling}, F. and {van de Brink}, R.~H. and {Broderick}, J. and {Broekema}, P.~C. and {Brouw}, W.~N. and {Br{\"u}ggen}, M. and {Butcher}, H.~R. and {van Cappellen}, W. and {Ciardi}, B. and {Coenen}, T. and {Conway}, J. and {Coolen}, A. and {Corstanje}, A. and {Damstra}, S. and {Davies}, O. and {Deller}, A.~T. and {Dettmar}, R.-J. and {van Diepen}, G. and {Dijkstra}, K. and {Donker}, P. and {Doorduin}, A. and {Dromer}, J. and {Drost}, M. and {van Duin}, A. and {Eisl{\"o}ffel}, J. and {van Enst}, J. and {Ferrari}, C. and {Frieswijk}, W. and {Gankema}, H. and {Garrett}, M.~A. and {de Gasperin}, F. and {Gerbers}, M. and {de Geus}, E. and {Grie{\ss}meier}, J.-M. and {Grit}, T. and {Gruppen}, P. and {Hamaker}, J.~P. and {Hassall}, T. and {Hoeft}, M. and {Holties}, H.~A. and {Horneffer}, A. and {van der Horst}, A. and {van Houwelingen}, A. and {Huijgen}, A. and {Iacobelli}, M. and {Intema}, H. and {Jackson}, N. and {Jelic}, V. and {de Jong}, A. and {Juette}, E. and {Kant}, D. and {Karastergiou}, A. and {Koers}, A. and {Kollen}, H. and {Kondratiev}, V.~I. and {Kooistra}, E. and {Koopman}, Y. and {Koster}, A. and {Kuniyoshi}, M. and {Kramer}, M. and {Kuper}, G. and {Lambropoulos}, P. and {Law}, C. and {van Leeuwen}, J. and {Lemaitre}, J. and {Loose}, M. and {Maat}, P. and {Macario}, G. and {Markoff}, S. and {Masters}, J. and {McFadden}, R.~A. and {McKay-Bukowski}, D. and {Meijering}, H. and {Meulman}, H. and {Mevius}, M. and {Middelberg}, E. and {Millenaar}, R. and {Miller-Jones}, J.~C.~A. and {Mohan}, R.~N. and {Mol}, J.~D. and {Morawietz}, J. and {Morganti}, R. and {Mulcahy}, D.~D. and {Mulder}, E. and {Munk}, H. and {Nieuwenhuis}, L. and {van Nieuwpoort}, R. and {Noordam}, J.~E. and {Norden}, M. and {Noutsos}, A. and {Offringa}, A.~R. and {Olofsson}, H. and {Omar}, A. and {Orr{\'u}}, E. and {Overeem}, R. and {Paas}, H. and {Pandey-Pommier}, M. and {Pandey}, V.~N. and {Pizzo}, R. and {Polatidis}, A. and {Rafferty}, D. and {Rawlings}, S. and {Reich}, W. and {de Reijer}, J.-P. and {Reitsma}, J. and {Renting}, G.~A. and {Riemers}, P. and {Rol}, E. and {Romein}, J.~W. and {Roosjen}, J. and {Ruiter}, M. and {Scaife}, A. and {van der Schaaf}, K. and {Scheers}, B. and {Schellart}, P. and {Schoenmakers}, A. and {Schoonderbeek}, G. and {Serylak}, M. and {Shulevski}, A. and {Sluman}, J. and {Smirnov}, O. and {Sobey}, C. and {Spreeuw}, H. and {Steinmetz}, M. and {Sterks}, C.~G.~M. and {Stiepel}, H.-J. and {Stuurwold}, K. and {Tagger}, M. and {Tang}, Y. and {Tasse}, C. and {Thomas}, I. and {Thoudam}, S. and {Toribio}, M.~C. and {van der Tol}, B. and {Usov}, O. and {van Veelen}, M. and {van der Veen}, A.-J. and {ter Veen}, S. and {Verbiest}, J.~P.~W. and {Vermeulen}, R. and {Vermaas}, N. and {Vocks}, C. and {Vogt}, C. and {de Vos}, M. and {van der Wal}, E. and {van Weeren}, R. and {Weggemans}, H. and {Weltevrede}, P. and {White}, S. and {Wijnholds}, S.~J. and {Wilhelmsson}, T. and {Wucknitz}, O. and {Yatawatta}, S. and {Zarka}, P. and {Zensus}, A.},
        title = "{LOFAR: The LOw-Frequency ARray}",
      journal = {\aap},
         year = 2013,
        month = aug,
       volume = {556},
          eid = {A2},
        pages = {A2},
          doi = {10.1051/0004-6361/201220873},
archivePrefix = {arXiv},
       eprint = {1305.3550},
 primaryClass = {astro-ph.IM}
}

@article{gordon2021,
	archiveprefix = {arXiv},
	author = {{Gordon}, Yjan A. and {Boyce}, Michelle M. and {O Dea}, Christopher P. and {Rudnick}, Lawrence and {Andernach}, Heinz and {Vantyghem}, Adrian N. and {Baum}, Stefi A. and {Bui}, Jean-Paul and {Dionyssiou}, Mathew and {Safi-Harb}, Samar and {Sander}, Isabel},
	doi = {10.3847/1538-4365/ac05c0},
	eid = {30},
	eprint = {2102.11753},
	journal = {ApJS},
	month = aug,
	number = {2},
	pages = {30},
	primaryclass = {astro-ph.GA},
	title = {{A Quick Look at the 3 GHz Radio Sky. I. Source Statistics from the Very Large Array Sky Survey}},
	volume = {255},
	year = 2021}

@article{LoTSS_catalog,
	author = {{Shimwell}, T.~W. and {R{\"o}ttgering}, H.~J.~A. and {Best}, P.~N. and {Williams}, W.~L. and {Dijkema}, T.~J. and {de Gasperin}, F. and {Hardcastle}, M.~J. and {Heald}, G.~H. and {Hoang}, D.~N. and {Horneffer}, A. and {Intema}, H. and {Mahony}, E.~K. and {Mandal}, S. and {Mechev}, A.~P. and {Morabito}, L. and {Oonk}, J.~B.~R. and {Rafferty}, D. and {Retana-Montenegro}, E. and {Sabater}, J. and {Tasse}, C. and {van Weeren}, R.~J. and {Br{\"u}ggen}, M. and {Brunetti}, G. and {Chy{\.z}y}, K.~T. and {Conway}, J.~E. and {Haverkorn}, M. and {Jackson}, N. and {Jarvis}, M.~J. and {McKean}, J.~P. and {Miley}, G.~K. and {Morganti}, R. and {White}, G.~J. and {Wise}, M.~W. and {van Bemmel}, I.~M. and {Beck}, R. and {Brienza}, M. and {Bonafede}, A. and {Calistro Rivera}, G. and {Cassano}, R. and {Clarke}, A.~O. and {Cseh}, D. and {Deller}, A. and {Drabent}, A. and {van Driel}, W. and {Engels}, D. and {Falcke}, H. and {Ferrari}, C. and {Fr{\"o}hlich}, S. and {Garrett}, M.~A. and {Harwood}, J.~J. and {Heesen}, V. and {Hoeft}, M. and {Horellou}, C. and {Israel}, F.~P. and {Kapi{\'n}ska}, A.~D. and {Kunert-Bajraszewska}, M. and {McKay}, D.~J. and {Mohan}, N.~R. and {Orr{\'u}}, E. and {Pizzo}, R.~F. and {Prandoni}, I. and {Schwarz}, D.~J. and {Shulevski}, A. and {Sipior}, M. and {Smith}, D.~J.~B. and {Sridhar}, S.~S. and {Steinmetz}, M. and {Stroe}, A. and {Varenius}, E. and {van der Werf}, P.~P. and {Zensus}, J.~A. and {Zwart}, J.~T.~L.},
	doi = {10.1051/0004-6361/201629313},
	eid = {A104},
	eprint = {1611.02700},
	journal = {\aap},
	month = feb,
	pages = {A104},
	primaryclass = {astro-ph.IM},
	title = {{The LOFAR Two-metre Sky Survey. I. Survey description and preliminary data release}},
	volume = {598},
	year = 2017,
	}

@article{LoTSSDR2,
	archiveprefix = {arXiv},
	author = {{Shimwell}, T.~W. and {Hardcastle}, M.~J. and {Tasse}, C. and {Best}, P.~N. and {R{\"o}ttgering}, H.~J.~A. and {Williams}, W.~L. and {Botteon}, A. and {Drabent}, A. and {Mechev}, A. and {Shulevski}, A. and {van Weeren}, R.~J. and {Bester}, L. and {Br{\"u}ggen}, M. and {Brunetti}, G. and {Callingham}, J.~R. and {Chy{\.z}y}, K.~T. and {Conway}, J.~E. and {Dijkema}, T.~J. and {Duncan}, K. and {de Gasperin}, F. and {Hale}, C.~L. and {Haverkorn}, M. and {Hugo}, B. and {Jackson}, N. and {Mevius}, M. and {Miley}, G.~K. and {Morabito}, L.~K. and {Morganti}, R. and {Offringa}, A. and {Oonk}, J.~B.~R. and {Rafferty}, D. and {Sabater}, J. and {Smith}, D.~J.~B. and {Schwarz}, D.~J. and {Smirnov}, O. and {O'Sullivan}, S.~P. and {Vedantham}, H. and {White}, G.~J. and {Albert}, J.~G. and {Alegre}, L. and {Asabere}, B. and {Bacon}, D.~J. and {Bonafede}, A. and {Bonnassieux}, E. and {Brienza}, M. and {Bilicki}, M. and {Bonato}, M. and {Calistro Rivera}, G. and {Cassano}, R. and {Cochrane}, R. and {Croston}, J.~H. and {Cuciti}, V. and {Dallacasa}, D. and {Danezi}, A. and {Dettmar}, R.~J. and {Di Gennaro}, G. and {Edler}, H.~W. and {En{\ss}lin}, T.~A. and {Emig}, K.~L. and {Franzen}, T.~M.~O. and {Garc{\'\i}a-Vergara}, C. and {Grange}, Y.~G. and {G{\"u}rkan}, G. and {Hajduk}, M. and {Heald}, G. and {Heesen}, V. and {Hoang}, D.~N. and {Hoeft}, M. and {Horellou}, C. and {Iacobelli}, M. and {Jamrozy}, M. and {Jeli{\'c}}, V. and {Kondapally}, R. and {Kukreti}, P. and {Kunert-Bajraszewska}, M. and {Magliocchetti}, M. and {Mahatma}, V. and {Ma{\l}ek}, K. and {Mandal}, S. and {Massaro}, F. and {Meyer-Zhao}, Z. and {Mingo}, B. and {Mostert}, R.~I.~J. and {Nair}, D.~G. and {Nakoneczny}, S.~J. and {Nikiel-Wroczy{\'n}ski}, B. and {Orr{\'u}}, E. and {Pajdosz-{\'S}mierciak}, U. and {Pasini}, T. and {Prandoni}, I. and {van Piggelen}, H.~E. and {Rajpurohit}, K. and {Retana-Montenegro}, E. and {Riseley}, C.~J. and {Rowlinson}, A. and {Saxena}, A. and {Schrijvers}, C. and {Sweijen}, F. and {Siewert}, T.~M. and {Timmerman}, R. and {Vaccari}, M. and {Vink}, J. and {West}, J.~L. and {Wo{\l}owska}, A. and {Zhang}, X. and {Zheng}, J.},
	doi = {10.1051/0004-6361/202142484},
	eid = {A1},
	eprint = {2202.11733},
	journal = {\aap},
	month = mar,
	pages = {A1},
	primaryclass = {astro-ph.GA},
	title = {{The LOFAR Two-metre Sky Survey. V. Second data release}},
	volume = {659},
	year = 2022}

@article{LoTSSopt,
	archiveprefix = {arXiv},
	author = {{Hardcastle}, M.~J. and {Horton}, M.~A. and {Williams}, W.~L. and {Duncan}, K.~J. and {Alegre}, L. and {Barkus}, B. and {Croston}, J.~H. and {Dickinson}, H. and {Osinga}, E. and {R{\"o}ttgering}, H.~J.~A. and {Sabater}, J. and {Shimwell}, T.~W. and {Smith}, D.~J.~B. and {Best}, P.~N. and {Botteon}, A. and {Br{\"u}ggen}, M. and {Drabent}, A. and {de Gasperin}, F. and {G{\"u}rkan}, G. and {Hajduk}, M. and {Hale}, C.~L. and {Hoeft}, M. and {Jamrozy}, M. and {Kunert-Bajraszewska}, M. and {Kondapally}, R. and {Magliocchetti}, M. and {Mahatma}, V.~H. and {Mostert}, R.~I.~J. and {O'Sullivan}, S.~P. and {Pajdosz-{\'S}mierciak}, U. and {Petley}, J. and {Pierce}, J.~C.~S. and {Prandoni}, I. and {Schwarz}, D.~J. and {Shulewski}, A. and {Siewert}, T.~M. and {Stott}, J.~P. and {Tang}, H. and {Vaccari}, M. and {Zheng}, X. and {Bailey}, T. and {Desbled}, S. and {Goyal}, A. and {Gonano}, V. and {Hanset}, M. and {Kurtz}, W. and {Lim}, S.~M. and {Mielle}, L. and {Molloy}, C.~S. and {Roth}, R. and {Terentev}, I.~A. and {Torres}, M.},
	doi = {10.1051/0004-6361/202347333},
	eid = {A151},
	eprint = {2309.00102},
	journal = {\aap},
	month = oct,
	pages = {A151},
	primaryclass = {astro-ph.GA},
	title = {{The LOFAR Two-Metre Sky Survey. VI. Optical identifications for the second data release}},
	volume = {678},
	year = 2023}

@article{unWISE,
	archiveprefix = {arXiv},
	author = {{Schlafly}, Edward F. and {Meisner}, Aaron M. and {Green}, Gregory M.},
	doi = {10.3847/1538-4365/aafbea},
	eid = {30},
	eprint = {1901.03337},
	journal = {\apjs},
	month = feb,
	number = {2},
	pages = {30},
	primaryclass = {astro-ph.IM},
	title = {{The unWISE Catalog: Two Billion Infrared Sources from Five Years of WISE Imaging}},
	volume = {240},
	year = 2019}

@article{legacy,
	archiveprefix = {arXiv},
	author = {{Dey}, Arjun and {Schlegel}, David J. and {Lang}, Dustin and {Blum}, Robert and {Burleigh}, Kaylan and {Fan}, Xiaohui and {Findlay}, Joseph R. and {Finkbeiner}, Doug and {Herrera}, David and {Juneau}, St{\'e}phanie and {Landriau}, Martin and {Levi}, Michael and {McGreer}, Ian and {Meisner}, Aaron and {Myers}, Adam D. and {Moustakas}, John and {Nugent}, Peter and {Patej}, Anna and {Schlafly}, Edward F. and {Walker}, Alistair R. and {Valdes}, Francisco and {Weaver}, Benjamin A. and {Y{\`e}che}, Christophe and {Zou}, Hu and {Zhou}, Xu and {Abareshi}, Behzad and {Abbott}, T.~M.~C. and {Abolfathi}, Bela and {Aguilera}, C. and {Alam}, Shadab and {Allen}, Lori and {Alvarez}, A. and {Annis}, James and {Ansarinejad}, Behzad and {Aubert}, Marie and {Beechert}, Jacqueline and {Bell}, Eric F. and {BenZvi}, Segev Y. and {Beutler}, Florian and {Bielby}, Richard M. and {Bolton}, Adam S. and {Brice{\~n}o}, C{\'e}sar and {Buckley-Geer}, Elizabeth J. and {Butler}, Karen and {Calamida}, Annalisa and {Carlberg}, Raymond G. and {Carter}, Paul and {Casas}, Ricard and {Castander}, Francisco J. and {Choi}, Yumi and {Comparat}, Johan and {Cukanovaite}, Elena and {Delubac}, Timoth{\'e}e and {DeVries}, Kaitlin and {Dey}, Sharmila and {Dhungana}, Govinda and {Dickinson}, Mark and {Ding}, Zhejie and {Donaldson}, John B. and {Duan}, Yutong and {Duckworth}, Christopher J. and {Eftekharzadeh}, Sarah and {Eisenstein}, Daniel J. and {Etourneau}, Thomas and {Fagrelius}, Parker A. and {Farihi}, Jay and {Fitzpatrick}, Mike and {Font-Ribera}, Andreu and {Fulmer}, Leah and {G{\"a}nsicke}, Boris T. and {Gaztanaga}, Enrique and {George}, Koshy and {Gerdes}, David W. and {Gontcho}, Satya Gontcho A. and {Gorgoni}, Claudio and {Green}, Gregory and {Guy}, Julien and {Harmer}, Diane and {Hernandez}, M. and {Honscheid}, Klaus and {Huang}, Lijuan Wendy and {James}, David J. and {Jannuzi}, Buell T. and {Jiang}, Linhua and {Joyce}, Richard and {Karcher}, Armin and {Karkar}, Sonia and {Kehoe}, Robert and {Kneib}, Jean-Paul and {Kueter-Young}, Andrea and {Lan}, Ting-Wen and {Lauer}, Tod R. and {Le Guillou}, Laurent and {Le Van Suu}, Auguste and {Lee}, Jae Hyeon and {Lesser}, Michael and {Perreault Levasseur}, Laurence and {Li}, Ting S. and {Mann}, Justin L. and {Marshall}, Robert and {Mart{\'\i}nez-V{\'a}zquez}, C.~E. and {Martini}, Paul and {du Mas des Bourboux}, H{\'e}lion and {McManus}, Sean and {Meier}, Tobias Gabriel and {M{\'e}nard}, Brice and {Metcalfe}, Nigel and {Mu{\~n}oz-Guti{\'e}rrez}, Andrea and {Najita}, Joan and {Napier}, Kevin and {Narayan}, Gautham and {Newman}, Jeffrey A. and {Nie}, Jundan and {Nord}, Brian and {Norman}, Dara J. and {Olsen}, Knut A.~G. and {Paat}, Anthony and {Palanque-Delabrouille}, Nathalie and {Peng}, Xiyan and {Poppett}, Claire L. and {Poremba}, Megan R. and {Prakash}, Abhishek and {Rabinowitz}, David and {Raichoor}, Anand and {Rezaie}, Mehdi and {Robertson}, A.~N. and {Roe}, Natalie A. and {Ross}, Ashley J. and {Ross}, Nicholas P. and {Rudnick}, Gregory and {Safonova}, Sasha and {Saha}, Abhijit and {S{\'a}nchez}, F. Javier and {Savary}, Elodie and {Schweiker}, Heidi and {Scott}, Adam and {Seo}, Hee-Jong and {Shan}, Huanyuan and {Silva}, David R. and {Slepian}, Zachary and {Soto}, Christian and {Sprayberry}, David and {Staten}, Ryan and {Stillman}, Coley M. and {Stupak}, Robert J. and {Summers}, David L. and {Sien Tie}, Suk and {Tirado}, H. and {Vargas-Maga{\~n}a}, Mariana and {Vivas}, A. Katherina and {Wechsler}, Risa H. and {Williams}, Doug and {Yang}, Jinyi and {Yang}, Qian and {Yapici}, Tolga and {Zaritsky}, Dennis and {Zenteno}, A. and {Zhang}, Kai and {Zhang}, Tianmeng and {Zhou}, Rongpu and {Zhou}, Zhimin},
	doi = {10.3847/1538-3881/ab089d},
	eid = {168},
	eprint = {1804.08657},
	journal = {\aj},
	month = may,
	number = {5},
	pages = {168},
	primaryclass = {astro-ph.IM},
	title = {{Overview of the DESI Legacy Imaging Surveys}},
	volume = {157},
	year = 2019,
}

@article{moc,
	archiveprefix = {arXiv},
	author = {{Ochsenbein}, F. and {Bauer}, P. and {Marcout}, J.},
	doi = {10.1051/aas:2000169},
	eprint = {astro-ph/0002122},
	journal = {\aaps},
	month = apr,
	pages = {23-32},
	primaryclass = {astro-ph},
	title = {{The VizieR database of astronomical catalogues}},
	volume = {143},
	year = 2000}

@article{1.4,
	adsurl = {https://ui.adsabs.harvard.edu/abs/2003AJ....125..383A},
	archiveprefix = {arXiv},
	author = {{Alexander}, D.~M. and {Bauer}, F.~E. and {Brandt}, W.~N. and {Hornschemeier}, A.~E. and {Vignali}, C. and {Garmire}, G.~P. and {Schneider}, D.~P. and {Chartas}, G. and {Gallagher}, S.~C.},
	doi = {10.1086/346088},
	eprint = {astro-ph/0211267},
	journal = {AJ},
	month = feb,
	number = {2},
	pages = {383-397},
	primaryclass = {astro-ph},
	title = {{The Chandra Deep Field North Survey. XIV. X-Ray-Detected Obscured AGNs and Starburst Galaxies in the Bright Submillimeter Source Population}},
	volume = {125},
	year = 2003}

@ARTICLE{Patil2022RadioQuasars,
       author = {{Patil}, Pallavi and {Whittle}, Mark and {Nyland}, Kristina and {Lonsdale}, Carol and {Lacy}, Mark and {Kimball}, Amy E. and {Lonsdale}, Colin and {Peters}, Wendy and {Clarke}, Tracy E. and {Efstathiou}, Andreas and {Giacintucci}, Simona and {Kim}, Minjin and {Lanz}, Lauranne and {Mukherjee}, Dipanjan and {Polisensky}, Emil},
        title = "{Radio Spectra of Luminous, Heavily Obscured WISE-NVSS Selected Quasars}",
      journal = {\apj},
         year = 2022,
        month = jul,
       volume = {934},
       number = {1},
          eid = {26},
        pages = {26},
          doi = {10.3847/1538-4357/ac71b0},
archivePrefix = {arXiv},
       eprint = {2201.07349},
 primaryClass = {astro-ph.GA}
}

@article{Rakshit2020SpectralCatalog,
    title = {{Spectral Properties of Quasars from Sloan Digital Sky Survey Data Release 14: The Catalog}},
    year = {2020},
    journal = {\apjs},
    author = {Rakshit, Suvendu and Stalin, C. S. and Kotilainen, Jari},
    number = {1},
    month = {7},
    pages = {17},
    volume = {249},
    publisher = {American Astronomical Society},
    url = {https://ui.adsabs.harvard.edu/abs/2020ApJS..249...17R/abstract},
    doi = {10.3847/1538-4365/ab99c5},
    issn = {0067-0049},
    arxivId = {1910.10395}
}

@article{Richards2006TheThree,
    title = {{The SDSS Quasar Survey: Quasar Luminosity Function from Data Release Three}},
    year = {2006},
    journal = {AJ},
    author = {Richards, Gordon T. and Strauss, Michael A. and Fan, Xiaohui and Hall, Patrick B. and Jester, Sebastian and Schneider, Donald P. and Berk, Daniel E. Vanden and Stoughton, Chris and Anderson, Scott F. and Brunner, Robert J. and Gray, Jim and Gunn, James E. and Ivezic, Zeljko and Kirkland, Margaret K. and Knapp, G. R. and Loveday, Jon and Meiksin, Avery and Pope, Adrian and Szalay, Alexander S. and Thakar, Anirudda R. and Yanny, Brian and York, Donald G.},
    number = {6},
    month = {2},
    pages = {2766--2787},
    volume = {131},
    publisher = {American Astronomical Society},
    url = {http://arxiv.org/abs/astro-ph/0601434 http://dx.doi.org/10.1086/503559},
    doi = {10.1086/503559},
    arxivId = {astro-ph/0601434}
}

@article{Harrison2012EnergeticActivity,
    title = {{Energetic galaxy-wide outflows in high-redshift ultraluminous infrared galaxies hosting AGN activity}},
    year = {2012},
    journal = {\mnras},
    author = {Harrison, C. M. and Alexander, D. M. and Swinbank, A. M. and Smail, Ian and Alaghband-Zadeh, S. and Bauer, F. E. and Chapman, S. C. and Del Moro, A. and Hickox, R. C. and Ivison, R. J. and Men{\'{e}}ndez-Delmestre, Karín and Mullaney, J. R. and Nesvadba, N. P.H.},
    number = {2},
    month = {10},
    pages = {1073--1096},
    volume = {426},
    publisher = {Oxford University Press},
    url = {https://ui.adsabs.harvard.edu/abs/2012MNRAS.426.1073H/abstract},
    doi = {10.1111/j.1365-2966.2012.21723.x},
    issn = {13652966},
    arxivId = {1205.1801}
}

@article{Sun2017SizesImages,
    title = {{Sizes and Kinematics of Extended Narrow-line Regions in Luminous Obscured AGN Selected by Broadband Images}},
    year = {2017},
    journal = {\apj},
    author = {Sun, Ai-Lei and Greene, Jenny E. and Zakamska, Nadia L.},
    number = {2},
    month = {2},
    pages = {222},
    volume = {835},
    publisher = {American Astronomical Society},
    url = {https://ui.adsabs.harvard.edu/abs/2017ApJ...835..222S/abstract},
    doi = {10.3847/1538-4357/835/2/222},
    issn = {0004-637X},
    arxivId = {1611.04469}
}

@article{Liu2013ObservationsNebulae,
    title = {{Observations of feedback from radio-quiet quasars – II. Kinematics of ionized gas nebulae}},
    year = {2013},
    journal = {\mnras},
    author = {Liu, Guilin and Zakamska, Nadia L. and Greene, Jenny E. and Nesvadba, Nicole P.H. and Liu, Xin},
    number = {3},
    month = {12},
    pages = {2576--2597},
    volume = {436},
    publisher = {Oxford Academic},
    url = {https://dx.doi.org/10.1093/mnras/stt1755},
    doi = {10.1093/MNRAS/STT1755},
    issn = {0035-8711},
    arxivId = {1305.6922}
}

@article{Andonie2025AnGoals,
    title = {{An obscured quasar census with the 4MOST IR AGN survey: design, predicted properties, and scientific goals}},
    year = {2025},
    journal = {\mnras},
    author = {Andonie, Carolina and Alexander, David M. and Greenwell, Claire and Fotopoulou, Sotiria and Hickox, Ryan and Rosario, David J. and Villforth, Carolin and Buchner, Johannes and Krogager, Jens Kristian and Laloux, Brivael and Merloni, Andrea and Salvato, Mara and Streicher, Ole and Yan, Wei},
    number = {3},
    month = {5},
    pages = {2202},
    volume = {539},
    publisher = {Oxford University Press},
    url = {https://ui.adsabs.harvard.edu/abs/2025MNRAS.539.2202A/abstract},
    doi = {10.1093/MNRAS/STAF624},
    issn = {0035-8711}
}

@article{Faucher-Giguere2012TheNuclei,
    title = {{The physics of galactic winds driven by active galactic nuclei}},
    year = {2012},
    journal = {\mnras},
    author = {Faucher-Giguere, C. -A. and Quataert, E.},
    number = {1},
    month = {6},
    pages = {605--622},
    volume = {425},
    publisher = {Oxford University Press},
    url = {http://arxiv.org/abs/1204.2547 http://dx.doi.org/10.1111/j.1365-2966.2012.21512.x},
    doi = {10.1111/j.1365-2966.2012.21512.x},
    arxivId = {1204.2547v2}
}

@article{Molyneux2019ExtremeGalaxies,
    title = {{Extreme ionised outflows are more common when the radio emission is compact in AGN host galaxies}},
    year = {2019},
    journal = {\aap},
    author = {Molyneux, S. J. and Harrison, C. M. and Jarvis, M. E.},
    month = {11},
    pages = {A132},
    volume = {631},
    publisher = {EDP Sciences},
    url = {https://ui.adsabs.harvard.edu/abs/2019A%26A...631A.132M/abstract},
    doi = {10.1051/0004-6361/201936408},
    issn = {14320746},
    arxivId = {1909.05260}
}

@article{Sarangi2019DustWinds,
    title = {{Dust formation in AGN winds}},
    year = {2019},
    journal = {\apj},
    author = {Sarangi, Arkaprabha and Dwek, Eli and Kazanas, Demos},
    number = {2},
    month = {9},
    pages = {126},
    volume = {885},
    publisher = {American Astronomical Society},
    url = {http://arxiv.org/abs/1909.10426 http://dx.doi.org/10.3847/1538-4357/ab46a9},
    doi = {10.3847/1538-4357/ab46a9},
    arxivId = {1909.10426v1}
}

@article{Petley2022ConnectingQuasars,
    title = {{Connecting radio emission to AGN wind properties with broad absorption line quasars}},
    year = {2022},
    journal = {\mnras},
    author = {Petley, J. W. and Morabito, L. K. and Alexander, D. M. and Rankine, A. L. and Fawcett, V. A. and Rosario, D. J. and Matthews, J. H. and Shimwell, T. M. and Drabent, A.},
    number = {4},
    month = {10},
    pages = {5159--5174},
    volume = {515},
    publisher = {Oxford University Press},
    url = {https://ui.adsabs.harvard.edu/abs/2022MNRAS.515.5159P/abstract},
    doi = {10.1093/mnras/stac2067},
    issn = {13652966},
    arxivId = {2207.10102}
}

@article{Richards2003RedSurvey,
    title = {{Red and Reddened Quasars in the Sloan Digital Sky Survey}},
    year = {2003},
    journal = {AJ},
    author = {Richards, Gordon T. and Hall, Patrick B. and Berk, Daniel E. Vanden and Strauss, Michael A. and Schneider, Donald P. and Weinstein, Michael A. and Reichard, Timothy A. and York, Donald G. and Knapp, G. R. and Fan, Xiaohui and Ivezic, Zeljko and Brinkmann, J. and Budavari, Tamas and Csabai, Istvan and Nichol, R. C.},
    number = {3},
    month = {5},
    pages = {1131--1147},
    volume = {126},
    publisher = {American Astronomical Society},
    url = {http://arxiv.org/abs/astro-ph/0305305 http://dx.doi.org/10.1086/377014},
    doi = {10.1086/377014},
    arxivId = {astro-ph/0305305}
}

@article{Urrutia2009TheQuasars,
    title = {{The first-2mass red quasar survey. II. an anomalously high fraction of lobals in searches for dust-reddened quasars}},
    year = {2009},
    journal = {\apj},
    author = {Urrutia, Tanya and Becker, Robert H. and White, Richard L. and Glikman, Eilat and Lacy, Mark and Hodge, Jacqueline and Gregg, Michael D.},
    number = {2},
    pages = {1095--1109},
    volume = {698},
    publisher = {Institute of Physics Publishing},
    url = {https://ui.adsabs.harvard.edu/abs/2009ApJ...698.1095U/abstract},
    doi = {10.1088/0004-637X/698/2/1095},
    issn = {15384357},
    arxivId = {0808.3668}
}

@article{Morabito2019TheSurvey,
    title = {{The origin of radio emission in broad absorption line quasars: Results from the LOFAR Two-metre Sky Survey}},
    year = {2019},
    journal = {\aap},
    author = {Morabito, L. K. and Matthews, J. H. and Best, P. N. and G{\"{u}}rkan, G. and Jarvis, M. J. and Prandoni, I. and Duncan, K. J. and Hardcastle, M. J. and Kunert-Bajraszewska, M. and Mechev, A. P. and Mooney, S. and Sabater, J. and R{\"{o}}ttgering, H. J.A. and Shimwell, T. W. and Smith, D. J.B. and Tasse, C. and Williams, W. L.},
    month = {2},
    pages = {A15},
    volume = {622},
    publisher = {EDP Sciences},
    url = {https://ui.adsabs.harvard.edu/abs/2019A%26A...622A..15M/abstract},
    doi = {10.1051/0004-6361/201833821},
    issn = {14320746},
    arxivId = {1811.07931}
}

@article{Tombesi2014UltrafastNuclei,
    title = {{Ultrafast outflows in radio-loud active galactic nuclei}},
    year = {2014},
    journal = {\mnras},
    author = {Tombesi, F. and Tazaki, F. and Mushotzky, R. F. and Ueda, Y. and Cappi, M. and Gofford, J. and Reeves, J. N. and Guainazzi, M.},
    number = {3},
    pages = {2154--2182},
    volume = {443},
    publisher = {Oxford University Press},
    url = {https://ui.adsabs.harvard.edu/abs/2014MNRAS.443.2154T/abstract},
    doi = {10.1093/mnras/stu1297},
    issn = {13652966},
    arxivId = {1406.7252}
}

@article{Mehdipour2019RelationAGN,
    title = {{Relation between winds and jets in radio-loud AGN}},
    year = {2019},
    journal = {\aap},
    author = {Mehdipour, Missagh and Costantini, Elisa},
    pages = {A25},
    volume = {625},
    publisher = {EDP Sciences},
    url = {https://ui.adsabs.harvard.edu/abs/2019A&A...625A..25M/abstract},
    doi = {10.1051/0004-6361/201935205},
    issn = {14320746},
    arxivId = {1903.11605}
}

@article{Kellermann1989VLASurvey,
    title = {{VLA Observations of Objects in the Palomar Bright Quasar Survey}},
    year = {1989},
    journal = {AJ},
    author = {Kellermann, K. I. and Sramek, R. and Schmidt, M. and Shaffer, D. B. and Green, R. and Kellermann, K. I. and Sramek, R. and Schmidt, M. and Shaffer, D. B. and Green, R.},
    number = {4},
    month = {10},
    pages = {1195},
    volume = {98},
    publisher = {American Astronomical Society},
    url = {https://ui.adsabs.harvard.edu/abs/1989AJ.....98.1195K/abstract},
    doi = {10.1086/115207},
    issn = {0004-6256}
}

@article{Jarvis2021TheGas,
    title = {{The quasar feedback survey: Discovering hidden Radio-AGN and their connection to the host galaxy ionized gas}},
    year = {2021},
    journal = {\mnras},
    author = {Jarvis, M. E. and Harrison, C. M. and Mainieri, V. and Alexander, D. M. and Battaia, F. Arrigoni and Rivera, G. Calistro and Circosta, C. and Costa, T. and de Breuck, C. and Edge, A. C. and Girdhar, A. and Kakkad, D. and Kharb, P. and Lansbury, G. B. and Molyneux, S. J. and Mukherjee, D. and Mullaney, J. R. and Farina, E. P. and Silpa, S. and Thomson, A. P. and Ward, S. R.},
    number = {2},
    month = {5},
    pages = {1780},
    volume = {503},
    publisher = {Oxford University Press},
    url = {https://ui.adsabs.harvard.edu/abs/2021MNRAS.503.1780J/abstract},
    doi = {10.1093/mnras/stab549},
    issn = {13652966},
    arxivId = {2103.00014}
}

@ARTICLE{Blundell2018TheQuasars,
       author = {{Blundell}, Katherine M. and {Beasley}, Anthony J.},
        title = "{The central engines of radio-quiet quasars}",
      journal = {\mnras},
         year = 1998,
        month = aug,
       volume = {299},
       number = {1},
        pages = {165-170},
          doi = {10.1046/j.1365-8711.1998.01752.x},
archivePrefix = {arXiv},
       eprint = {astro-ph/9805169},
 primaryClass = {astro-ph},
}

@article{ODea1998TheSources,
    title = {{The Compact Steep-Spectrum and Gigahertz Peaked-Spectrum Radio Sources}},
    year = {1998},
    journal = {\pasp},
    author = {O'Dea, Christopher P.},
    number = {747},
    month = {5},
    pages = {493},
    volume = {110},
    publisher = {IOP Publishing},
    url = {https://ui.adsabs.harvard.edu/abs/1998PASP..110..493O/abstract},
    doi = {10.1086/316162},
    issn = {0004-6280}
}

@article{Jarvis2019PrevalenceQuasars,
    title = {{Prevalence of radio jets associated with galactic outflows and feedback from quasars}},
    year = {2019},
    journal = {\mnras},
    author = {Jarvis, M. E. and Harrison, C. M. and Thomson, A. P. and Circosta, C. and Mainieri, V. and Alexander, D. M. and Edge, A. C. and Lansbury, G. B. and Molyneux, S. J. and Mullaney, J. R.},
    number = {2},
    month = {5},
    pages = {2710--2730},
    volume = {485},
    publisher = {Oxford Academic},
    url = {https://dx.doi.org/10.1093/mnras/stz556},
    doi = {10.1093/MNRAS/STZ556},
    issn = {0035-8711},
    arxivId = {1902.07727}
}

@ARTICLE{Yamada2024DecipheringNuclei,
       author = {{Yamada}, Tomoya and {Sakai}, Nobuyuki and {Inoue}, Yoshiyuki and {Michiyama}, Tomonari},
        title = "{Deciphering Radio Emissions from Accretion Disk Winds in Radio-quiet Active Galactic Nuclei}",
      journal = {\apj},
         year = 2024,
        month = jun,
       volume = {968},
       number = {2},
          eid = {116},
        pages = {116},
          doi = {10.3847/1538-4357/ad3a63},
archivePrefix = {arXiv},
       eprint = {2404.04632},
 primaryClass = {astro-ph.HE}
}

@ARTICLE{Xia2025RadioObservations,
       author = {{Xia}, Haojie and {Yuan}, Feng and {Li}, Zhiyuan and {Zhu}, Bocheng},
        title = "{Radio signatures of AGN-wind-driven shocks in elliptical galaxies: From simulations to observations}",
      journal = {\aap},
         year = 2025,
        month = jul,
          eid = {arXiv:2507.19716},
    volume = {704},
        pages = {A148},
          doi = {10.48550/arXiv.2507.19716},
archivePrefix = {arXiv},
       eprint = {2507.19716},
 primaryClass = {astro-ph.GA}
}

@ARTICLE{Meenakshi2024APolarization,
       author = {{Meenakshi}, Moun and {Mukherjee}, Dipanjan and {Bodo}, Gianluigi and {Rossi}, Paola and {Harrison}, Chris M.},
        title = "{A comparative study of radio signatures from winds and jets: modelling synchrotron emission and polarization}",
      journal = {\mnras},
         year = 2024,
        month = sep,
       volume = {533},
       number = {2},
        pages = {2213-2231},
          doi = {10.1093/mnras/stae1890},
archivePrefix = {arXiv},
       eprint = {2408.00099},
 primaryClass = {astro-ph.HE},
}

@article{Molyneux2025Evidence2,
    title = {{Evidence for universal gas depletion in a sample of 41 luminous Type 1 quasars at z ∼ 2}},
    year = {2025},
    journal = {\mnras},
    author = {Molyneux, S. J. and Banerji, M. and Temple, M. J. and Aravena, M. and Assef, R. J. and Hewett, P. and Jones, G. C. and Puglisi, A. and Rankine, A. L. and Ricci, C. and Stepney, M. and Tang, S.},
    number = {1},
    month = {6},
    pages = {1163--1184},
    volume = {540},
    publisher = {Oxford University Press},
    url = {https://ui.adsabs.harvard.edu/abs/2025MNRAS.540.1163M/abstract},
    doi = {10.1093/mnras/staf769},
    issn = {13652966},
    arxivId = {2505.03884}
}

@ARTICLE{Sun2024PhysicalALMA,
       author = {{Sun}, Weibin and {Fan}, Lulu and {Han}, Yunkun and {Knudsen}, Kirsten K. and {Chen}, Guangwen and {Zhang}, Hong-Xin},
        title = "{Physical Properties of Hyperluminous, Dust-obscured Quasars at z {\ensuremath{\sim}} 3: Multiwavelength Spectral Energy Distribution Analysis and Cold Gas Content Revealed by ALMA}",
      journal = {\apj},
         year = 2024,
        month = mar,
       volume = {964},
       number = {1},
          eid = {95},
        pages = {95},
          doi = {10.3847/1538-4357/ad22e3},
archivePrefix = {arXiv},
       eprint = {2402.15306},
 primaryClass = {astro-ph.GA}
}

@article{Zakamska2014QuasarQuasars,
    title = {{Quasar feedback and the origin of radio emission in radio-quiet quasars}},
    year = {2014},
    journal = {\mnras},
    author = {Zakamska, Nadia L. and Greene, Jenny E.},
    number = {1},
    month = {7},
    pages = {784--804},
    volume = {442},
    publisher = {Oxford University Press},
    url = {https://ui.adsabs.harvard.edu/abs/2014MNRAS.442..784Z/abstract},
    doi = {10.1093/mnras/stu842},
    issn = {13652966},
    arxivId = {1402.6736}
}

@article{Blecha2018TheStudy,
    title = {{The power of infrared AGN selection in mergers: A theoretical study}},
    year = {2018},
    journal = {\mnras},
    author = {Blecha, Laura and Snyder, Gregory F. and Satyapal, Shobita and Ellison, Sara L.},
    number = {3},
    month = {8},
    pages = {3056--3071},
    volume = {478},
    publisher = {Oxford University Press},
    url = {https://ui.adsabs.harvard.edu/abs/2018MNRAS.478.3056B/abstract},
    doi = {10.1093/MNRAS/STY1274},
    issn = {13652966},
    arxivId = {1711.02094}
}

@article{Ishibashi2016AGN-starburstFeedback,
    title = {{AGN-starburst evolutionary connection: A physical interpretation based on radiative feedback}},
    year = {2016},
    journal = {\mnras},
    author = {Ishibashi, W. and Fabian, A. C.},
    number = {2},
    month = {12},
    pages = {1291--1296},
    volume = {463},
    publisher = {Oxford University Press},
    url = {https://ui.adsabs.harvard.edu/abs/2016MNRAS.463.1291I/abstract},
    doi = {10.1093/mnras/stw2063},
    issn = {13652966},
    arxivId = {1609.08963}
}

@article{Glikman2022TheRegime,
    title = {{The WISE-2MASS Survey: Red Quasars Into the Radio Quiet Regime}},
    year = {2022},
    journal = {\apj},
    author = {Glikman, E. and Lacy, M. and LaMassa, S. and Bradley, C. and Djorgovski, S. G. and Urrutia, T. and Gates, E. L. and Graham, M. J. and Urry, M. and Yoon, I.},
    number = {2},
    month = {8},
    pages = {119},
    volume = {934},
    publisher = {American Astronomical Society},
    url = {https://ui.adsabs.harvard.edu/abs/2022ApJ...934..119G/abstract},
    doi = {10.3847/1538-4357/ac6bee},
    issn = {0004-637X},
    arxivId = {2204.13745}
}

@article{Webster1995EvidenceQuasars,
    title = {{Evidence for a large undetected population of dust-reddened quasars}},
    year = {1995},
    journal = {Nature},
    author = {Webster, Rachel L. and Francis, Paul J. and Petersont, Bruce A. and Drinkwater, Michael J. and Masci, Frank J.},
    number = {6531},
    month = {6},
    pages = {469--471},
    volume = {375},
    publisher = {Nature Publishing Group},
    url = {https://www.nature.com/articles/375469a0},
    doi = {10.1038/375469a0},
    issn = {1476-4687}
}

@article{Hwang2018WindsQuasars,
    title = {{Winds as the origin of radio emission in z = 2.5 radio-quiet extremely red quasars}},
    year = {2018},
    journal = {\mnras},
    author = {Hwang, Hsiang Chih and Zakamska, Nadia L. and Alexandroff, Rachael M. and Hamann, Fred and Greene, Jenny E. and Perrotta, Serena and Richards, Gordon T.},
    number = {1},
    month = {6},
    pages = {830--844},
    volume = {477},
    publisher = {Oxford Academic},
    url = {https://dx.doi.org/10.1093/mnras/sty742},
    doi = {10.1093/MNRAS/STY742},
    issn = {0035-8711},
    arxivId = {1803.02821}
}

@article{Morganti2014ExtragalacticGas, 
title={Radio jets clearing the way through galaxies: the view from Hi and molecular gas}, 
volume={10}, 
DOI={10.1017/S1743921315002331}, 
number={S313}, 
journal={Proc. IAU}, 
author={Morganti, Raffaella}, 
year={2014}, pages={283–288}}

@article{Sprayberry1992ExtinctionObjects,
    title = {{Extinction in Low-Ionization Broad Absorption Line Quasi-stellar Objects}},
    year = {1992},
    journal = {\apj},
    author = {Sprayberry, David and Foltz, Craig B.},
    month = {5},
    pages = {39},
    volume = {390},
    publisher = {American Astronomical Society},
    url = {https://ui.adsabs.harvard.edu/abs/1992ApJ...390...39S/abstract},
    doi = {10.1086/171257},
    issn = {0004-637X}
}

@article{Kim2018WhatAnalysis,
    title = {{What makes red quasars red? Observational evidence for dust extinction from line ratio analysis}},
    year = {2018},
    journal = {\aap},
    author = {Kim, Dohyeong and Im, Myungshin},
    month = {2},
    pages = {A31},
    volume = {610},
    publisher = {EDP Sciences},
    url = {https://ui.adsabs.harvard.edu/abs/2018A%26A...610A..31K/abstract},
    doi = {10.1051/0004-6361/201731963},
    issn = {14320746},
    arxivId = {1712.01851}
}

@article{Foltz1987The1303+308,
    title = {{The Complex Absorption Spectrum of the Broad Absorption Line QSO 1303+308}},
    year = {1987},
    journal = {\apj},
    author = {Foltz, Craig B. and Weymann, Ray J. and Morris, Simon L. and Turnshek, David A.},
    month = {6},
    pages = {450},
    volume = {317},
    publisher = {American Astronomical Society},
    url = {https://ui.adsabs.harvard.edu/abs/1987ApJ...317..450F/abstract},
    doi = {10.1086/165290},
    issn = {0004-637X}
}

@article{Weymann1991ComparisonsObjects,
    title = {{Comparisons of the Emission-Line and Continuum Properties of Broad Absorption Line and Normal Quasi-stellar Objects}},
    year = {1991},
    journal = {ApJ},
    author = {Weymann, Ray J. and Morris, Simon L. and Foltz, Craig B. and Hewett, Paul C. and Weymann, Ray J. and Morris, Simon L. and Foltz, Craig B. and Hewett, Paul C.},
    month = {5},
    pages = {23},
    volume = {373},
    publisher = {American Astronomical Society},
    url = {https://ui.adsabs.harvard.edu/abs/1991ApJ...373...23W/abstract},
    doi = {10.1086/170020},
    issn = {0004-637X}
}

@article{Fawcett2021HowProperties,
    title = {{How are red and blue quasars different? The radio properties}},
    year = {2021},
    journal = {Galaxies},
    author = {Fawcett, Victoria A. and Alexander, David M. and Rosario, David J. and Klindt, Lizelke},
    number = {4},
    month = {12},
    pages = {107},
    volume = {9},
    publisher = {MDPI},
    url = {https://ui.adsabs.harvard.edu/abs/2021Galax...9..107F/abstract},
    doi = {10.3390/galaxies9040107},
    issn = {20754434},
    arxivId = {2111.10384}
}

@ARTICLE{Kukreti2023IonisedCycle,
       author = {{Kukreti}, Pranav and {Morganti}, Raffaella and {Tadhunter}, Clive and {Santoro}, Francesco},
        title = "{Ionised gas outflows over the radio AGN life cycle}",
      journal = {\aap},
         year = 2023,
        month = jun,
       volume = {674},
          eid = {A198},
        pages = {A198},
          doi = {10.1051/0004-6361/202245691},
archivePrefix = {arXiv},
       eprint = {2305.03725},
 primaryClass = {astro-ph.GA},
}

@article{Shimwell2019TheRelease,
    title = {{The LOFAR Two-metre Sky Survey: II. First data release}},
    year = {2019},
    journal = {\aap},
    author = {Shimwell, T. W. and Tasse, C. and Hardcastle, M. J. and Mechev, A. P. and Williams, W. L. and Best, P. N. and R{\"{o}}ttgering, H. J.A. and Callingham, J. R. and Dijkema, T. J. and De Gasperin, F. and Hoang, D. N. and Hugo, B. and Mirmont, M. and Oonk, J. B.R. and Prandoni, I. and Rafferty, D. and Sabater, J. and Smirnov, O. and Van Weeren, R. J. and White, G. J. and Atemkeng, M. and Bester, L. and Bonnassieux, E. and Br{\"{u}}ggen, M. and Brunetti, G. and Chy, K. T. and Cochrane, R. and Conway, J. E. and Croston, J. H. and Danezi, A. and Duncan, K. and Haverkorn, M. and Heald, G. H. and Iacobelli, M. and Intema, H. T. and Jackson, N. and Jamrozy, M. and Jarvis, M. J. and Lakhoo, R. and Mevius, M. and Miley, G. K. and Morabito, L. and Morganti, R. and Nisbet, D. and Orr{\'{u}}, E. and Perkins, S. and Pizzo, R. F. and Schrijvers, C. and Smith, D. J.B. and Vermeulen, R. and Wise, M. W. and Alegre, L. and Bacon, D. J. and Van Bemmel, I. M. and Beswick, R. J. and Bonafede, A. and Botteon, A. and Bourke, S. and Brienza, M. and Calistro Rivera, G. and Cassano, R. and Clarke, A. O. and Conselice, C. J. and Dettmar, R. J. and Drabent, A. and Dumba, C. and Emig, K. L. and En{\ss}lin, T. A. and Ferrari, C. and Garrett, M. A. and G{\'{e}}nova-Santos, R. T. and Goyal, A. and G{\"{u}}rkan, G. and Hale, C. and Harwood, J. J. and Heesen, V. and Hoeft, M. and Horellou, C. and Jackson, C. and Kokotanekov, G. and Kondapally, R. and Kunert-Bajraszewska, M. and Mahatma, V. and Mahony, E. K. and Mandal, S. and McKean, J. P. and Merloni, A. and Mingo, B. and Miskolczi, A. and Mooney, S. and Nikiel-Wroczy{\'{n}}ski, B. and O'Sullivan, S. P. and Quinn, J. and Reich, W. and Roskowi{\'{n}}ski, C. and Rowlinson, A. and Savini, F. and Saxena, A. and Schwarz, D. J. and Shulevski, A. and Sridhar, S. S. and Stacey, H. R. and Urquhart, S. and Van Der Wiel, M. H.D. and Varenius, E. and Webster, B. and Wilber, A.},
    month = {2},
    pages = {A1},
    volume = {622},
    publisher = {EDP Sciences},
    url = {https://ui.adsabs.harvard.edu/abs/2019A&A...622A...1S/abstract},
    doi = {10.1051/0004-6361/201833559},
    issn = {14320746},
    arxivId = {1811.07926}
}

@software{Mohan2015PyBDSF:Finder,
       author = {{Mohan}, Niruj and {Rafferty}, David},
        title = "{PyBDSF: Python Blob Detection and Source Finder}",
 howpublished = {Astrophysics Source Code Library, record ascl:1502.007},
         year = 2015,
        month = feb,
          eid = {ascl:1502.007},
archivePrefix = {ascl},
       eprint = {1502.007},
       adsurl = {https://ui.adsabs.harvard.edu/abs/2015ascl.soft02007M}
}

@article{Gordon2020AIdentifications,
    title = {{A Catalog of Very Large Array Sky Survey Epoch 1 Quick Look Components, Sources, and Host Identifications}},
    year = {2020},
    journal = {RNAAS},
    author = {Gordon, Yjan A. and Boyce, Michelle M. and O'Dea, Christopher P. and Rudnick, Lawrence and Andernach, Heinz and Vantyghem, Adrian N. and Baum, Stefi A. and Bui, Jean-Paul and Dionyssiou, Mathew and Gordon, Yjan A. and Boyce, Michelle M. and O'Dea, Christopher P. and Rudnick, Lawrence and Andernach, Heinz and Vantyghem, Adrian N. and Baum, Stefi A. and Bui, Jean-Paul and Dionyssiou, Mathew},
    number = {10},
    month = {10},
    pages = {175},
    volume = {4},
    publisher = {American Astronomical Society},
    url = {https://ui.adsabs.harvard.edu/abs/2020RNAAS...4..175G/abstract},
    doi = {10.3847/2515-5172/ABBE23},
    issn = {2515-5172}
}

@article{An2012TheSources,
    title = {{The dynamic evolution of young extragalactic radio sources}},
    year = {2012},
    journal = {\apj},
    author = {An, Tao and Baan, Willem A.},
    number = {1},
    month = {11},
    pages = {77},
    volume = {760},
    publisher = {Institute of Physics Publishing},
    url = {https://ui.adsabs.harvard.edu/abs/2012ApJ...760...77A/abstract},
    doi = {10.1088/0004-637X/760/1/77},
    issn = {15384357},
    arxivId = {1211.1760}
}

@article{Kharb2016AActivity,
    title = {{A GMRT study of Seyfert galaxies NGC 4235 and NGC 4594: Evidence of episodic activity?}},
    year = {2016},
    journal = {\mnras},
    author = {Kharb, P. and Srivastava, S. and Singh, V. and Gallimore, J. F. and Ishwara-Chandra, C. H. and Ananda, Hota},
    number = {2},
    month = {6},
    pages = {1310--1326},
    volume = {459},
    publisher = {Oxford University Press},
    url = {https://ui.adsabs.harvard.edu/abs/2016MNRAS.459.1310K/abstract},
    doi = {10.1093/mnras/stw699},
    issn = {13652966},
    arxivId = {1603.08364}
}

@article{Silpa2020ProbingI,
    title = {{Probing the origin of low-frequency radio emission in PG quasars with the uGMRT - I}},
    year = {2020},
    journal = {\mnras},
    author = {Silpa, S. and Kharb, P. and Ho, L. C. and Ishwara-Chandra, C. H. and Jarvis, M. E. and Harrison, C.},
    number = {4},
    pages = {5826--5839},
    volume = {499},
    publisher = {Oxford University Press},
    url = {https://ui.adsabs.harvard.edu/abs/2020MNRAS.499.5826S/abstract},
    doi = {10.1093/MNRAS/STAA2970},
    issn = {13652966},
    arxivId = {2009.12264}
}

@article{Nyland2020QuasarsFIRST,
    title = {{Quasars That Have Transitioned from Radio-quiet to Radio-loud on Decadal Timescales Revealed by VLASS and FIRST}},
    year = {2020},
    journal = {\apj},
    author = {Nyland, Kristina and Dong, Dillon Z. and Patil, Pallavi and Lacy, Mark and van Velzen, Sjoert and Kimball, Amy E. and Sarbadhicary, Sumit K. and Hallinan, Gregg and Baldassare, Vivienne and Clarke, Tracy E. and Goulding, Andy D. and Greene, Jenny and Hughes, Andrew and Kassim, Namir and Kunert-Bajraszewska, Magdalena and Maccarone, Thomas J. and Mooley, Kunal and Mukherjee, Dipanjan and Peters, Wendy and Petrov, Leonid and Polisensky, Emil and Rujopakarn, Wiphu and Whittle, Mark and Vaccari, Mattia},
    number = {1},
    month = {12},
    pages = {74},
    volume = {905},
    publisher = {American Astronomical Society},
    url = {https://ui.adsabs.harvard.edu/abs/2020ApJ...905...74N/abstract},
    doi = {10.3847/1538-4357/abc341},
    issn = {0004-637X},
    arxivId = {2011.08872}
}

@article{Kimball2011TheAGN,
    title = {{The Two-Component Radio Luminosity Function of QSOs: Star Formation and AGN}},
    year = {2011},
    journal = {\apjl},
    author = {Kimball, Amy E. and Kellermann, Kenneth I. and Condon, James J. and Ivezic, Zeljko and Perley, Richard A.},
    number = {1},
    month = {7},
    pages = {L29},
    volume = {739},
    url = {http://arxiv.org/abs/1107.3551 http://dx.doi.org/10.1088/2041-8205/739/1/L29},
    doi = {10.1088/2041-8205/739/1/L29},
    arxivId = {1107.3551}
}

@article{Condon2013AGNQSOs,
    title = {{AGN and Starburst Radio Emission from Optically Selected QSOs}},
    year = {2013},
    journal = {ApJ},
    author = {Condon, J. J. and Kellermann, K. I. and Kimball, Amy E. and Ivezic, Zeljko and Perley, R. A.},
    number = {1},
    month = {3},
    pages = {37},
    volume = {768},
    publisher = {Institute of Physics Publishing},
    url = {http://arxiv.org/abs/1303.3448 http://dx.doi.org/10.1088/0004-637X/768/1/37},
    doi = {10.1088/0004-637X/768/1/37},
    arxivId = {1303.3448}
}

@ARTICLE{Rivera2002The2.5,
       author = {{Calistro Rivera}, G. and {Williams}, W.~L. and {Hardcastle}, M.~J. and {Duncan}, K. and {R{\"o}ttgering}, H.~J.~A. and {Best}, P.~N. and {Br{\"u}ggen}, M. and {Chy{\.z}y}, K.~T. and {Conselice}, C.~J. and {de Gasperin}, F. and {Engels}, D. and {G{\"u}rkan}, G. and {Intema}, H.~T. and {Jarvis}, M.~J. and {Mahony}, E.~K. and {Miley}, G.~K. and {Morabito}, L.~K. and {Prandoni}, I. and {Sabater}, J. and {Smith}, D.~J.~B. and {Tasse}, C. and {van der Werf}, P.~P. and {White}, G.~J.},
        title = "{The LOFAR window on star-forming galaxies and AGNs - curved radio SEDs and IR-radio correlation at 0<z<2.5}",
      journal = {\mnras},
         year = 2017,
        month = aug,
       volume = {469},
       number = {3},
        pages = {3468-3488},
          doi = {10.1093/mnras/stx1040},
archivePrefix = {arXiv},
       eprint = {1704.06268},
 primaryClass = {astro-ph.GA},
}

@ARTICLE{Kukreti2024ConnectingAGN,
       author = {{Kukreti}, Pranav and {Morganti}, Raffaella},
        title = "{Connecting the radio AGN life cycle to feedback: Ionised gas is more disturbed in young radio AGN}",
      journal = {\aap},
         year = 2024,
        month = oct,
       volume = {690},
          eid = {A140},
        pages = {A140},
          doi = {10.1051/0004-6361/202450454},
archivePrefix = {arXiv},
       eprint = {2407.06265},
 primaryClass = {astro-ph.GA},
}

@article{Alexander2025WhatProgress,
    title = {{What drives the growth of black holes: A decade of progress}},
    year = {2025},
    journal = {\nar},
    author = {Alexander, D. M. and Hickox, R. C. and Aird, J. and Combes, F. and Costa, T. and Habouzit, M. and Harrison, C. M. and Leng, R. I. and Morabito, L. K. and Uckelman, S. L. and Vickers, P.},
    month = {12},
    pages = {101733},
    volume = {101},
    publisher = {Elsevier B.V.},
    url = {https://ui.adsabs.harvard.edu/abs/2025NewAR.10101733A/abstract},
    doi = {10.1016/j.newar.2025.101733},
    issn = {13876473},
    arxivId = {2506.19166}
}

@article{Blandford1987ParticleOrigin,
    title = {{Particle acceleration at astrophysical shocks: A theory of cosmic ray origin}},
    year = {1987},
    journal = {\physrep},
    author = {Blandford, Roger and Eichler, David},
    number = {1},
    month = {10},
    pages = {1--75},
    volume = {154},
    publisher = {North-Holland},
    doi = {10.1016/0370-1573(87)90134-7},
    issn = {0370-1573}
}

@article{Ward2024AGN-driRelations,
    title = {{AGN-dri v en outflo ws in clumpy media: multiphase structure and scaling relations}},
    year = {2024},
    journal = {MNRAS},
    author = {Ward, S R and Costa, T and Harrison, C M and Mainieri, V},
    pages = {1733--1755},
    volume = {533},
    url = {https://doi.org/10.1093/mnras/stae1816},
    doi = {10.1093/mnras/stae1816}
}

@article{Harrison2018AGNOn,
    title = {{AGN outflows and feedback twenty years on}},
    year = {2018},
    journal = {Nature Astronomy},
    author = {Harrison, C. M. and Costa, T. and Tadhunter, C. N. and Fl{\"{u}}tsch, A. and Kakkad, D. and Perna, M. and Vietri, G.},
    number = {3},
    month = {3},
    pages = {198--205},
    volume = {2},
    publisher = {Nature Publishing Group},
    doi = {10.1038/S41550-018-0403-6},
    issn = {23973366},
    arxivId = {1802.10306}
}

@article{Athreya1998TheGalaxies.,
    title = {{The redshift dependence of spectral index in powerful radio galaxies.}},
    year = {1998},
    journal = {\japa},
    author = {Athreya, Ramana M. and Kapahi, Vijay K.},
    number = {3-4},
    month = {12},
    pages = {63},
    volume = {19},
    publisher = {Indian Academy of Sciences},
    url = {https://ui.adsabs.harvard.edu/abs/1998JApA...19...63A/abstract},
    doi = {10.1007/BF02714911},
    issn = {0250-6335}
}

@article{Kirk1987OnFronts,
    title = {{On the Acceleration of Charged Particles at Relativistic Shock Fronts}},
    year = {1987},
    journal = {ApJ},
    author = {Kirk, J. G. and Schneider, P. and Kirk, J. G. and Schneider, P.},
    month = {4},
    pages = {425},
    volume = {315},
    publisher = {American Astronomical Society},
    url = {https://ui.adsabs.harvard.edu/abs/1987ApJ...315..425K/abstract},
    doi = {10.1086/165147},
    issn = {0004-637X}
}

@ARTICLE{Kardashev1962NonstationarityEmission,
       author = {{Kardashev}, N.~S.},
        title = "{Nonstationarity of Spectra of Young Sources of Nonthermal Radio Emission}",
      journal = {\sovast},
         year = 1962,
        month = dec,
       volume = {6},
        pages = {317},
}

@article{Banerji2012HeavilyEvolution,
    title = {{Heavily reddened quasars at z ∼ 2 in the UKIDSS Large Area Survey: A transitional phase in AGN evolution}},
    year = {2012},
    journal = {\mnras},
    author = {Banerji, Manda and Mcmahon, Richard G. and Hewett, Paul C. and Alaghband-Zadeh, Susannah and Gonzalez-Solares, Eduardo and Venemans, Bram P. and Hawthorn, Melanie J.},
    number = {3},
    month = {12},
    pages = {2275--2291},
    volume = {427},
    publisher = {Oxford University Press},
    url = {https://ui.adsabs.harvard.edu/abs/2012MNRAS.427.2275B/abstract},
    doi = {10.1111/j.1365-2966.2012.22099.x},
    issn = {13652966},
    arxivId = {1203.5530}
}

@article{Njeri2025TheE-MERLIN,
    title = {{The Quasar Feedback Survey: zooming into the origin of radio emission with e-MERLIN}},
    year = {2025},
    journal = {\mnras},
    author = {Njeri, Ann and Harrison, Chris M. and Kharb, Preeti and Beswick, Robert and Calistro-Rivera, Gabriela and Circosta, Chiara and Mainieri, Vincenzo and Molyneux, Stephen and Mullaney, James and Sasikumar, Silpa},
    number = {2},
    month = {1},
    pages = {705},
    volume = {537},
    publisher = {Oxford University Press},
    url = {http://arxiv.org/abs/2501.03433 http://dx.doi.org/10.1093/mnras/staf020},
    doi = {10.1093/mnras/staf020},
    arxivId = {2501.03433}
}

@ARTICLE{Meenakshi2023AFields,
       author = {{Meenakshi}, Moun and {Mukherjee}, Dipanjan and {Bodo}, Gianluigi and {Rossi}, Paola},
        title = "{A polarization study of jets interacting with turbulent magnetic fields}",
      journal = {\mnras},
         year = 2023,
        month = dec,
       volume = {526},
       number = {4},
        pages = {5418-5440},
          doi = {10.1093/mnras/stad3092},
archivePrefix = {arXiv},
       eprint = {2310.03139},
 primaryClass = {astro-ph.HE},
}

@article{Padovani2017ActiveName,
    title = {{Active galactic nuclei: what’s in a name?}},
    year = {2017},
    journal = {\aapr},
    author = {Padovani, P. and Alexander, D. M. and Assef, R. J. and De Marco, B. and Giommi, P. and Hickox, R. C. and Richards, G. T. and Smol{\v{c}}i{\'{c}}, V. and Hatziminaoglou, E. and Mainieri, V. and Salvato, M.},
    number = {1},
    month = {11},
    pages = {2},
    volume = {25},
    publisher = {Springer Verlag},
    url = {https://ui.adsabs.harvard.edu/abs/2017A%26ARv..25....2P/abstract},
    doi = {10.1007/s00159-017-0102-9},
    issn = {09354956},
    arxivId = {1707.07134}
}

@article{Andonie2022ASystems,
    title = {{A panchromatic view of infrared quasars: excess star formation and radio emission in the most heavily obscured systems}},
    year = {2022},
    journal = {MNRAS},
    author = {Andonie, Carolina and Alexander, David M and Rosario, David and ael Laloux, Bri and Georgakakis, Antonis and Morabito, Leah K and Villforth, Carolin and Avirett-Mackenzie, Mathilda and Calistro Rivera, Gabriela and Del Moro, Agnese and Fotopoulou, Sotiria and Harrison, Chris and Lapi, Andrea and Petley, James and Petter, Grayson and Shankar, Francesco},
    pages = {2577--2598},
    volume = {517},
    url = {https://doi.org/10.1093/mnras/stac2800},
    doi = {10.1093/mnras/stac2800}
}

@article{Reyes2008SpaceQuasars,
    title = {{Space density of optically selected type 2 quasars}},
    year = {2008},
    journal = {\aj},
    author = {Reyes, Reinabelle and Zakamska, Nadia L. and Strauss, Michael A. and Green, Joshua and Krolik, Julian H. and Shen, Yue and Richards, Gordon T. and Anderson, Scott F. and Schneider, Donald P.},
    number = {6},
    pages = {2373--2390},
    volume = {136},
    url = {https://ui.adsabs.harvard.edu/abs/2008AJ....136.2373R/abstract},
    doi = {10.1088/0004-6256/136/6/2373},
    issn = {00046256},
    arxivId = {0801.1115}
}

@article{Bicknell2018RelativisticGalaxies,
    title = {{Relativistic jet feedback - II. Relationship to gigahertz peak spectrum and compact steep spectrum radio galaxies}},
    year = {2018},
    journal = {\mnras},
    author = {Bicknell, Geoffrey V. and Mukherjee, Dipanjan and Wagner, Alexander Y. and Sutherland, Ralph S. and Nesvadba, Nicole P.H.},
    number = {3},
    month = {4},
    pages = {3493--3501},
    volume = {475},
    publisher = {Oxford University Press},
    url = {https://ui.adsabs.harvard.edu/abs/2018MNRAS.475.3493B/abstract},
    doi = {10.1093/mnras/sty070},
    issn = {13652966},
    arxivId = {1801.06518}
}

@article{Costa2018DrivingSimulations,
    title = {{Driving gas shells with radiation pressure on dust in radiation-hydrodynamic simulations}},
    year = {2018},
    journal = {\mnras},
    author = {Costa, Tiago and Rosdahl, Joakim and Sijacki, Debora and Haehnelt, Martin G.},
    number = {3},
    month = {5},
    pages = {4197--4219},
    volume = {473},
    publisher = {Oxford University Press},
    url = {http://arxiv.org/abs/1703.05766 http://dx.doi.org/10.1093/mnras/stx2598},
    doi = {10.1093/mnras/stx2598},
    arxivId = {1703.05766v2}
}

@article{Costa2018QuenchingRadiation,
    title = {{Quenching star formation with quasar outflows launched by trapped IR radiation}},
    year = {2018},
    journal = {\mnras},
    author = {Costa, Tiago and Rosdahl, Joakim and Sijacki, Debora and Haehnelt, Martin G.},
    number = {2},
    month = {9},
    pages = {2079--2111},
    volume = {479},
    publisher = {Oxford Academic},
    url = {https://dx.doi.org/10.1093/mnras/sty1514},
    doi = {10.1093/MNRAS/STY1514},
    issn = {0035-8711},
    arxivId = {1709.08638}
}

@ARTICLE{SKA,
       author = {{Dewdney}, P.~E. and {Hall}, P.~J. and {Schilizzi}, R.~T. and {Lazio}, T.~J.~L.~W.},
        title = "{The Square Kilometre Array}",
      journal = {IEEE Proc.},
         year = 2009,
        month = aug,
       volume = {97},
       number = {8},
        pages = {1482-1496},
          doi = {10.1109/JPROC.2009.2021005}
}

@INPROCEEDINGS{ngVLA,
       author = {{Di Francesco}, James and {Chalmers}, Dean and {Denman}, Nolan and {Fissel}, Laura and {Friesen}, Rachel and {Gaensler}, Bryan and {Hlavacek-Larrondo}, Julie and {Kirk}, Helen and {Matthews}, Brenda and {O'Dea}, Christopher and {Robishaw}, Tim and {Rosolowsky}, Erik and {Rupen}, Michael and {Sadavoy}, Sarah and {Sa-Harb}, Samar and {Sivakoff}, Greg and {Tahani}, Mehrnoosh and {van der Marel}, Nienke and {White}, Jacob and {Wilson}, Christine},
        title = "{The Next Generation Very Large Array}",
    booktitle = {Canadian Long Range Plan for Astronomy and Astrophysics White Papers},
         year = 2019,
       volume = {2020},
        month = oct,
          eid = {32},
        pages = {32},
 primaryClass = {astro-ph.IM}
}

@ARTICLE{unstablejets1,
       author = {{Bromberg}, Omer and {Tchekhovskoy}, Alexander},
        title = "{Relativistic MHD simulations of core-collapse GRB jets: 3D instabilities and magnetic dissipation}",
      journal = {\mnras},
         year = 2016,
        month = feb,
       volume = {456},
       number = {2},
        pages = {1739-1760},
          doi = {10.1093/mnras/stv2591},
archivePrefix = {arXiv},
       eprint = {1508.02721},
 primaryClass = {astro-ph.HE}
}

@ARTICLE{unstablejets2,
       author = {{Bodo}, G. and {Mamatsashvili}, G. and {Rossi}, P. and {Mignone}, A.},
        title = "{Current-driven kink instabilities in relativistic jets: dissipation properties}",
      journal = {\mnras},
         year = 2022,
        month = feb,
       volume = {510},
       number = {2},
        pages = {2391-2406},
          doi = {10.1093/mnras/stab3492},
archivePrefix = {arXiv},
       eprint = {2111.14575},
 primaryClass = {astro-ph.HE}
}




\appendix
\section{Non-detected sources}
\label{sec:undetected}
Of the 2171 FIRST-detected cQSOs and 1175 FIRST-detected rQSOs, we find that 277 cQSOs and 126 rQSOs have no LoTSS catalogue counterpart, and 294 cQSOs and 233 rQSOs have no VLASS catalogue counterpart, within 5$''$ of the optical (SDSS) position. We use the following procedure to search for 3$\sigma$ detections or upper limits:

\noindent \textbf{LoTSS 3$\sigma$ search:}
\begin{itemize}[align=parleft,left=0pt]
    \item We use $2'\times 2'$ images centred on each QSO position, and use the Python Blob Detection and Source Finder \citep[PyBDSF;][]{Mohan2015PyBDSF:Finder}, which we set to detect islands of flux at $3\sigma$ above the image mean, and then fit Gaussian components within these islands at $3\sigma$ above the image mean. We use the default PyBDSF parameters. The output is a list of individual radio components including radio flux density and position information, for which we cross match to our remaining QSOs using a 5$''$ radius, finding 3$\sigma$ radio counterparts for 105 cQSOs and 61 rQSOs.
    \item For QSOs still undetected in the radio images (172 cQSOs and 65 rQSOs), we use rms image cutouts to calculate the median rms within a 2$'$ radius of the QSO position, and calculate a $3\sigma$ upper limit on the flux density as 3 $\times$ rms.
\end{itemize}
\textbf{VLASS 3$\sigma$ search:}
\begin{itemize}[align=parleft,left=0pt]
    \item We again use PyBDSF in the same way as above, except using slightly different parameter settings for calculating the rms in the fitting process, to be consistent with the settings used to create the VLASS component catalogue  \citep{Gordon2020AIdentifications}. We find $3\sigma$ detections for 24 cQSOs and 10 rQSOs.
    \item We calculate $3\sigma$ flux density upper limits for the remaining 270 cQSOs and 223 rQSOs. Our high proportion of VLASS upper limits is perhaps not surprising, given the similar sensitivities of FIRST and VLASS. The higher frequency of VLASS means a source only just detected at the catalogue  limit of FIRST (1 mJy) at 1.4 GHz, with a typical spectral slope of $\alpha = -0.7$, would have a flux density of 0.6 mJy at 3 GHz in VLASS, equivalent to the $5\sigma$ threshold of 0.6 mJy. 
\end{itemize}

As mentioned in Section~\ref{sec:selection}, only a small fraction of non-detections in LoTSS and VLASS are actually included in the final bright ($F_{1.4\text{GHz}}>3$ mJy) sample used in our analyses. An even smaller fraction remain after the visual inspection and quality cuts (see Section~\ref{sec:VI} and Appendix~\ref{sec:app_VI}).

\section{Visual inspection \& quality cuts}
\label{sec:app_VI}
\begin{figure*}
\centering
\includegraphics[width=0.8\textwidth]{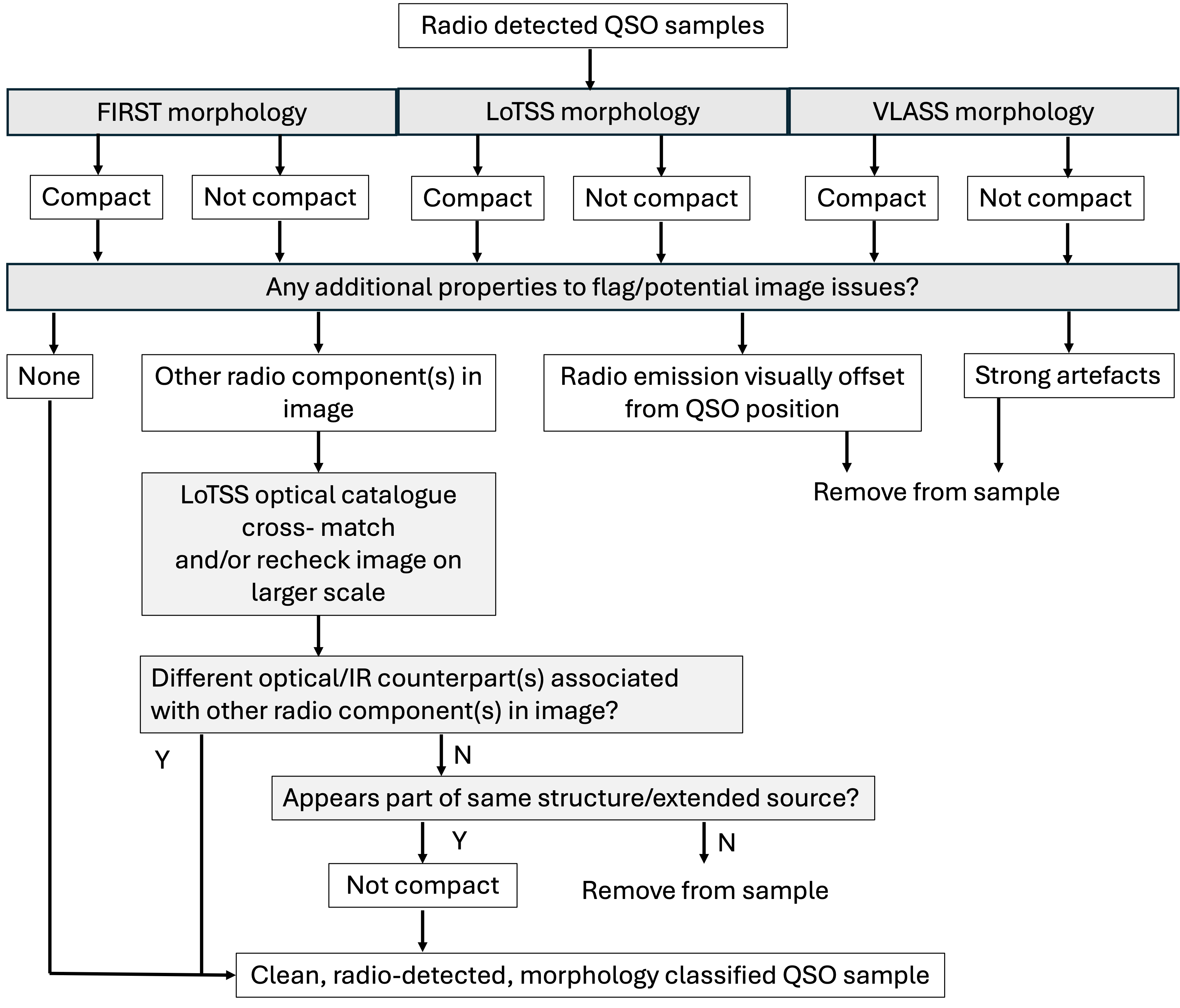}
\caption{A flowchart showing our visual inspection procedure for radio-detected QSOs with FIRST peak flux densities $>3$ mJy. We simultaneously visually inspect the $2' \times 2'$ radio images of each QSO in all three surveys (FIRST; LoTSS; VLASS), and assign a classification in each image as either compact or not compact (showing some level of extended emission). If the image shows strong artefacts or the radio emission is offset from the QSO position, we remove the QSO from our sample. If the image shows other radio counterparts that appear unassociated with the QSO, we check whether other optical/IR sources are associated with them, or recheck our image on a larger scale. If the other radio components appear unassociated to our QSO, or they appear as part of an extended structure around our QSO, we reclassify accordingly and add the QSO to our clean morphology-classified sample. If the other radio components are not associated to other sources, and do not appear as part of the same extended structure, we remove the QSO from our sample. If there are no flags, we are confident of our classification and add the QSO to our clean morphology-classified sample.}
\label{fig:VI_flowchart}
\end{figure*}

The flowchart in Figure \ref{fig:VI_flowchart} outlines our visual inspection and radio morphology classification procedure. The primary aim was to produce a clean sample of compact sources, so the visual inspection (and particularly the discarding of sources with uncertain classifications or strong artefacts) was constructed with that in mind. For each QSO, we classified the radio morphology in each of the three radio images by simultaneously viewing the $2' \times 2'$ cutout images in each survey, centered on the QSO optical positions. Whilst each QSO was given a classification in each of the three radio images, we used information from each image to influence the classification in certain cases. For a scenario where we saw contiguous extended emission in the LoTSS image from a radio lobe connected to the core emission at the QSO position, but which appears in the FIRST image as more than one component where the additional component(s) is/are not connected, we classified this as non-compact in both FIRST and LoTSS, since the additional information from LoTSS indicates they are associated. An example of this case is shown in the bottom panel of Figure~\ref{fig:VI_example}. However, we only did this if we could actually see the emission in FIRST and it broadly followed the same shape as in LoTSS. 

We flagged up cases where the images made it difficult for us to classify, or required additional checks to ensure our classification was correct. These included a flag for images where there appeared to be strong imaging artefacts (often around very bright sources), that limited our ability to determine whether or not the source was compact, and so we removed these from the sample. A flag was also given if the radio emission appeared offset from the centre of the image (centred on the QSO position), which could indicate an issue with the position, or could be an extended source. Since we cannot confidently distinguish between these, we removed them from the sample. 

We also flagged images that had another radio component(s) in the image that did not appear to be associated with the radio emission from the QSO, but we couldn't be sure that it was not. In the images that we identified these other radio components they always appeared in the LoTSS image (due to the better sensitivity), but not necessarily in the FIRST or VLASS images. In order to determine if the other radio component was associated with the QSO or not, we either looked at the image on a larger scale to see if it was associated with another source outside our image region, and/or we made use of the LoTSS-optical catalogue \citep{LoTSSopt} to see if there was an optical/infrared detection associated with the other radio component(s). If there was, we considered the source to be unassociated with our QSO, and included it in our clean sample. If there was no optical detection associated with the other radio component, we used the resolved flag and angular size information from the LoTSS-optical catalogue to determine if the components were associated. If they were, we classified the source as non-compact and included it in our clean sample. If they were not associated according to this criteria, we could not confidently determine the morphology classification of the source and we did not include it in our clean sample.

The fractions of rQSOs and cQSOs with the various flags is consistent, at 148/1426 ($10.1\%$) of the cQSOS and 59/632 ($9.3\%$) of the rQSOs sent for visual inspection, suggesting this approach is not biasing our morphology classification results. If an image was not flagged, this meant we were confident with the classification and it was included in the clean sample that we used for our morphology and radio spectral slope analysis, which consisted of 90\% of the original sample selected for visual inspection.

\begin{figure*}
\centering
\includegraphics[width=1\textwidth]{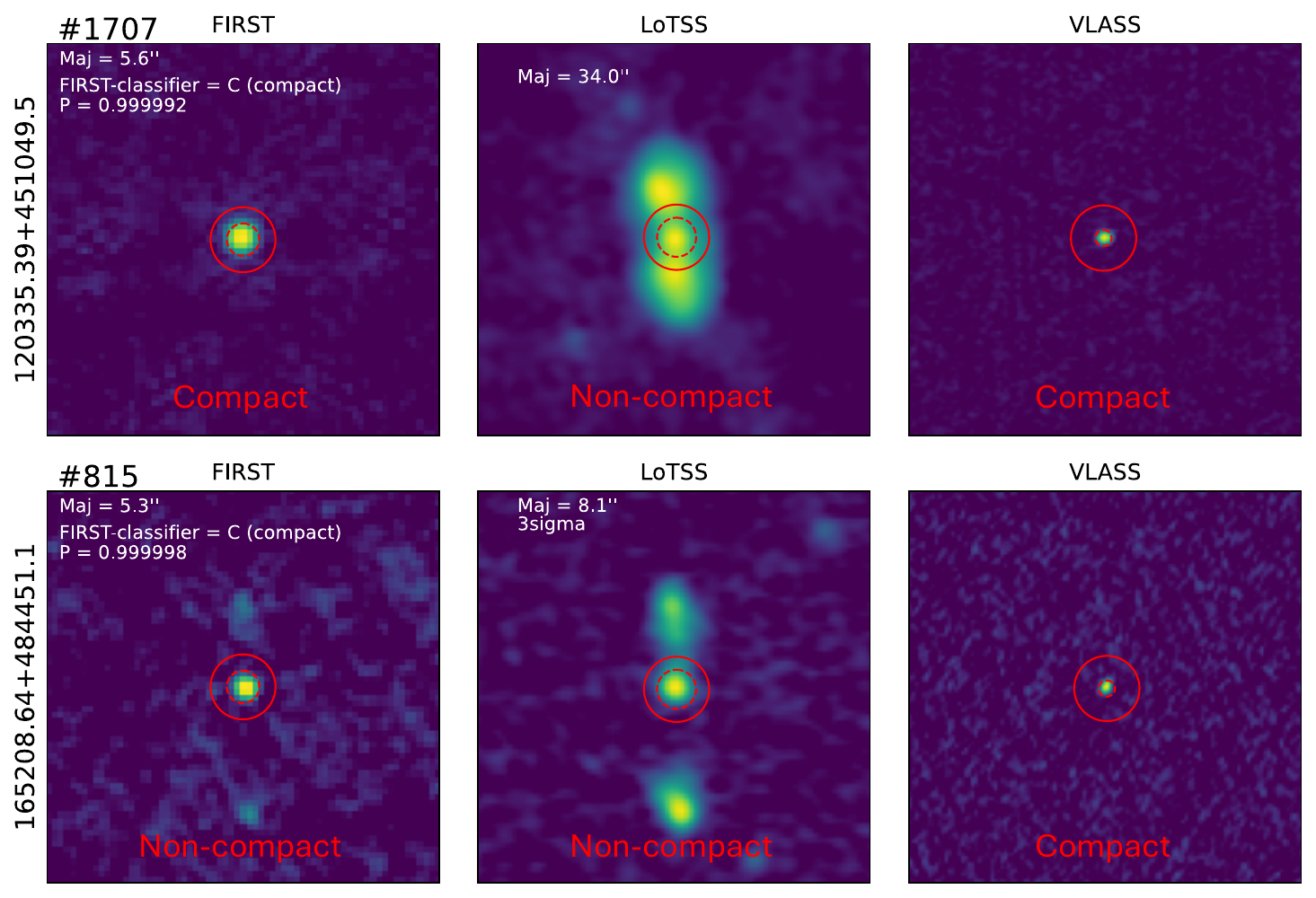}
\caption{An example of two different sources from our visual inspection procedure (top and bottom panels). We show the $2' \times 2'$ cutout images, centred on the QSO position, in each survey simultaneously, and classify the source in each as either compact or non-compact. The inner dashed circles show the resolutions of the respective surveys, of $5''$ (FIRST), $6''$ (LoTSS), and $2''.5$ (VLASS). The outer solid circles are a radius of $10''$. The “Maj” label in the FIRST and LoTSS panels represents the catalogue value for the major axis of the fitted Gaussian to the source. \textit{Top panel:} A QSO that is classified as compact in FIRST and VLASS, but clearly shows extended emission in LoTSS, and is therefore classified as non-compact in LoTSS, i.e., a "fake compact" QSO. \textit{Bottom panel:} A QSO that is classified as non-compact in both FIRST and LoTSS. The additional radio components in the FIRST image do not appear connected to the central QSO. However, using the information from the LoTSS image, we see contiguous extended emission connecting the components, and thus we know the components are connected and classify it as non-compact in both FIRST and LoTSS. We do not see any additional radio components in the VLASS image, thus it is classified as compact.}
\label{fig:VI_example}
\end{figure*}

\section{Impact of sensitivity on fake compact classification}
\label{sec:app_fake}
\begin{figure*}
\centering
\includegraphics[width=0.9\textwidth]{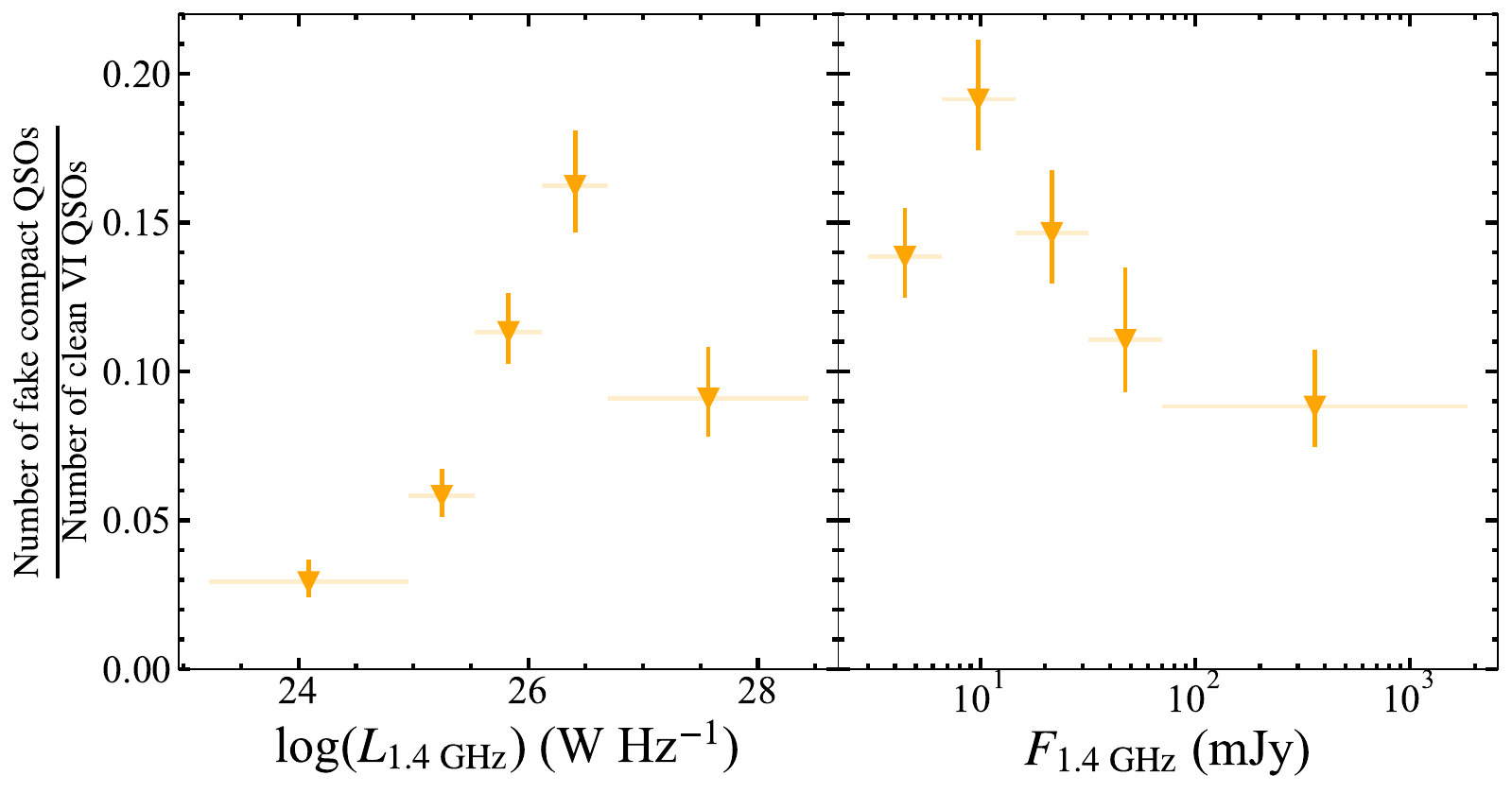}
\caption{The fraction of fake compact QSOs (divided by the total number of QSOs we visually inspected that were in the clean sample) as a function of 1.4 GHz radio luminosity (left panel) and 1.4 GHz peak flux (right panel). Vertical error bars are $1\sigma$ binomial uncertainties, and horizontal error bars show the bin range. \textit{Left panel:} The largest fraction of fake compact sources occurs at intermediate luminosities, not at the lowest or highest luminosities. Since it should be easier to identify extended radio emission in high-luminosity sources than in low-luminosity sources, this suggests that the fake compact sample does not have a significant contamination of truly extended radio sources (i.e., those extended at both low and high frequencies) at low luminosities. \textit{Right panel:} 
The fraction of fake compact sources in the faintest flux bin is broadly consistent with that found for the brighter flux bins, except for the brightest flux bin. This suggests that the fake compact sources do not suffer significant contamination from truly extended sources at the faintest radio fluxes.}
\label{fig:fake_compact}
\end{figure*}

Our visual inspection and the classification of sources into truly compact and fake compact is potentially limited by the sensitivity of our radio observations. In particular, for the fake compact QSOs, it is possible that extended emission exists at higher frequencies but is too faint to be detected in FIRST or VLASS. Figure~\ref{fig:fake_compact} shows the fraction of fake compact sources (compared to all visually inspected, clean sources) as a function of both the radio luminosity (left panel) and radio flux (right panel). If sensitivity effects have a large impact on the fake compact definition then we would expect the lowest-luminosity sources to have the largest fraction of fake compact sources and the highest-luminosity sources to have the smallest fraction. The left panel of Figure~\ref{fig:fake_compact} shows that the largest fraction of fake compact sources occurs at intermediate luminosities, not at the lowest or highest luminosities. Since it should be easier to identify extended radio emission in high-luminosity sources than in low-luminosity sources, this suggests that the fake compact sample does not have a significant contamination of truly extended radio sources (i.e., those extended at both low and high frequencies) at low luminosities. 

The right panel of Figure~\ref{fig:fake_compact} shows the fraction of fake compact sources as a function of flux, which will remove some potential luminosity effects (e.g., due to the wide redshift ranges of our radio-detected QSOs). The fraction of fake compact sources in the faintest flux bin, which would be most impacted by sensitivity, is broadly consistent with that found for the brighter flux bins, except for the brightest flux bin. This further supports the view that the fake compact sources do not suffer significant contamination from truly extended sources at the faintest radio fluxes. This result is perhaps not surprising given we only investigate sources in this study with FIRST fluxes of $F_{1.4\text{ GHz}}>3$~mJy, which corresponds to a S/N $\gtrsim 15$ for the nominal rms ($\sim 0.15$ mJy) of the FIRST images.

\section{Faint sources and spectral slope incompleteness}
\label{sec:app_faint}
Due to the different sensitivities of the three radio surveys, there are regions across the radio spectral slope plane that cannot be directly probed at the faintest FIRST fluxes. Figure~\ref{fig:alpha_vs_flux} shows the radio spectral slopes between both FIRST$-$VLASS (left panel) and LoTSS$-$FIRST (right panel), as a function of the FIRST 1.4 GHz peak flux density, for the truly compact QSOs and the faint QSOs. The black solid lines and grey regions show the minimum (maximum) spectral slope value we can measure at the 3$\sigma$ flux density limit of 0.36 mJy (0.25 mJy) for VLASS (LoTSS). As we go to lower FIRST flux densities, the range of spectral slope values we can directly probe through source detection decreases. Although we can still calculate an upper/lower limit on the spectral slope for these sources, this can introduce some biases into the average slopes we infer at these flux densities, as the upper/lower limits are not necessarily representative of the true range of spectral slopes. The impact of this limitation is reduced in our study due to the bright flux density cut of $F_{1.4\text{GHz}} > 3$ mJy, which is set to allow for visual inspection but also ensures we can probe most of the available FIRST-VLASS and FIRST-LoTSS spectral slopes even for our faintest sources. At the VLASS (LoTSS) limit, a QSO with a FIRST flux density of 3 mJy must have an extreme spectral slope of $\alpha <-2.78$ ($\alpha > 1.09$) to be undetected in VLASS (LoTSS). This corresponds to the 0.1st (98.9th) percentile of all measured spectral slopes, and so above this 3 mJy flux density cut, we are essentially complete to the full distribution of spectral slopes.

The differences in the survey sensitivities has a more severe impact on the directly measurable spectral slopes of the faint population ($F_{1.4\text{GHz}} < 3$ mJy). However, we test whether the spectral slopes of the faint sources are different than expected from the bright population. This is estimated from the spectral slope distributions of the bright sources, taking into account the amount of incompleteness (grey regions) toward lower flux densities. The red and blue solid lines in the faint regions in Figure~\ref{fig:alpha_vs_flux} are the predicted median spectral slope values as a function of flux density. The diamond markers in the faint region show the observed median spectral slope values in three flux density bins. The observed medians are generally in agreement with the predicted medians, indicating that the spectral slopes of the rQSOs remain steeper than the cQSOs even in the faint regime. We note however that the faint classification are too faint to visually inspect and so there is potentially contamination from  extended sources.

\begin{figure*}
\centering
\includegraphics[width=1\textwidth]{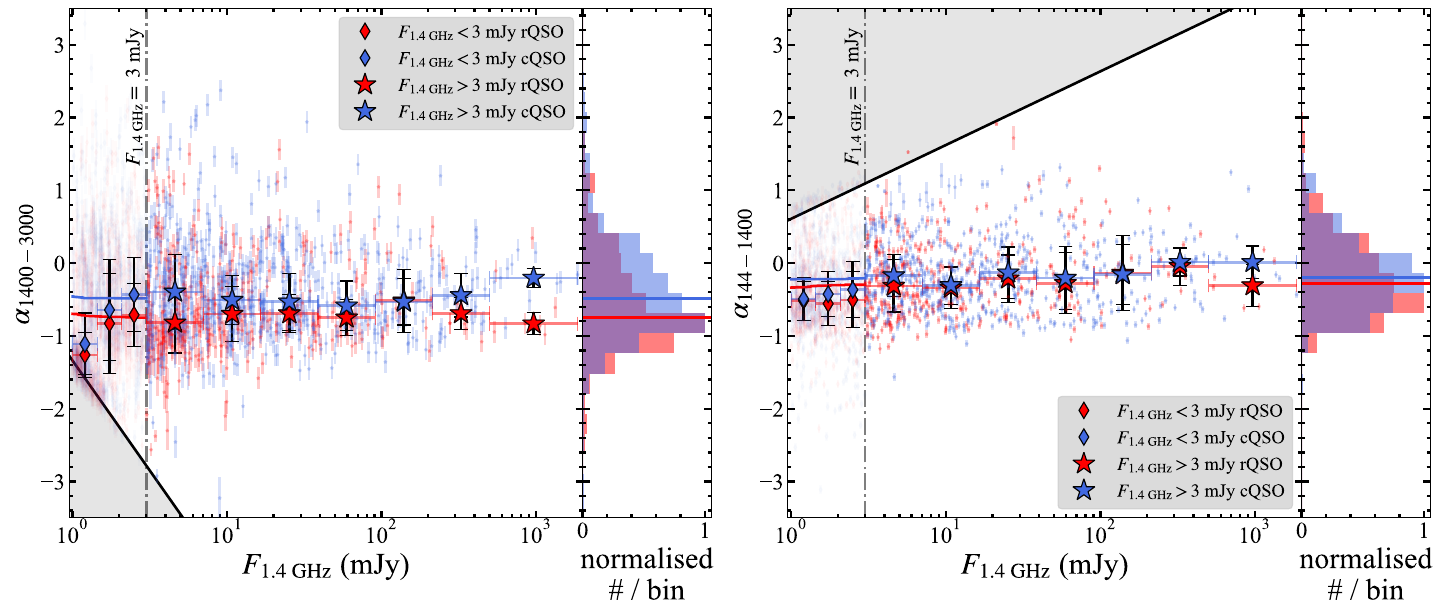}
\caption{The radio spectral slope between (left panel) FIRST and VLASS and (right panel) LoTSS and FIRST as a function of the FIRST peak flux density, for the rQSOs and cQSOs. Above 3 mJy all plotted QSOs are truly compact, and below 3 mJy they are all faint QSOs, due to the lack of reliable FIRST morphological information. The black solid lines indicate the maximum (minimum) detectable spectral slope value at
the canonical 3$\sigma$ catalogue depth of 0.25 mJy for LoTSS (0.36 mJy for VLASS), and the grey regions highlight the values we therefore cannot (in theory) measure (assuming the catalogue depth is the same across the whole of the survey area). The large star markers indicate the median spectral slope values in seven flux density bins above $F_{1.4\text{GHz}} > 3$ mJy, with error bars indicating $\pm1$ MAD. The diamond markers below $F_{1.4\text{GHz}} < 3$ mJy show the measured median spectral slope value in three flux density bins, and the solid lines in this region represent the predicted medians. The histograms on each plot
show the distributions of the spectral slopes above $F_{1.4\text{GHz}} > 3$ mJy only, with the median values indicated by the solid lines.}
\label{fig:alpha_vs_flux}
\end{figure*}

\section{Radio colour colour plots}
\label{sec:app_CC}
\begin{figure*}
\centering
\includegraphics[width=1\textwidth]{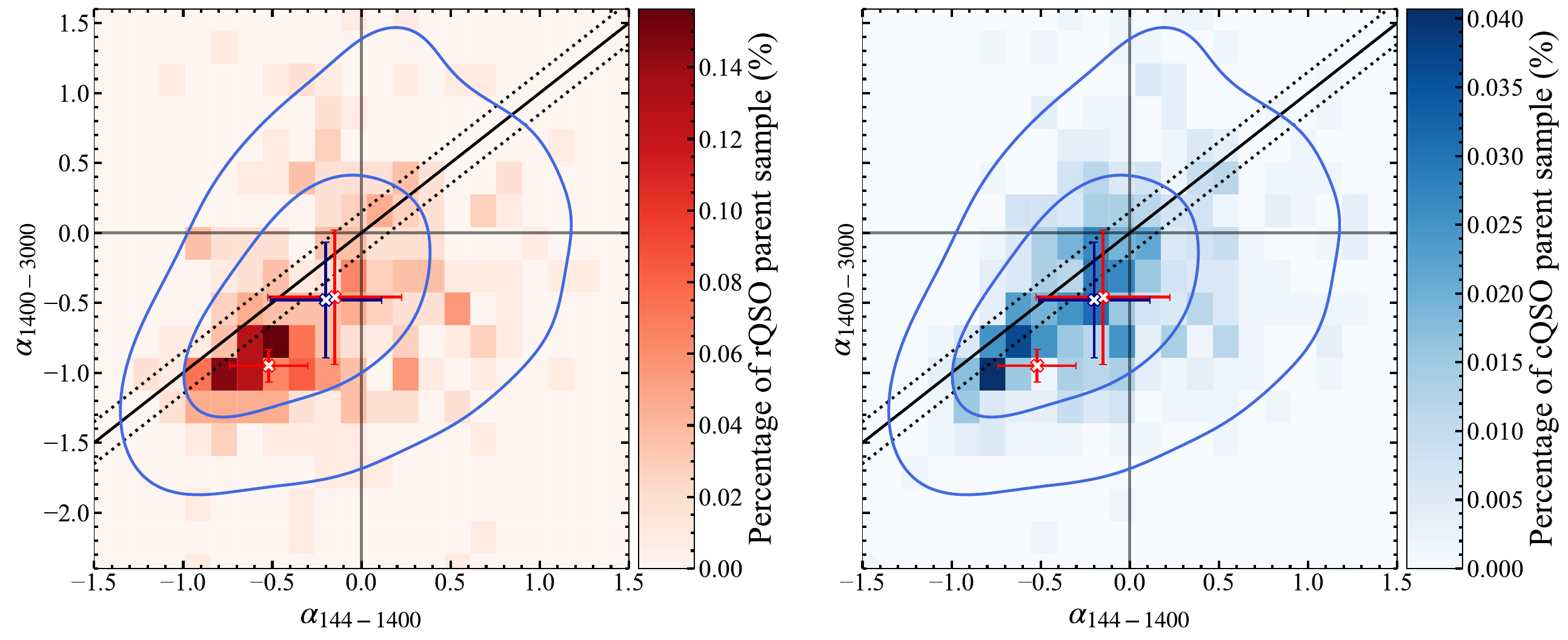}
\caption{The percentage of truly compact radio-detected rQSOs (left panel) and cQSOs (right panel) across the $\alpha_{1400-3000}$--$\alpha_{144-1400}$ radio spectral slope plane, calculated from the respective colour-selected parent samples. Essentially a 2-dimensional illustration of the Figures~\ref{fig:comb_hists} and~\ref{fig:alphaLF} histograms is shown; the red/white crosses indicate the average $\alpha_{1400-3000}$--$\alpha_{144-1400}$ spectral slopes of the narrow and broad histograms for the rQSOs, while the blue/white cross shows the average $\alpha_{1400-3000}$--$\alpha_{144-1400}$ spectral slopes of the cQSO histograms. The crosses are the same in both panels, and the error bars represent the median absolute deviations on the distributions. Note the colour scales for the detection fraction of rQSOs and cQSOs are not the same. The blue contours indicate a detection rate (percentage of parent sample) of 0.1\% and 0.05\% for the truly compact cQSOs. The black lines indicate a 1:1 power-law relation (solid line) with a tolerance determined from the median spectral slope error (dotted lines). The majority of the truly compact radio-detected rQSOs and cQSOs reside in the bottom left quadrant, corresponding to steep radio spectral slopes. However, the rQSOs deviate from the power-law relationship and have an excess of sources with steeper radio spectral slopes than the cQSOs.}
\label{fig:cc_det}
\end{figure*}

To demonstrate the combined spectral slope differences between truly compact rQSOs and cQSOs in Figure~\ref{fig:cc_det} we show the radio-detection rate across the $\alpha_{1400-3000}$--$\alpha_{144-1400}$ spectral slope plane. This gives an indication of the most common shapes of the broad-band radio SED, though we note that the spectral slope from LoTSS (144 MHz) to FIRST (1.4 GHz) covers a decade in frequency, and so the true radio SED shape could be more complex and deviate from the simplistic power-law shape implied from this basic characterisation. For example this analysis would not reveal a peak in the radio SED between 144 MHz to 1.4 GHz. The solid diagonal line shows a 1:1 relation for slopes at both frequencies (ie. a power-law), and the dotted lines either side indicate the tolerance on this line of equality, taken from the median uncertainties on $\alpha_{144-1400}$ and $\alpha_{1400-3000}$. Significant deviations from the equality line suggest clear spectral complexity (e.g., a curved SED).

The left panel in the figure shows that the truly compact rQSOs are preferentially detected near to the steep power law relation, but with some curvature towards steeper high-frequency slopes. The white-red crosses indicate the mean of the narrow (lower left point) and broad fitted Gaussians in $\alpha_{1400-3000}$, and the weighted medians of the $\alpha_{144-1400}$ values to those with steep $\alpha_{1400-3000}$ and those not. The white-blue cross indicates the mean of Gaussian fit to $\alpha_{1400-3000}$ and the median $\alpha_{144-1400}$ for the cQSOs. The "base" of the truly compact rQSOs is similar to the cQSOs (where the crosses overlap), but there is the clear extra population of truly compact rQSOs (indicated by the lower left red cross) with high detection rates around a broken power-law type model. The blue contours on both panels indicate a detection rate of 0.1\% and 0.05\% for the cQSOs. The cQSOs have a much broader spread of values across the observed parameter space, and are generally more consistent with a power-law SED, showing less curvature. However, it it is quite possible that there is a wide range of different radio SED shapes within each QSO population and intermediate-frequency radio observations (e.g., using uGMRT at 400 and 650 MHz) are required to reliably characterise the SED shape (e.g., \citealp{Fawcett2025ConnectionQSOs}).

\section{Estimating decadal radio variability}
\label{sec:app_var}
Since the three radio surveys used in this work were observed at different epochs, this means the observed spectral slope values will potentially include both effects due to the variability of the fluxes between the surveys, and the "intrinsic" spectral slopes of the sources. 

To check that the differences in the average radio spectral slopes of rQSOs and cQSOs are not driven by differences in average variability, we estimate the variability in the tens of years between FIRST (1995-2011) and VLASS (2020$-$) surveys. Single epoch continuum images have so far been
produced for $\sim1000$ deg$^2$ of Epoch 2 of VLASS. The single epoch continuum images are of better quality than the Quick Look images and include in-band ($2-4$ GHz) spectral slope measurements. This provides an instantaneous measure of the spectral slope, which we can use to estimate 1.4 GHz flux densities to compare to our measured FIRST 1.4 GHz flux densities. We are only able to perform this analysis for a small subset of our sample: 22 rQSOs (19 truly compact, 3 fake compact) and 47 cQSOs (35 truly compact, 12 fake compact) that are FIRST and VLASS compact and have an in-band spectral slope measurement that is reliable, using \texttt{Alpha\char`_quality\char`_flag == 1}.

We estimate the 1.4 GHz flux density at the epoch of VLASS, by assuming that the spectral slope is the same from $2-4$ GHz to 1.4 GHz, via \begin{equation} F_{\text{1.4 GHz, estimated}} = F_{\text{3 GHz}}\left(\frac{1.4}{3.0}\right)^{\alpha_{2-4}}. \end{equation} Figure~\ref{fig:var} shows this estimated flux density against the actual 1.4 GHz flux density measured in FIRST, with the solid black line indicating a 1:1 ratio (no change in flux density between the two surveys), and the grey lines showing changes in flux density from $0.25\times$ to $4\times$ the original flux density. We do not find any significant offsets in the average ratio of the flux densities, with a median ratio ($\pm 1$ MAD) for the rQSOs of $0.99\pm0.21$, and for the cQSOs of $0.98\pm0.18$, and a KS test between the rQSO and cQSO distributions resulting in a $p$-value = 0.45, indicating no significant differences. Separating into the truly and fake compact morphology classifications, we find they are all in agreement with a ratio of unity other than the fake compact rQSOs which have a median ratio of $0.73\pm0.07$, but we note there are only three fake compact rQSOs in this analysis. Although we can only perform this analysis for a small subset of our samples, we find no indication that any differences in the variability of rQSOs and cQSOs impacts our spectral slope of spectral shape analysis. We do note however that any individual source may show significant variability.

\begin{figure}
\centering
\includegraphics[width=1\columnwidth]{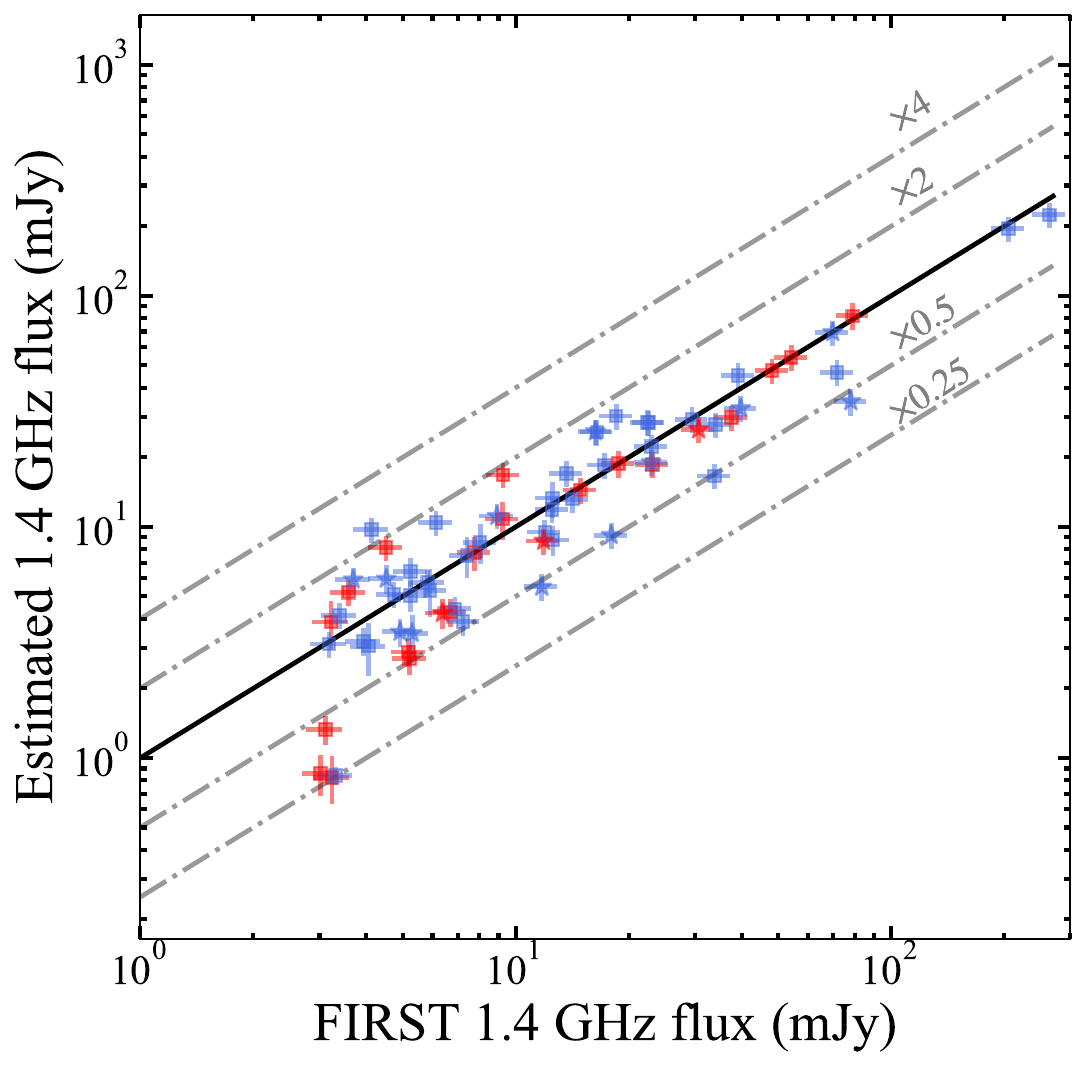}
\caption{The estimated 1.4 GHz flux at the epoch of VLASS, calculated using the 3 GHz VLASS flux and in-band ($2-4$ GHz) VLASS spectral slope, plotted against the actual 1.4 GHz flux from FIRST. The dash-dot lines show the fractional change in flux between the two measurements ($\sim$ a decade apart), and the solid line is a 1:1 ratio (no estimated change in flux over this timescale). There are no significant differences in the variability of rQSOs and cQSOs over a decadal timescale.}
\label{fig:var}
\end{figure}


\bsp	
\label{lastpage}
\end{document}